\documentclass[12pt, preprint]{aastex}
\usepackage{epsfig}
\usepackage{natbib}
\usepackage{amsmath}
\shorttitle{HR 2142}
\shortauthors{Peters et al.}

\newcommand{\noprint}[1]{}
\newcommand{\figsetstart}{{\bf Fig. Set} }
\newcommand{\figsetend}{}
\newcommand{\figsetgrpstart}{}
\newcommand{\figsetgrpend}{}
\newcommand{\figsetnum}[1]{{\bf #1.}}
\newcommand{\figsettitle}[1]{ {\bf #1} }
\newcommand{\figsetgrpnum}[1]{\noprint{#1}}
\newcommand{\figsetgrptitle}[1]{\noprint{#1}}
\newcommand{\figsetplot}[1]{\noprint{#1}}
\newcommand{\figsetgrpnote}[1]{\noprint{#1}}
\begin{document}

\received{}
\accepted{}
\setcounter{footnote}{0}

\title{The Hot Companion and Circumbinary Disk \\ of the Be Star HR 2142}

\author{Geraldine J. Peters\altaffilmark{1,2}} 
\affil{Space Sciences Center, University of Southern California, Los Angeles, CA 90089-1341, USA;  
gpeters@usc.edu} 
 
\author{Luqian Wang, Douglas R. Gies} 
\affil{Center for High Angular Resolution Astronomy and  
 Department of Physics and Astronomy,\\ 
 Georgia State University, P. O. Box 5060, Atlanta, GA 30302-5060, USA; \\ 
 lwang@chara.gsu.edu, gies@chara.gsu.edu} 
 
\author{Erika D. Grundstrom\altaffilmark{2}}
\affil{Physics and Astronomy Department,
 Vanderbilt University, 6301 Stevenson Center, Nashville, TN 37235, USA; \\
 erika.grundstrom@vanderbilt.edu}

\altaffiltext{1}{Guest Observer with the {\it International Ultraviolet Explorer 
Satellite}.} 
 
\altaffiltext{2}{Visiting Astronomer, Kitt Peak National Observatory,
National Optical Astronomy Observatory, operated by the Association
of Universities for Research in Astronomy, Inc., under contract with
the National Science Foundation.}

\slugcomment{Version 2; 05/31/2016} 
\paperid{AAS01042}

%%%%%%%%%%%%%%%%%%%%%%%%%%%%%%%%%%%%%%%%%%%%%%%%%%%%%%%%%%%%%%%%%%%%%%%%%%%%

\begin{abstract}
We present a spectroscopic investigation of the Be+sdO binary system HR~2142
that is based upon large sets of ultraviolet observations from the 
{\it International Ultraviolet Explorer} and ground-based H$\alpha$ observations. 
We measured radial velocities for the Be star component from these spectra, 
and computed a revised orbit.   In order to search for the spectral signature
of the hot subdwarf, we cross-correlated the short wavelength end of each 
{\it IUE} spectrum with a model hot star spectrum, and then we used the 
predicted Doppler shifts of the subdwarf to shift-and-add all the 
cross-correlation functions to the frame of the subdwarf.   This merged 
function shows the weak signal from the spectral lines of the hot star, 
and a best fit is obtained with a mass ratio $M_2/M_1 = 0.07 \pm 0.02$, 
companion temperature $T_{\rm eff} \geq 43\pm5$~kK, projected rotational 
velocity $V\sin i < 30$ km~s$^{-1}$, and a monochromatic flux ratio 
near 1170 \AA ~of $f_2/f_1 > 0.009 \pm 0.001$. 
This hot subdwarf creates a one-armed spiral, tidal wake in the disk of the Be star, 
and we present a circumbinary disk model that can explain the occurrence 
of shell absorption lines by gas enhancements that occur where gas crossing 
the gap created by the subdwarf strikes the disk boundaries.   
The faint companion of HR~2142 may be representative of a significant fraction 
of Be stars with undetected {\bf former} mass donor companion stars. 
\end{abstract}

 
\setcounter{footnote}{2}

%%%%%%%%%%%%%%%%%%%%%%%%%%%%%%%%%%%%%%%%%%%%%%%%%%%%%%%%%%%%%%%%%%%%%%%%%%%%

\section{Introduction}                                           % Section 1

Close binary stars are progressively more common among stars of higher mass
\citep{duchene2013}, so we expect that binary interactions will play 
a fundamental role in the evolution of a significant fraction of massive stars
\citep{demink2014}.  Roche lobe overflow (RLOF) in close binaries with 
extreme mass ratios may lead to mergers or a common envelope stage, 
but in systems with comparable mass stars, RLOF can proceed in a stable 
manner in which the mass donor star eventually becomes stripped down 
to a much lower mass \citep{wellstein2001}.  RLOF provides mass and 
angular momentum to the mass gainer star, so that products of binary 
interaction may appear as a population of rapidly rotating stars 
\citep{demink2014}.  \citet{pols1991} first suggested that such mass
gainers might be associated with the rapidly rotating Be stars, emission-line 
stars surrounded by out-flowing circumstellar disks \citep{rivinius2013}. 
This idea is supported by the fact that most of the massive X-ray binaries 
consist of a Be star with an orbiting neutron star companion, the remnant 
of a massive donor star \citep{reig2011}.  However, mass donors below the 
Chandrasekhar limit will appear as hot subdwarfs or white dwarfs, and 
such systems with faint companions may account for a large fraction of 
Be stars \citep{shao2014}. 

Detection of such faint, subdwarf companions to Be stars is difficult because 
of the huge flux contrast between the Be star and companion.  The stripped down 
donor stars are expected to be much hotter than the Be stars, so detection is
favored in the ultraviolet part of the spectrum where the flux ratio increases. 
In fact, the first {\bf direct spectroscopic} detection of a hot subdwarf companion of the Be star 
$\phi$~Per was made by \citet{thaller1995} through investigation of 
ultraviolet (UV) spectra made with the {\it International Ultraviolet Explorer (IUE)}
satellite.  The nature of the companion was revealed in subsequent UV  
spectroscopy from {\it Hubble Space Telescope} \citep{gies1998}, and recent 
interferometric observations with the CHARA Array have led to an astrometric 
orbit and masses for the $\phi$~Per system \citep{mourard2015}. 
Encouraged by the UV detection of the companion of $\phi$~Per, we subsequently 
launched an investigation of the UV spectra of Be binaries that were 
frequently observed by {\it IUE}, and our work led to the 
discovery of the hot subdwarf companions of FY~CMa \citep{peters2008} and 
59~Cyg \citep{peters2013}.  The presence of a hot companion in $\phi$~Per
was first suggested by emission line variations caused by heating of the 
Be star's disk gas by the hot companion \citep{poeckert1981,hummel2001}, and 
observations of similar emission line variations may indicate the presence of 
hot subdwarf companions to the Be stars $o$~Pup \citep{koubsky2012} and 
HD~161306 \citep{koubsky2014}. 

Here we turn our attention to the Be binary HR~2142, another system with an
excellent set of UV spectra in the {\it IUE} archive.  
The spectroscopic emission features of HR~2142 (HD~41335, V696~Mon; B1.5 IV-Vnne) 
were first noted by \citet{campbell1895}, and it has since attracted much attention 
among spectroscopists because of its strong Balmer emission lines
that vary periodically in a predictable manner. 
\citet{peters1971} observed the development of strong absorption components 
in the Balmer lines, and their cyclic appearance led \citet{peters1972} to propose
that these shell-type absorptions occur during a short part of the 
orbit in a binary system.  Subsequent optical spectroscopy by \citet{peters1983}
confirmed the binary nature of HR~2142 through measurement of the radial velocity
variations associated with the orbital period of $P=80.86$ days.   Early observations
with {\it IUE} demonstrated that many transitions in the UV also displayed the 
shell line variations found in the Balmer lines \citep{peters2001}.  
\citet{waters1991} examined the spectral energy distribution of HR~2142 in the infrared 
and showed that the companion cannot be a cool giant (as found in Algol binaries).
Instead, they proposed that the companion is a hot subdwarf or ``helium star'', 
the stripped down remains of the donor star in this interacting binary. 
Thus, HR~2142 is a prime candidate for detection of a hot companion through
a search for its flux contribution in the UV. 

This paper presents the results of our analysis of the collection of the 
{\it IUE} spectra of HR~2142.  Section 2 describes the UV spectra and 
a large sample of complementary H$\alpha$ spectroscopy.  We present 
radial velocities of the bright Be star component measured from these spectra,
and we calculate an updated radial velocity curve from these data. 
Section 3 describes our cross-correlation analysis of the far-UV part 
of the spectrum that provides the first positive evidence of the 
existence of a hot subdwarf companion.  In Section 4, we document the 
appearance of the shell line variations in the UV and H$\alpha$, and
we argue in Section 5 that these shell episodes result from tidal wakes 
in the disk of the Be star that are caused by the gravitational force
of the companion.  Our results are summarized in Section 6.  
Preliminary results of this study appeared earlier in several 
conference proceedings \citep{peters2001,peters2002,peters2015}. 

%%%%%%%%%%%%%%%%%%%%%%%%%%%%%%%%%%%%%%%%%%%%%%%%%%%%%%%%%%%%%%%%%%%%%%%%%%%%

\section{Radial Velocities and Orbital Solution}                 % Section 2

The first step in this investigation was to revisit the orbital radial 
velocity curve of the Be star in order to establish a contemporary 
ephemeris and to verify the orbital semiamplitude.   Radial velocity 
measurements were made using the set of spectra summarized in Table~1, 
which lists the source, number of spectra $N$, spectral resolving 
power and range, dates of observation, and primary observer. 
The main focus of this work is a set of 88 high resolution, SWP HIRES
FUV spectra acquired over the lifetime of the {\it IUE} observatory.  
These were downloaded from MAST\footnote{https://archive.stsci.edu/iue/}
and resampled on a uniform grid of heliocentric wavelength 
in $\log \lambda$ and rectified to a unit pseudo-continuum. 
We also collected a set of 49 LWR and LWP near-UV spectra that were 
used to inspect the orbital variations in the \ion{Mg}{2}
$\lambda\lambda 2796, 2803$ feature.  The UV spectra were supplemented
with a large collection of H$\alpha$ spectra that we secured with the
KPNO Coude Feed telescope and that were obtained by amateur astronomers
participating in the Be Star Spectra database project
\citep{pollmann2007,neiner2011}. 

%%%%%%%%%%%%%%%%%%%%%%%%%%%%%%%%%%%%%%%%%%%%%%%%%%%%%%%%%%%%%%%

% Table 1 - Sources of Spectroscopy

%\newpage
\begin{deluxetable}{clccccc}
  %\tabletypesize{\scriptsize}
  %\rotate
  \tablenum{1}
  \tablecaption{Spectroscopic Observations of HR 2142}
  \tablewidth{0pt}
  \tablehead{
  \colhead{Code}     &
  \colhead{Source}   & 
  \colhead{$N$}      & 
  \colhead{$\lambda/\Delta\lambda$}   & 
  \colhead{$\lambda$ Range} &
  \colhead{Dates} &
  \colhead{Observer} \\
  \colhead{ }      &
  \colhead{ }      & 
  \colhead{ }      & 
  \colhead{ }      & 
  \colhead{(\AA )} &
  \colhead{(BY)}   &
  \colhead{} 
  }
  \startdata
1 &  Lick 3 m &     50 & 10000--15000 & 3200--4700 & 1969--1975 & Peters \\
2 &  IUE      &     88 &     12000    & 1150--1950 & 1978--1995 & Peters {\bf \& others} \\
3 &  KPNO CFT &     84 & 12000--25000 & 6400--6700 & 1985--2001 & Peters \\
4 &  KPNO CFT & \phn 8 & \phn 9500    & 6430--7140 & 2004--2008 & Grundstrom \\
5 &  BeSS     &     41 &   $>10000$   & 6563:      & 2006--2015 & various \\     
  \enddata
  \end{deluxetable}

Measurement of the radial velocity for individual line transitions is 
very difficult in the UV because the line spectrum is dense and strongly
blended due to the large projected rotational broadening of the Be star's
spectrum ($V\sin i = 358$ km~s$^{-1}$; \citealt{fremat2005}).  
Instead, we followed our well-honed method of determining the 
radial velocity through cross-correlation with a spectral template
\citep{peters2013}.  We used the average spectrum as the template 
to form the cross-correlation function (CCF) in order to derive 
relative radial velocity shifts.  The calculation omitted wavelength
regions at the extreme wavelength ends, regions with broad wind lines, 
and sections with deep, narrow lines (formed in the interstellar medium 
and/or circumbinary disk).   The derived CCFs were very broad as 
expected for the photospheric spectrum of the Be star, but still 
displayed a narrow peak at the core due to correlations among 
some remaining sharp lines.  Consequently, we measured a bisector 
velocity of the wings of the CCF using a convolution with oppositely-signed 
Gaussian functions to sample the wing distribution \citep{shafter1986}. 

The H$\alpha$ emission line is strong and formed over a large volume 
of the Be star disk.  However, we can safely assume that the extreme 
wings of the emission line are formed in the part of the disk 
closest to the Be star, so that radial velocity variations of the 
wings reflect the motion of the Be star itself.  Once again we 
determined a bisector velocity for the H$\alpha$ emission line 
wings using a convolution with oppositely-signed Gaussian functions.  
These H$\alpha$ emission line velocities are listed in Table~2 
(given in full as a Machine Readable Table in the electronic version)
that lists heliocentric Julian date of mid-exposure, the corresponding 
orbital phase (see below), the radial velocity, its uncertainty, the
observed minus calculated velocity residual from the orbital solution, 
and a spectrum source code (associated with Table~1).  

%%%%%%%%%%%%%%%%%%%%%%%%%%%%%%%%%%%%%%%%%%%%%%%%%%%%%%%%%%%%%%%

% Table 2 - Be star radial velocity measurements

%\newpage
\begin{deluxetable}{cccccc}
  \tabletypesize{\scriptsize}
  \tablenum{2}
  \tablecaption{Be Star Radial Velocities}
  \tablewidth{0pt}
  \tablehead{
  \colhead{Date} & 
  \colhead{Orbital} & 
  \colhead{$V_{r}$} & 
  \colhead{$\sigma$} & 
  \colhead{$O-C$} &
  \colhead{Source}  \\
  \colhead{(HJD-2400000)} & 
  \colhead{Phase} & 
  \colhead{(km\ s$^{-1}$)} & 
  \colhead{(km\ s$^{-1}$)} & 
  \colhead{(km\ s$^{-1}$)} & 
  \colhead{Code\tablenotemark{a}}}
 \startdata
40520.03\phn\phn   &  0.830 & 
 57.1 & \phn $ 6.0$ &\phn\phs $  5.0$ & 1 \\
40527.05\phn\phn   &  0.917 & 
 56.8 &      $12.2$ &\phn\phs $  7.4$ & 1 \\
40544.94\phn\phn   &  0.138 & 
 41.7 &      $14.0$ &\phn\phs $  1.3$ & 1 \\
40572.94\phn\phn   &  0.484 & 
 47.9 &      $23.3$ &\phn\phs $  2.8$ & 1 \\
40611.80\phn\phn   &  0.964 & 
 49.9 &      $13.4$ &\phn\phs $  2.5$ & 1 \\
40637.73\phn\phn   &  0.285 & 
 37.5 & \phn $ 7.0$ &\phn     $ -1.4$ & 1 \\
40638.63\phn\phn   &  0.296 & 
 36.2 &      $16.0$ &\phn     $ -2.8$ & 1 \\
40659.61\phn\phn   &  0.555 & 
 45.8 &      $15.9$ &\phn     $ -2.5$ & 1 \\
40850.04\phn\phn   &  0.908 & 
 55.2 &      $10.7$ &\phn\phs $  5.5$ & 1 \\
40877.96\phn\phn   &  0.254 & 
 41.9 &      $11.6$ &\phn\phs $  3.2$ & 1 \\
40903.92\phn\phn   &  0.574 & 
 55.8 & \phn $ 4.5$ &\phn\phs $  6.7$ & 1 \\
40996.68\phn\phn   &  0.721 & 
 58.0 &      $11.0$ &\phn\phs $  5.1$ & 1 \\
41200.04\phn\phn   &  0.234 & 
 34.9 &      $14.7$ &\phn     $ -3.8$ & 1 \\
41228.06\phn\phn   &  0.580 & 
 48.5 &      $14.0$ &\phn     $ -0.8$ & 1 \\
41236.04\phn\phn   &  0.679 & 
 60.5 & \phn $ 9.6$ &\phn\phs $  8.2$ & 1 \\
41256.98\phn\phn   &  0.938 & 
 49.1 & \phn $ 9.5$ &\phn\phs $  0.5$ & 1 \\
41256.98\phn\phn   &  0.938 & 
 51.5 & \phn $ 8.4$ &\phn\phs $  2.9$ & 1 \\
41261.00\phn\phn   &  0.987 & 
 47.6 & \phn $ 6.2$ &\phn\phs $  1.2$ & 1 \\
41261.00\phn\phn   &  0.987 & 
 42.2 &      $10.7$ &\phn     $ -4.2$ & 1 \\
41346.70\phn\phn   &  0.047 & 
 44.8 & \phn $ 8.2$ &\phn\phs $  1.0$ & 1 \\
41346.70\phn\phn   &  0.047 & 
 42.3 &      $11.3$ &\phn     $ -1.5$ & 1 \\
41374.67\phn\phn   &  0.392 & 
 35.3 &      $19.3$ &\phn     $ -6.1$ & 1 \\
41404.62\phn\phn   &  0.762 & 
 56.0 &      $12.8$ &\phn\phs $  3.0$ & 1 \\
41406.63\phn\phn   &  0.787 & 
 50.7 &      $10.3$ &\phn     $ -2.1$ & 1 \\
41433.65\phn\phn   &  0.121 & 
 34.5 &      $17.0$ &\phn     $ -6.4$ & 1 \\
41590.04\phn\phn   &  0.054 & 
 38.9 & \phn $ 6.5$ &\phn     $ -4.6$ & 1 \\
41591.00\phn\phn   &  0.066 & 
 40.2 &      $14.1$ &\phn     $ -2.8$ & 1 \\
41614.94\phn\phn   &  0.362 & 
 38.0 &      $18.0$ &\phn     $ -2.4$ & 1 \\
41616.96\phn\phn   &  0.387 & 
 34.6 &      $18.4$ &\phn     $ -6.6$ & 1 \\
41643.98\phn\phn   &  0.721 & 
 50.3 &      $11.1$ &\phn     $ -2.6$ & 1 \\
41644.89\phn\phn   &  0.732 & 
 47.4 &      $11.3$ &\phn     $ -5.5$ & 1 \\
41644.90\phn\phn   &  0.732 & 
 49.8 &      $12.0$ &\phn     $ -3.1$ & 1 \\
41936.02\phn\phn   &  0.330 & 
 34.4 & \phn $ 6.9$ &\phn     $ -5.2$ & 1 \\
41975.00\phn\phn   &  0.812 & 
 52.9 &      $25.0$ &\phn\phs $  0.5$ & 1 \\
41987.02\phn\phn   &  0.960 & 
 40.5 & \phn $ 8.5$ &\phn     $ -7.1$ & 1 \\
41987.97\phn\phn   &  0.972 & 
 51.0 & \phn $ 8.6$ &\phn\phs $  3.9$ & 1 \\
41990.95\phn\phn   &  0.009 & 
 48.8 &      $16.7$ &\phn\phs $  3.4$ & 1 \\
41991.90\phn\phn   &  0.021 & 
 44.1 &      $18.0$ &\phn     $ -0.8$ & 1 \\
41992.07\phn\phn   &  0.023 & 
 44.3 &      $14.5$ &\phn     $ -0.5$ & 1 \\
42036.87\phn\phn   &  0.576 & 
 49.2 &      $35.9$ &\phn\phs $  0.1$ & 1 \\
42083.73\phn\phn   &  0.156 & 
 36.8 &      $14.5$ &\phn     $ -3.1$ & 1 \\
42149.65\phn\phn   &  0.970 & 
 44.9 & \phn $ 6.9$ &\phn     $ -2.3$ & 1 \\
42312.02\phn\phn   &  0.977 & 
 40.0 & \phn $ 8.8$ &\phn     $ -6.9$ & 1 \\
42316.97\phn\phn   &  0.038 & 
 52.4 & \phn $ 8.9$ &\phn\phs $  8.3$ & 1 \\
42317.03\phn\phn   &  0.039 & 
 45.7 &      $11.1$ &\phn\phs $  1.6$ & 1 \\
42317.98\phn\phn   &  0.051 & 
 45.7 & \phn $ 9.0$ &\phn\phs $  2.1$ & 1 \\
42317.98\phn\phn   &  0.051 & 
 44.0 &      $11.3$ &\phn\phs $  0.4$ & 1 \\
42318.97\phn\phn   &  0.063 & 
 37.4 & \phn $ 3.5$ &\phn     $ -5.7$ & 1 \\
42682.01\phn\phn   &  0.550 & 
 54.4 &      $17.9$ &\phn\phs $  6.4$ & 1 \\
42733.94\phn\phn   &  0.191 & 
 40.9 &      $24.0$ &\phn\phs $  1.7$ & 1 \\
43742.3180   &  0.654 & 
 54.7 & \phn $ 6.6$ &\phn\phs $  3.0$ & 2355 \\
43808.2337   &  0.468 & 
 56.7 & \phn $ 7.6$ &\phs     $ 12.3$ & 3162 \\
43844.1164   &  0.912 & 
 63.0 & \phn $ 7.5$ &\phs     $ 13.4$ & 3505 \\
43848.1808   &  0.962 & 
 46.0 & \phn $ 8.9$ &\phn     $ -1.5$ & 3537 \\
44167.1164   &  0.904 & 
 43.9 & \phn $ 9.2$ &\phn     $ -6.0$ & 6929 \\
44169.9608   &  0.939 & 
 52.4 & \phn $ 9.8$ &\phn\phs $  3.9$ & 6961 \\
44169.9955   &  0.939 & 
 49.2 & \phn $ 6.5$ &\phn\phs $  0.8$ & 6962 \\
44172.0051   &  0.964 & 
 41.3 & \phn $ 5.3$ &\phn     $ -6.1$ & 6991 \\
44173.7810   &  0.986 & 
 49.3 & \phn $ 5.6$ &\phn\phs $  2.9$ & 7007 \\
44328.5142   &  0.899 & 
 57.2 & \phn $ 5.5$ &\phn\phs $  7.1$ & 8596 \\
44328.5478   &  0.899 & 
 58.8 & \phn $ 5.6$ &\phn\phs $  8.7$ & 8597 \\
44329.2100   &  0.907 & 
 52.1 & \phn $ 5.7$ &\phn\phs $  2.3$ & 8604 \\
44330.1423   &  0.919 & 
 53.5 & \phn $ 6.4$ &\phn\phs $  4.2$ & 8616 \\
44332.4697   &  0.947 & 
 43.0 & \phn $ 6.9$ &\phn     $ -5.2$ & 8637 \\
44483.8703   &  0.819 & 
 51.2 & \phn $ 5.5$ &\phn     $ -1.1$ & 9978 \\
44485.8594   &  0.843 & 
 42.5 & \phn $ 5.4$ &\phn     $ -9.2$ & 9992 \\
44511.0631   &  0.155 & 
 47.6 & \phn $ 6.1$ &\phn\phs $  7.6$ & 10240 \\
44533.7852   &  0.436 & 
 40.2 & \phn $ 5.5$ &\phn     $ -2.8$ & 10436 \\
44612.3243   &  0.406 & 
 37.1 & \phn $ 5.1$ &\phn     $ -4.8$ & 10993 \\
44616.4285   &  0.457 & 
 39.0 & \phn $ 5.1$ &\phn     $ -4.9$ & 11026 \\
44661.2357   &  0.011 & 
 47.9 & \phn $ 5.0$ &\phn\phs $  2.6$ & 13359 \\
44867.9669   &  0.566 & 
 63.0 & \phn $ 5.2$ &\phs     $ 14.3$ & 15054 \\
44891.3648   &  0.855 & 
 59.2 & \phn $ 6.3$ &\phn\phs $  7.7$ & 15255 \\
44920.7181   &  0.218 & 
 37.4 & \phn $ 5.7$ &\phn     $ -1.4$ & 15477 \\
44972.5984   &  0.859 & 
 57.5 & \phn $ 6.1$ &\phn\phs $  6.1$ & 15932 \\
44974.5335   &  0.883 & 
 51.2 & \phn $ 5.7$ &\phn\phs $  0.6$ & 15955 \\
44975.6692   &  0.897 & 
 48.4 & \phn $ 5.4$ &\phn     $ -1.8$ & 15977 \\
45042.6617   &  0.725 & 
 64.5 & \phn $ 6.9$ &\phs     $ 11.6$ & 16526 \\
46121.7891   &  0.062 & 
 46.8 & \phn $ 1.1$ &\phn\phs $  3.7$ & 3 \\
46122.6951   &  0.073 & 
 41.5 & \phn $ 1.1$ &\phn     $ -1.1$ & 3 \\
46122.7103   &  0.073 & 
 42.0 & \phn $ 1.3$ &\phn     $ -0.6$ & 3 \\
46152.6593   &  0.443 & 
 41.6 & \phn $ 7.5$ &\phn     $ -1.7$ & 3 \\
46338.9463   &  0.745 & 
 64.7 & \phn $ 5.5$ &\phs     $ 11.8$ & 26809 \\
46734.9250   &  0.639 & 
 55.5 & \phn $ 4.2$ &\phn\phs $  4.2$ & 3 \\
46737.7984   &  0.675 & 
 54.0 & \phn $ 2.3$ &\phn\phs $  1.8$ & 3 \\
46820.6364   &  0.699 & 
 51.0 & \phn $ 5.2$ &\phn     $ -1.6$ & 30156 \\
46823.6438   &  0.736 & 
 47.4 & \phn $ 5.5$ &\phn     $ -5.6$ & 30179 \\
46824.5968   &  0.747 & 
 51.5 & \phn $ 5.1$ &\phn     $ -1.5$ & 30188 \\
46826.6402   &  0.773 & 
 63.1 & \phn $ 6.0$ &\phs     $ 10.2$ & 30207 \\
46826.6638   &  0.773 & 
 55.7 & \phn $ 5.1$ &\phn\phs $  2.8$ & 30208 \\
46826.6978   &  0.773 & 
 54.7 & \phn $ 5.7$ &\phn\phs $  1.8$ & 30209 \\
46826.7272   &  0.774 & 
 58.4 & \phn $ 5.7$ &\phn\phs $  5.5$ & 30210 \\
46826.7616   &  0.774 & 
 65.8 & \phn $ 5.6$ &\phs     $ 12.9$ & 30211 \\
46826.7876   &  0.775 & 
 57.4 & \phn $ 5.4$ &\phn\phs $  4.5$ & 30212 \\
46828.5744   &  0.797 & 
 55.4 & \phn $ 4.9$ &\phn\phs $  2.7$ & 30228 \\
46828.6136   &  0.797 & 
 53.1 & \phn $ 5.1$ &\phn\phs $  0.5$ & 30229 \\
46828.6442   &  0.797 & 
 44.7 & \phn $ 6.5$ &\phn     $ -7.9$ & 30230 \\
46828.6690   &  0.798 & 
 52.2 & \phn $ 5.3$ &\phn     $ -0.4$ & 30231 \\
46828.6952   &  0.798 & 
 44.9 & \phn $ 5.2$ &\phn     $ -7.7$ & 30232 \\
46830.5249   &  0.821 & 
 52.8 & \phn $ 5.5$ &\phn\phs $  0.6$ & 30250 \\
46832.5591   &  0.846 & 
 60.0 & \phn $ 6.7$ &\phn\phs $  8.3$ & 30263 \\
46875.5064   &  0.377 & 
 48.7 & \phn $ 5.3$ &\phn\phs $  7.8$ & 30580 \\
46875.5362   &  0.377 & 
 42.5 & \phn $ 4.6$ &\phn\phs $  1.6$ & 30581 \\
46875.5655   &  0.377 & 
 43.6 & \phn $ 5.0$ &\phn\phs $  2.7$ & 30582 \\
46875.5925   &  0.378 & 
 49.8 & \phn $ 4.8$ &\phn\phs $  8.9$ & 30583 \\
46901.6395   &  0.700 & 
 57.3 & \phn $ 3.0$ &\phn\phs $  4.7$ & 3 \\
46901.6452   &  0.700 & 
 59.1 & \phn $ 3.4$ &\phn\phs $  6.5$ & 3 \\
46902.6503   &  0.712 & 
 59.4 & \phn $ 1.5$ &\phn\phs $  6.6$ & 3 \\
46903.6545   &  0.725 & 
 59.6 & \phn $ 3.0$ &\phn\phs $  6.7$ & 3 \\
46905.6139   &  0.749 & 
 56.7 & \phn $ 1.1$ &\phn\phs $  3.7$ & 3 \\
46918.6402   &  0.910 & 
 47.8 & \phn $ 9.7$ &\phn     $ -1.9$ & 3 \\
46919.6238   &  0.922 & 
 53.4 & \phn $ 3.5$ &\phn\phs $  4.2$ & 3 \\
46919.6287   &  0.922 & 
 54.4 & \phn $ 6.8$ &\phn\phs $  5.2$ & 3 \\
46920.6564   &  0.935 & 
 54.2 & \phn $ 6.5$ &\phn\phs $  5.6$ & 3 \\
47030.9707   &  0.298 & 
 28.4 & \phn $ 7.9$ &         $-10.6$ & 3 \\
47030.9854   &  0.298 & 
 51.5 & \phn $ 7.8$ &\phs     $ 12.4$ & 3 \\
47032.9747   &  0.323 & 
 50.3 & \phn $ 6.8$ &\phs     $ 10.9$ & 3 \\
47033.9595   &  0.335 & 
 50.2 & \phn $ 5.7$ &\phs     $ 10.6$ & 3 \\
47033.9629   &  0.335 & 
 39.3 & \phn $ 6.5$ &\phn     $ -0.4$ & 3 \\
47103.9017   &  0.199 & 
 29.2 & \phn $ 4.8$ &\phn     $ -9.8$ & 32224 \\
47141.6472   &  0.666 & 
 52.6 & \phn $ 6.1$ &\phn\phs $  0.6$ & 32507 \\
47169.4943   &  0.010 & 
 44.1 & \phn $ 5.8$ &\phn     $ -1.2$ & 32687 \\
47172.4692   &  0.047 & 
 47.8 & \phn $ 5.8$ &\phn\phs $  4.1$ & 32701 \\
47198.4327   &  0.368 & 
 48.4 & \phn $ 5.3$ &\phn\phs $  7.8$ & 32874 \\
47215.4720   &  0.578 & 
 43.4 & \phn $ 5.4$ &\phn     $ -5.8$ & 32970 \\
47258.3965   &  0.109 & 
 32.6 & \phn $ 5.8$ &\phn     $ -8.7$ & 33220 \\
47414.0776   &  0.033 & 
 38.2 & \phn $ 5.0$ &\phn     $ -6.1$ & 34212 \\
47422.9379   &  0.142 & 
 33.9 & \phn $ 5.2$ &\phn     $ -6.3$ & 34267 \\
47464.9085   &  0.661 & 
 51.3 & \phn $ 5.6$ &\phn     $ -0.6$ & 34633 \\
47469.8194   &  0.722 & 
 53.0 & \phn $ 9.8$ &\phn\phs $  0.1$ & 3 \\
47470.0047   &  0.724 & 
 49.0 & \phn $ 9.4$ &\phn     $ -3.9$ & 3 \\
47470.8213   &  0.734 & 
 55.4 & \phn $ 2.0$ &\phn\phs $  2.5$ & 3 \\
47471.0573   &  0.737 & 
 51.7 & \phn $ 2.0$ &\phn     $ -1.2$ & 3 \\
47471.8378   &  0.747 & 
 56.1 & \phn $ 2.0$ &\phn\phs $  3.2$ & 3 \\
47471.8404   &  0.747 & 
 54.9 & \phn $ 1.5$ &\phn\phs $  1.9$ & 3 \\
47473.0386   &  0.761 & 
 50.6 & \phn $ 2.5$ &\phn     $ -2.4$ & 3 \\
47484.8542   &  0.907 & 
 40.2 & \phn $ 6.9$ &\phn     $ -9.5$ & 34770 \\
47533.7509   &  0.512 & 
 45.1 & \phn $ 5.6$ &\phn     $ -1.3$ & 35282 \\
47550.5978\tablenotemark{b}   &  0.720 & 
 33.7 & \phn $ 5.5$ &         $-19.1$ & 35400 \\
47556.6147   &  0.794 & 
 44.3 & \phn $ 5.5$ &\phn     $ -8.4$ & 35456 \\
47560.6629   &  0.844 & 
 51.0 & \phn $ 5.5$ &\phn     $ -0.7$ & 3 \\
47560.8172   &  0.846 & 
 50.6 & \phn $ 5.7$ &\phn     $ -1.1$ & 3 \\
47561.6730   &  0.857 & 
 49.5 & \phn $ 5.5$ &\phn     $ -2.0$ & 3 \\
47572.3104   &  0.988 & 
 31.8 & \phn $ 5.0$ &         $-14.6$ & 35551 \\
47607.3133   &  0.421 & 
 34.0 & \phn $ 5.0$ &\phn     $ -8.4$ & 35833 \\
47636.6277   &  0.783 & 
 42.4 & \phn $ 5.6$ &         $-10.4$ & 3 \\
47636.6330   &  0.783 & 
 42.1 & \phn $ 5.5$ &         $-10.8$ & 3 \\
47636.6381   &  0.783 & 
 43.9 & \phn $ 5.4$ &\phn     $ -8.9$ & 3 \\
47638.6422   &  0.808 & 
 49.0 & \phn $ 5.6$ &\phn     $ -3.5$ & 3 \\
47639.6267   &  0.820 & 
 48.3 & \phn $ 5.8$ &\phn     $ -4.0$ & 3 \\
47640.6235   &  0.833 & 
 46.7 & \phn $ 5.5$ &\phn     $ -5.4$ & 3 \\
47640.6276   &  0.833 & 
 47.5 & \phn $ 5.4$ &\phn     $ -4.5$ & 3 \\
47789.9438   &  0.678 & 
 44.7 & \phn $ 5.8$ &\phn     $ -7.5$ & 37091 \\
47871.8559   &  0.690 & 
 52.9 & \phn $ 5.3$ &\phn\phs $  0.5$ & 37795 \\
47938.4644   &  0.514 & 
 45.4 & \phn $ 4.8$ &\phn     $ -1.1$ & 38198 \\
47939.6030   &  0.528 & 
 42.5 &      $11.8$ &\phn     $ -4.5$ & 3 \\
47940.7746   &  0.542 & 
 43.2 & \phn $ 6.2$ &\phn     $ -4.5$ & 3 \\
47941.6557   &  0.553 & 
 42.7 & \phn $ 8.1$ &\phn     $ -5.5$ & 3 \\
47941.6606   &  0.553 & 
 44.5 & \phn $ 7.5$ &\phn     $ -3.6$ & 3 \\
47945.4140   &  0.600 & 
 55.1 & \phn $ 5.2$ &\phn\phs $  5.1$ & 38227 \\
47981.7112   &  0.048 & 
 44.4 & \phn $ 5.1$ &\phn\phs $  0.7$ & 3 \\
47983.6008   &  0.071 & 
 38.6 & \phn $ 5.1$ &\phn     $ -4.1$ & 3 \\
48121.9663   &  0.782 & 
 57.4 & \phn $ 4.3$ &\phn\phs $  4.5$ & 3 \\
48123.9556   &  0.806 & 
 55.0 & \phn $ 8.7$ &\phn\phs $  2.5$ & 3 \\
48124.9817   &  0.819 & 
 47.3 & \phn $ 1.5$ &\phn     $ -5.1$ & 3 \\
48125.0065   &  0.819 & 
 45.5 & \phn $ 0.9$ &\phn     $ -6.8$ & 3 \\
48125.9833   &  0.831 & 
 55.7 & \phn $ 2.7$ &\phn\phs $  3.6$ & 3 \\
48126.0140   &  0.832 & 
 55.8 & \phn $ 2.3$ &\phn\phs $  3.7$ & 3 \\
48228.7021   &  0.101 & 
 46.5 & \phn $ 5.1$ &\phn\phs $  4.9$ & 40252 \\
48230.6937   &  0.125 & 
 42.3 & \phn $ 5.8$ &\phn\phs $  1.5$ & 40271 \\
48234.7019   &  0.175 & 
 48.0 & \phn $ 6.3$ &\phn\phs $  8.5$ & 40292 \\
48313.6160   &  0.150 & 
 51.3 & \phn $ 1.9$ &\phs     $ 11.3$ & 3 \\
48318.5967   &  0.212 & 
 39.0 & \phn $ 1.4$ &\phn\phs $  0.1$ & 3 \\
48318.6029   &  0.212 & 
 37.8 & \phn $ 1.2$ &\phn     $ -1.1$ & 3 \\
48319.7591   &  0.226 & 
 37.4 & \phn $ 1.7$ &\phn     $ -1.4$ & 3 \\
48320.6062   &  0.236 & 
 40.6 & \phn $ 2.3$ &\phn\phs $  1.9$ & 3 \\
48320.6093   &  0.237 & 
 39.4 & \phn $ 1.8$ &\phn\phs $  0.6$ & 3 \\
48321.6023   &  0.249 & 
 51.2 & \phn $ 1.2$ &\phs     $ 12.6$ & 3 \\
48351.4009   &  0.617 & 
 47.5 & \phn $ 5.5$ &\phn     $ -3.1$ & 41292 \\
48353.4552   &  0.642 & 
 38.0 & \phn $ 5.3$ &         $-13.4$ & 41312 \\
48356.5255   &  0.680 & 
 49.6 & \phn $ 5.2$ &\phn     $ -2.7$ & 41329 \\
48514.9599   &  0.638 & 
 58.0 & \phn $ 4.7$ &\phn\phs $  6.7$ & 3 \\
48514.9619   &  0.639 & 
 55.5 & \phn $ 2.9$ &\phn\phs $  4.2$ & 3 \\
48515.9288   &  0.650 & 
 53.6 & \phn $ 4.2$ &\phn\phs $  2.0$ & 3 \\
48515.9315   &  0.650 & 
 54.2 & \phn $ 3.9$ &\phn\phs $  2.5$ & 3 \\
48516.9660   &  0.663 & 
 61.1 & \phn $ 6.1$ &\phn\phs $  9.1$ & 3 \\
49058.7301   &  0.359 & 
 47.9 & \phn $ 1.4$ &\phn\phs $  7.6$ & 3 \\
49276.7565   &  0.053 & 
 46.4 & \phn $ 5.5$ &\phn\phs $  2.9$ & 48925 \\
49335.5789   &  0.780 & 
 51.5 & \phn $ 6.7$ &\phn     $ -1.3$ & 49589 \\
49374.4579   &  0.261 & 
 40.7 & \phn $ 5.8$ &\phn\phs $  2.0$ & 49862 \\
49379.4985   &  0.323 & 
 37.4 & \phn $ 4.8$ &\phn     $ -2.0$ & 49900 \\
49383.4999   &  0.373 & 
 32.6 & \phn $ 5.8$ &\phn     $ -8.1$ & 49918 \\
49443.6959   &  0.117 & 
 42.1 & \phn $ 3.5$ &\phn\phs $  1.1$ & 3 \\
49443.6977   &  0.117 & 
 42.8 & \phn $ 2.2$ &\phn\phs $  1.7$ & 3 \\
49444.6856   &  0.129 & 
 43.0 & \phn $ 0.8$ &\phn\phs $  2.4$ & 3 \\
49445.7002   &  0.141 & 
 43.0 & \phn $ 1.2$ &\phn\phs $  2.7$ & 3 \\
49658.8770   &  0.776 & 
 45.0 & \phn $ 5.7$ &\phn     $ -7.9$ & 52713 \\
49690.6816   &  0.169 & 
 39.1 & \phn $ 4.9$ &\phn     $ -0.4$ & 52974 \\
49694.8170   &  0.220 & 
 36.3 & \phn $ 6.3$ &\phn     $ -2.5$ & 53010 \\
49696.8457   &  0.245 & 
 41.9 & \phn $ 6.0$ &\phn\phs $  3.2$ & 53030 \\
49783.5320   &  0.317 & 
 28.9 & \phn $ 5.5$ &         $-10.4$ & 54053 \\
49816.5209   &  0.724 & 
 42.0 & \phn $ 5.2$ &         $-10.8$ & 54339 \\
49976.9670   &  0.707 & 
 52.5 & \phn $ 5.7$ &\phn     $ -0.2$ & 55942 \\
49976.9934   &  0.708 & 
 56.3 & \phn $ 5.5$ &\phn\phs $  3.6$ & 55943 \\
50445.7157   &  0.501 & 
 44.4 & \phn $ 1.1$ &\phn     $ -1.4$ & 3 \\
50446.7308   &  0.513 & 
 45.9 & \phn $ 1.5$ &\phn     $ -0.5$ & 3 \\
50449.9064   &  0.552 & 
 46.4 & \phn $ 1.9$ &\phn     $ -1.7$ & 3 \\
50449.9519   &  0.553 & 
 47.4 & \phn $ 1.6$ &\phn     $ -0.8$ & 3 \\
50450.6275   &  0.561 & 
 47.1 & \phn $ 1.0$ &\phn     $ -1.4$ & 3 \\
50451.6439   &  0.574 & 
 47.5 & \phn $ 0.9$ &\phn     $ -1.6$ & 3 \\
51191.8840   &  0.722 & 
 56.3 & \phn $ 0.6$ &\phn\phs $  3.5$ & 3 \\
51196.9206   &  0.785 & 
 53.8 & \phn $ 0.9$ &\phn\phs $  1.0$ & 3 \\
51502.9245   &  0.566 & 
 51.7 & \phn $ 1.8$ &\phn\phs $  3.0$ & 3 \\
51503.9888   &  0.580 & 
 52.2 & \phn $ 0.7$ &\phn\phs $  3.0$ & 3 \\
51504.9810   &  0.592 & 
 53.1 & \phn $ 0.6$ &\phn\phs $  3.4$ & 3 \\
51507.0622   &  0.618 & 
 52.5 & \phn $ 0.6$ &\phn\phs $  1.9$ & 3 \\
51509.0572   &  0.642 & 
 52.0 & \phn $ 0.9$ &\phn\phs $  0.6$ & 3 \\
51510.0615   &  0.655 & 
 53.2 & \phn $ 0.9$ &\phn\phs $  1.5$ & 3 \\
51510.9336   &  0.665 & 
 54.9 & \phn $ 0.7$ &\phn\phs $  2.9$ & 3 \\
51913.6433   &  0.643 & 
 48.4 & \phn $ 0.5$ &\phn     $ -3.1$ & 3 \\
51914.6872   &  0.655 & 
 47.2 & \phn $ 0.8$ &\phn     $ -4.5$ & 3 \\
51915.7806   &  0.669 & 
 46.6 & \phn $ 0.8$ &\phn     $ -5.5$ & 3 \\
53291.9354   &  0.677 & 
 46.4 & \phn $ 1.1$ &\phn     $ -5.9$ & 4 \\
53292.9944   &  0.690 & 
 46.4 & \phn $ 1.0$ &\phn     $ -6.1$ & 4 \\
53294.9640   &  0.714 & 
 46.5 & \phn $ 0.9$ &\phn     $ -6.3$ & 4 \\
53294.9696   &  0.714 & 
 46.4 & \phn $ 0.9$ &\phn     $ -6.3$ & 4 \\
54000.5828   &  0.435 & 
 34.6 & \phn $ 8.2$ &\phn     $ -8.4$ & 5 \\
54020.9788   &  0.687 & 
 50.8 & \phn $ 1.3$ &\phn     $ -1.7$ & 4 \\
54024.9341   &  0.736 & 
 46.8 & \phn $ 1.3$ &\phn     $ -6.1$ & 4 \\
54041.5544   &  0.941 & 
 46.3 & \phn $ 3.9$ &\phn     $ -2.1$ & 5 \\
54360.6253   &  0.885 & 
 49.1 & \phn $ 6.3$ &\phn     $ -1.5$ & 5 \\
54365.6499   &  0.947 & 
 56.4 & \phn $ 5.8$ &\phn\phs $  8.2$ & 5 \\
54366.6335   &  0.959 & 
 43.0 & \phn $ 3.1$ &\phn     $ -4.6$ & 5 \\
54379.6041   &  0.119 & 
 36.5 & \phn $ 3.2$ &\phn     $ -4.4$ & 5 \\
54387.5470   &  0.217 & 
 26.6 & \phn $ 7.9$ &         $-12.3$ & 5 \\
54390.5816   &  0.255 & 
 38.1 & \phn $ 3.7$ &\phn     $ -0.6$ & 5 \\
54396.5591   &  0.329 & 
 48.4 & \phn $ 5.2$ &\phn\phs $  8.9$ & 5 \\
54405.5336   &  0.440 & 
 49.1 & \phn $ 5.4$ &\phn\phs $  5.9$ & 5 \\
54468.3319   &  0.216 & 
 30.2 & \phn $ 8.0$ &\phn     $ -8.6$ & 5 \\
54478.2873   &  0.339 & 
 22.8 &      $18.2$ &         $-16.9$ & 5 \\
54504.2894   &  0.660 & 
 56.9 &      $12.0$ &\phn\phs $  5.0$ & 5 \\
54506.3162   &  0.685 & 
 45.5 & \phn $ 3.3$ &\phn     $ -6.9$ & 5 \\
54507.3105   &  0.697 & 
 54.6 & \phn $ 7.7$ &\phn\phs $  2.0$ & 5 \\
54521.3459   &  0.871 & 
 40.5 & \phn $ 5.2$ &         $-10.5$ & 5 \\
54527.4288   &  0.946 & 
 62.9 & \phn $ 5.4$ &\phs     $ 14.7$ & 5 \\
54555.3556   &  0.291 & 
 28.2 & \phn $ 5.2$ &         $-10.7$ & 5 \\
54562.3347   &  0.377 & 
 31.9 &      $33.3$ &\phn     $ -8.9$ & 5 \\
54563.3236   &  0.390 & 
 31.8 &      $28.5$ &\phn     $ -9.5$ & 5 \\
54764.9840   &  0.882 & 
 44.9 & \phn $ 0.6$ &\phn     $ -5.8$ & 4 \\
54774.0358   &  0.994 & 
 43.6 & \phn $ 0.8$ &\phn     $ -2.5$ & 4 \\
54822.4689   &  0.592 & 
 48.3 & \phn $ 8.9$ &\phn     $ -1.5$ & 5 \\
54879.3956   &  0.296 & 
 40.9 & \phn $ 1.7$ &\phn\phs $  2.0$ & 5 \\
54887.4120   &  0.395 & 
 42.6 &      $12.4$ &\phn\phs $  1.1$ & 5 \\
55576.4532   &  0.911 & 
 50.0 & \phn $ 2.2$ &\phn\phs $  0.4$ & 5 \\
55988.3921   &  0.002 & 
 49.1 & \phn $ 2.5$ &\phn\phs $  3.4$ & 5 \\
56000.3542   &  0.150 & 
 46.3 & \phn $ 1.4$ &\phn\phs $  6.3$ & 5 \\
56271.5006   &  0.501 & 
 44.4 & \phn $ 1.8$ &\phn     $ -1.4$ & 5 \\
56295.4159   &  0.796 & 
 50.0 & \phn $ 3.9$ &\phn     $ -2.7$ & 5 \\
56630.1881   &  0.934 & 
 43.3 & \phn $ 1.7$ &\phn     $ -5.4$ & 5 \\
56647.4494   &  0.147 & 
 43.7 & \phn $ 2.5$ &\phn\phs $  3.6$ & 5 \\
56685.3676   &  0.616 & 
 51.6 & \phn $ 1.6$ &\phn\phs $  1.0$ & 5 \\
56686.0369   &  0.624 & 
 45.9 & \phn $ 1.7$ &\phn     $ -4.9$ & 5 \\
56711.3368   &  0.937 & 
 44.5 & \phn $ 4.3$ &\phn     $ -4.1$ & 5 \\
56713.3043   &  0.961 & 
 44.0 & \phn $ 2.9$ &\phn     $ -3.6$ & 5 \\
56733.6663   &  0.213 & 
 41.5 & \phn $ 2.0$ &\phn\phs $  2.7$ & 5 \\
56739.3400   &  0.283 & 
 40.5 & \phn $ 3.0$ &\phn\phs $  1.7$ & 5 \\
56741.5539   &  0.310 & 
 41.3 & \phn $ 2.2$ &\phn\phs $  2.1$ & 5 \\
56745.6829   &  0.361 & 
 42.3 & \phn $ 1.3$ &\phn\phs $  1.9$ & 5 \\
56998.7604   &  0.489 & 
 62.1 & \phn $ 0.7$ &\phs     $ 16.8$ & 5 \\
57042.6163   &  0.031 & 
 46.4 & \phn $ 0.7$ &\phn\phs $  1.9$ & 5 \\
57061.3817   &  0.263 & 
 45.6 & \phn $ 2.1$ &\phn\phs $  6.9$ & 5 \\
\enddata
\tablenotetext{a}{Source row listed in Table 1 or IUE SWP number if greater than 5.}
\tablenotetext{b}{{\bf This discrepant datum was} assigned zero-weight in orbital solution.}
\end{deluxetable}

\newpage
\clearpage

In order to combine the {\it IUE} and H$\alpha$ measurements in a
single orbital solution, we need to transform the relative {\it IUE}
measurements to the frame of the H$\alpha$ measurements.  We did this
by making preliminary orbital solutions to both sets, and then we 
applied the difference in the derived systemic velocity 
$\Delta \gamma = +28.68$ km~s$^{-1}$ to the {\it IUE} relative
velocities, and these corrected velocities appear in Table~2. 
Likewise, we also wanted to include the original set of measurements
of blue range spectroscopy from \citet{peters1983} in the final solution, 
so we again determined a systemic velocity difference 
$\Delta \gamma = +24.16$ km~s$^{-1}$ to place these measurements 
in the H$\alpha$ frame.  These adjusted measurements from 
\citet{peters1983} are also included in Table~2 for the convenience
of the reader.  {\bf Neither shift should be interpreted in physical 
terms, but they are required to bring different kinds on measurements 
onto the same reference frame.}

We made an equally weighted fit of the radial velocities using the 
non-linear, least-squares method described by \citet{morbey1974}
to determine the spectroscopic orbital elements that are given 
in Table~3.  We adopted a circular orbital solution because 
elliptical solutions made no significant improvement in the 
residuals from the fit.   The elements are compared in Table~3
to those derived by \citet{peters1983}, and the agreement is 
reasonable.  The new period is somewhat larger ($2.8\sigma$ different), 
and the uncertainty in the period is larger than given by 
\citet{peters1983}.  The new estimate of semiamplitude $K_1$ is slightly lower
than found by \citet{peters1983} ($2.2\sigma$ different). 
The radial velocities and orbital solution are shown in Figure~1, 
and we see that there is good agreement between the UV (larger 
scatter) and H$\alpha$ results.  We will use the new orbital 
elements throughout this paper with the epoch of phase zero, $T_{sc}$,
defined as the time of superior conjunction of the Be star 
(as adopted by \citealt{peters1983}). 

%%%%%%%%%%%%%%%%%%%%%%%%%%%%%%%%%%%%%%%%%%%%%%%%%%%%%%%%%%%%%%%

% Table 3: Orbital parameters

%\newpage
\begin{deluxetable}{lcc}
  %\tabletypesize{\scriptsize}
  %\rotate
  \tablenum{3}
  \tablecaption{Circular Orbital Elements}
  \tablewidth{0pt}
  \tablehead{
  \colhead{Element}  & \colhead{This\ Study} & \colhead{Peters\ (1983)}}
  \startdata
  $P$\ (days)                        &  80.913 $\pm$ 0.018  &  80.860 $\pm$ 0.005 \\
  $T_{sc}$\ (HJD-2400000)            & 46845.0 $\pm$ 1.0    & 41990.5 $\pm$ 1.1   \\
  $K_1$\ (km\ s$^{-1}$)              &     7.1 $\pm$ 0.5    &     9.4 $\pm$ 0.9   \\ 
  $\gamma$(H$\alpha$) (km\ s$^{-1}$) &    45.8 $\pm$ 0.4    &          \nodata    \\
  $f(m)$\ ($M_\odot$)                &  0.0031 $\pm$ 0.0006 &   0.007 $\pm$ 0.002 \\
  $a_1 \sin\ {i}$ ($10^6$\ km)       &     7.9 $\pm$ 0.5    &    10.4 $\pm$ 1.0   \\
  rms (km\ s$^{-1}$)                 &          5.8         &          4.1        \\
  \enddata
  \end{deluxetable}

\clearpage

% Figure 1
\begin{figure} 
\begin{center} 
{\includegraphics[angle=90,height=12cm]{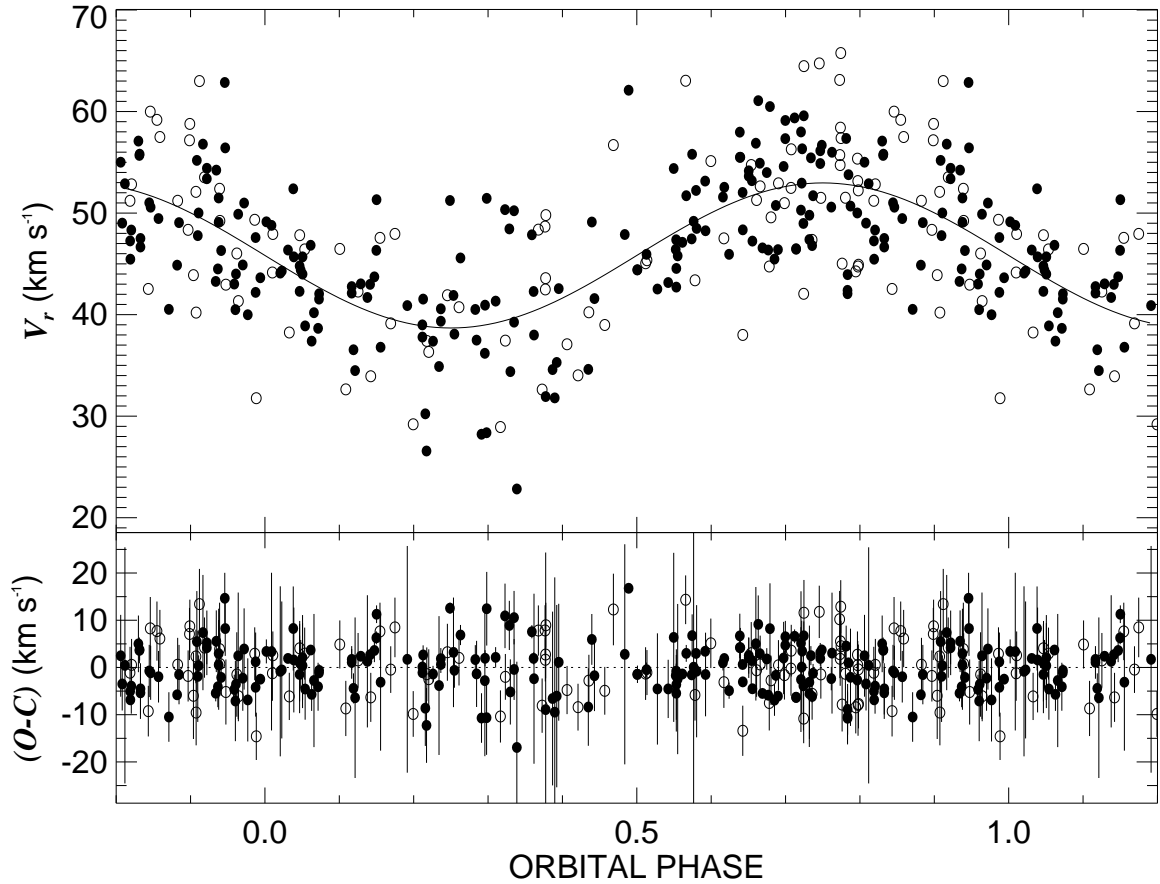}} 
\end{center} 
\caption{{\it Top panel:} Radial velocity curve for the Be star component of HR~2142. 
Orbital phase 0.0 corresponds to Be star superior conjunction.  
The solid circles represent the velocities from optical H$\alpha$ spectroscopy
and the open circles show the same from {\it IUE}. {\it Lower panel: 
the observed minus calculated residuals from the fit and their uncertainties.} 
\label{fig1}} 
\end{figure} 
 
\clearpage

%%%%%%%%%%%%%%%%%%%%%%%%%%%%%%%%%%%%%%%%%%%%%%%%%%%%%%%%%%%%%%%%%%%%%%%%%%%%

\section{Detection of the FUV Flux of the Companion}             % Section 3

We realized at the outset that the companion orbiting the Be star is 
probably too faint for direct detection of its lines in individual UV spectra 
because of the low S/N ($\approx 10$) of the {\it IUE} spectra. 
Consequently, we adopted a search method based on cross-correlation techniques 
that was optimized for detection of a hot companion.  First we selected 
a spectrum template from the non-LTE TLUSTY/SYNSPEC atmosphere 
models\footnote{http://nova.astro.umd.edu/} of \citet{lanz2003} for 
$T_{\rm eff} \approx 45$~kK, $\log g = 4.75$ (the largest gravity value 
in the grid, but probably lower than actual), solar abundances, and 
no rotational broadening.  Next, we restricted the wavelength 
range for the CCF to the far-UV, 1150 -- 1189 \AA\ (omitting the 
broad \ion{C}{3} $\lambda 1176$ feature of the Be star's spectrum), 
in order to focus on the part of the spectrum where the hot companion 
has a relatively larger flux contribution.   Once the CCFs were 
calculated, we subtracted the mean CCF from each to remove any signal 
from correlation between the Be star's spectral features and those
in the model template and to search for a component with the 
expected Doppler shifts of the companion.  

Visual inspection of the individual difference CCFs did not 
immediately reveal any obvious signal from a hot companion. 
Thus, in the final step we shifted each of the difference CCFs
to the {\bf expected} Doppler reference frame of the companion, and 
then formed the mean of all the difference CCFs to seek the signal of the 
companion.   We need to adopt a value of the mass ratio $q=M_2/M_1$ 
in order to estimate the Doppler shifts of the secondary star 
in each observation from the orbital radial velocity curve of 
the Be star (Table~3).  We assumed a Be star mass of $10.5 M_\odot$
(appropriate for its spectral classification; \citealt{peters2001}) 
and then used the observed spectroscopic mass function $f(m)=(M_2 \sin i)^3 /(M_1+M_2)^2$ 
(Table~3) to determine the mass ratio $q$ for a range in assumed 
orbital inclination $i$.   The result ranges from $q=0.077$ for $i=65^\circ$ 
to $q=0.070$ for $i=85^\circ$, and we adopted the latter value 
based upon large inclination needed to account for the appearance 
of the shell features (see Section 4).   

We show in Figure~2 the resulting CCF after shifting and 
averaging all 88 individual difference CCFs formed using the 
hot model spectrum as a template.  We find that there is indeed 
a peak near the expected rest frame zero velocity that corresponds
to correlation with the weak spectral features of the companion. 
The peak reaches a maximum that is $4\sigma$ above the background, 
so we suggest that this approach offers us the first tentative 
detection of the spectrum of the hot companion. 

%\clearpage

% Figure 2
\begin{figure} 
\begin{center} 
{\includegraphics[angle=90,height=12cm]{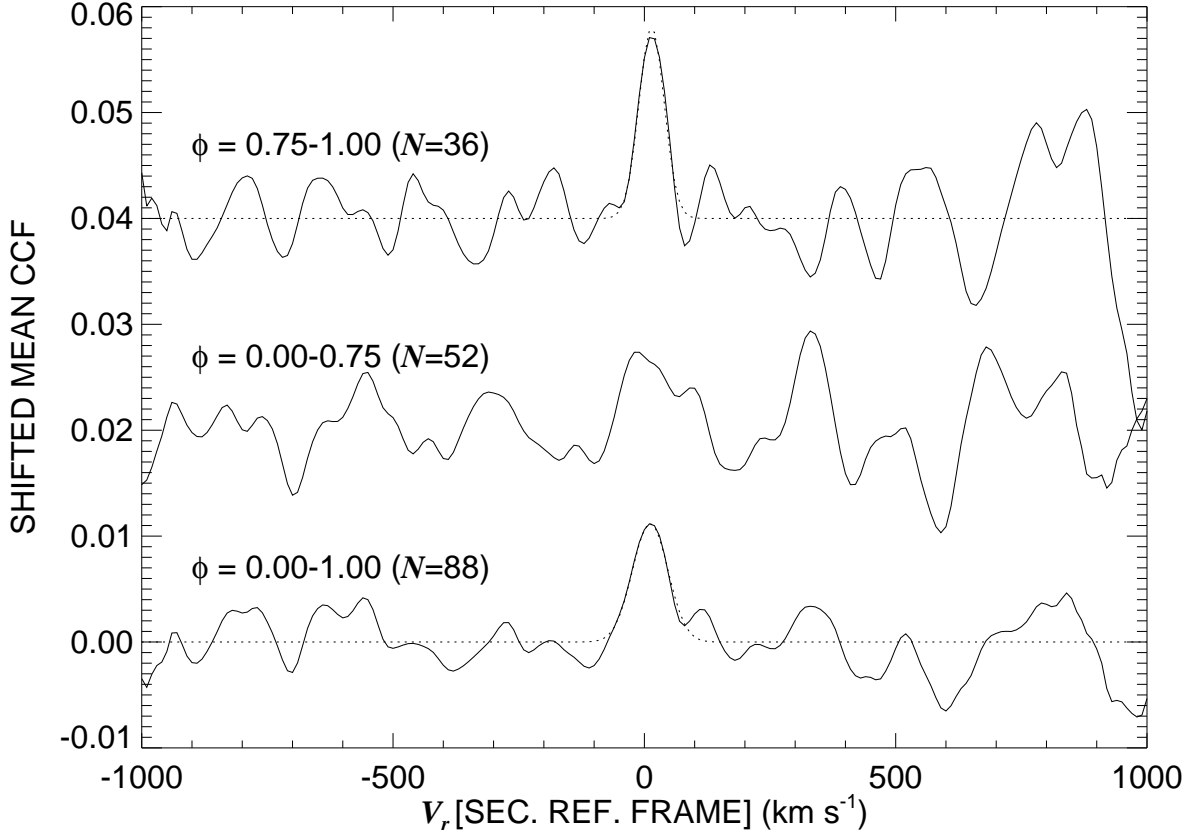}} 
\end{center} 
\caption{The lower plot shows the mean CCF of the FUV spectra (1150--1189 \AA ) 
with a hot model spectrum after subtraction of the overall mean CCF 
and shifting each to the adopted reference frame of the faint secondary.  
A Gaussian fit of the weak central peak is shown as a dotted line.
The middle and upper plots show the same for subsamples from orbital 
phases $\phi = 0.00 - 0.75$ and $0.75 - 1.00$ offset by $+0.02$ and 
$+0.04$, respectively, with the number of spectra $N$ noted in the label. 
The much weaker signal in the phase range $\phi = 0.00 - 0.75$ 
suggests phase-variable obscuration of the secondary spectrum.  
\label{fig2}} 
\end{figure}

Figure~2 also shows a Gaussian fit of the peak with parameters 
of peak height $p=0.0112 \pm 0.0012$, a central position of 
$V_r = 12 \pm 5$ km~s$^{-1}$, and a Gaussian dispersion of 
$\sigma = 35 \pm 5$ km~s$^{-1}$.  We think the small velocity 
offset is insignificant (probably resulting from model mis-match
and uncertainties in the physical value of the systemic velocity),
and the dispersion $\sigma$ is that expected from cross-correlations
of model spectra with no rotational broadening.  Thus, if we 
accept the detection as reliable, then the projected rotational 
velocity of the companion must be small, $V\sin i < 30$ km~s$^{-1}$.

We experimented with using a range in assumed model $T_{\rm eff}$ and 
mass ratio $q=M_2/M_1$ to find values that maximized the peak height of the 
CCF signal.  We took the uncertainty associated with the Gaussian 
fit of the peak height to find the range over which the resulting 
peak height exceeded $p({\rm max}) - 1 \sigma = 0.010$. 
The optimal results are $T_{\rm eff} = 43 \pm 5$~kK and 
$q = 0.072 \pm 0.021$.  The derived mass ratio $q$ agrees with our
estimate from the mass function (based upon $f(m)$, $M_1$, $i$). 
We caution that the $T_{\rm eff}$ estimate is probably a lower limit,
because the model template was based on a smaller than actual 
$\log g$ and the ionization balance reflected in the relative 
line strengths will lead to higher temperature at higher gravity. 

Figure~2 also shows the results of restricting the sample according
to orbital phase.  We found that the peak is weak or absent in 
the difference CCFs from orbital phases $\phi =0.00$ to 0.75 
(middle plot) but is relatively stronger in the range $\phi =0.75$ 
to 1.00 (upper plot), i.e., between the times of fastest secondary
approach and secondary inferior conjunction.  This is suggestive 
of phase-variable obscuration of the flux of the hot companion 
(perhaps due to circumstellar gas structures; see the discussion in Section 5). 
Because some obscuration may also be present in the phase range $\phi =0.75$ to 1.00,
we will assume that the peak height $p=0.0180 \pm 0.0023$
observed then represents a lower limit to the signal from the
flux contribution of the hot secondary star. 

Finally, we made a sequence of model difference CCFs using the 
mean spectrum to represent that of the Be star and the optimum 
model template for the secondary, and the flux contributions of
both stars were summed for a grid of assumed values of flux ratio 
$f_2/f_1$.   We measured the height of the resulting peak in the 
difference CCF to form a linear relation between $f_2/f_1$ and $p$. 
This relationship leads to a monochromatic flux ratio estimate of 
$f_2/f_1 > 0.0088 \pm 0.0011$ in the region from 1150 to 1189 \AA , 
i.e., the companion contributes at least $0.9\%$ of the flux in this
spectral range. 

The observed flux ratio is approximately related to radius ratio by
\begin{equation}
{f_2 \over f_1} = {F_2 \over F_1} \left({R_2 \over R_1}\right)^2
\end{equation}
where $F_2 / F_1$ is the monochromatic flux ratio per unit area. 
Using adopted temperatures of 21~kK and 43~kK for the Be star 
and companion, respectively, we estimate that 
$F_2 / F_1 = 12.8$ at 1170 \AA\ based upon the TLUSTY/SYNSPEC models. 
Then the radius ratio is $R_2 / R_1 > 0.026 \pm 0.002$, and if 
we assume $R_1 = 5 R_\odot$ based upon the spectral classification of 
the Be star, then the companion radius is $R_2 > 0.13 R_\odot$. 

%%%%%%%%%%%%%%%%%%%%%%%%%%%%%%%%%%%%%%%%%%%%%%%%%%%%%%%%%%%%%%%%%%%%%%%%%%%%

\section{Orbital Variations of Shell Lines}                      % Section 4

The shell absorption features that appear each orbit in the spectrum of HR~2142 
may originate in disk structures created by the gravity of the companion 
(Section 5), so it is important to document the appearance and orbital 
variability of these shell components.  The optical spectral variations 
of the shell lines were described by \citet{peters1983}, and they begin 
about six days prior to Be star superior conjunction displaying 
red-shifts in a {\it primary shell phase}, and this is followed by a 
short 1.5 day {\it secondary shell phase} of blue-shifted absorption 
immediately after the time of conjunction.  The UV spectrum is rich 
in similar shell line features, and in this section we present plots 
of orbital variations in the strongest of the observed shell features. 

Figure Set 3 (given in full in the electronic edition) documents the 
shell line variations for 18 representative and strong absorption features. 
Rest wavelengths are derived from \citet{morton1988} or the 
NIST database\footnote{http://physics.nist.gov/cgi-bin/ASD/lines1.pl}.
The top panel of each figure consists of line plots offset in flux according
to the orbital phase of observation, and the lower panel shows the same 
set of spectra represented as a gray scale image of spectral flux strength 
as a function of Doppler shift and orbital phase.  Many of these features 
also display a radial velocity constant component that is formed in the 
interstellar medium and/or Be disk.  Thus, it is informative to begin the
discussion with a representative case where the constant component is 
absent.  An excellent example is the \ion{Si}{2} $\lambda 1264$ line that
is shown in Figure 3.5.  Here the shell feature begins near $V_r = 0$ km~s$^{-1}$
shortly after orbital phase $\phi =0.5$ (Be star inferior conjunction), 
and it becomes wider and deeper to reach maximum absorption just before $\phi =0.0$.
Then the red-shifted absorption abruptly disappears and a short-lived 
blue-shifted component is briefly seen.   These two manifestations are 
the UV counterparts of the {\it primary} and {\it secondary shell phases} 
seen in optical spectral lines \citep{peters1983}.  The absorption maximum 
during the {\it primary shell phase} occurs near $\phi=0.95$ and spans a
velocity range of $+10$ to $+160$ km~s$^{-1}$.  The {\it secondary shell phase} 
occurs near $\phi=0.03$ and spans a velocity range of $-40$ to $+20$ km~s$^{-1}$.
This basic pattern of variation is repeated in many other lines including 
\ion{Si}{2} $\lambda 1193$ (Fig.\ 3.1), \ion{Si}{3} $\lambda 1299$ (Fig.\ 3.6), 
and \ion{S}{3} $\lambda 1200$ (Fig.\ 3.2). 

\clearpage

% Figure 3: in full in submitted ms

% Figure Set 3 
\figsetstart
\figsetnum{3}
\figsettitle{Shell Line Variations}

\figsetgrpstart
\figsetgrpnum{3.1}
\figsetgrptitle{r1}
\figsetplot{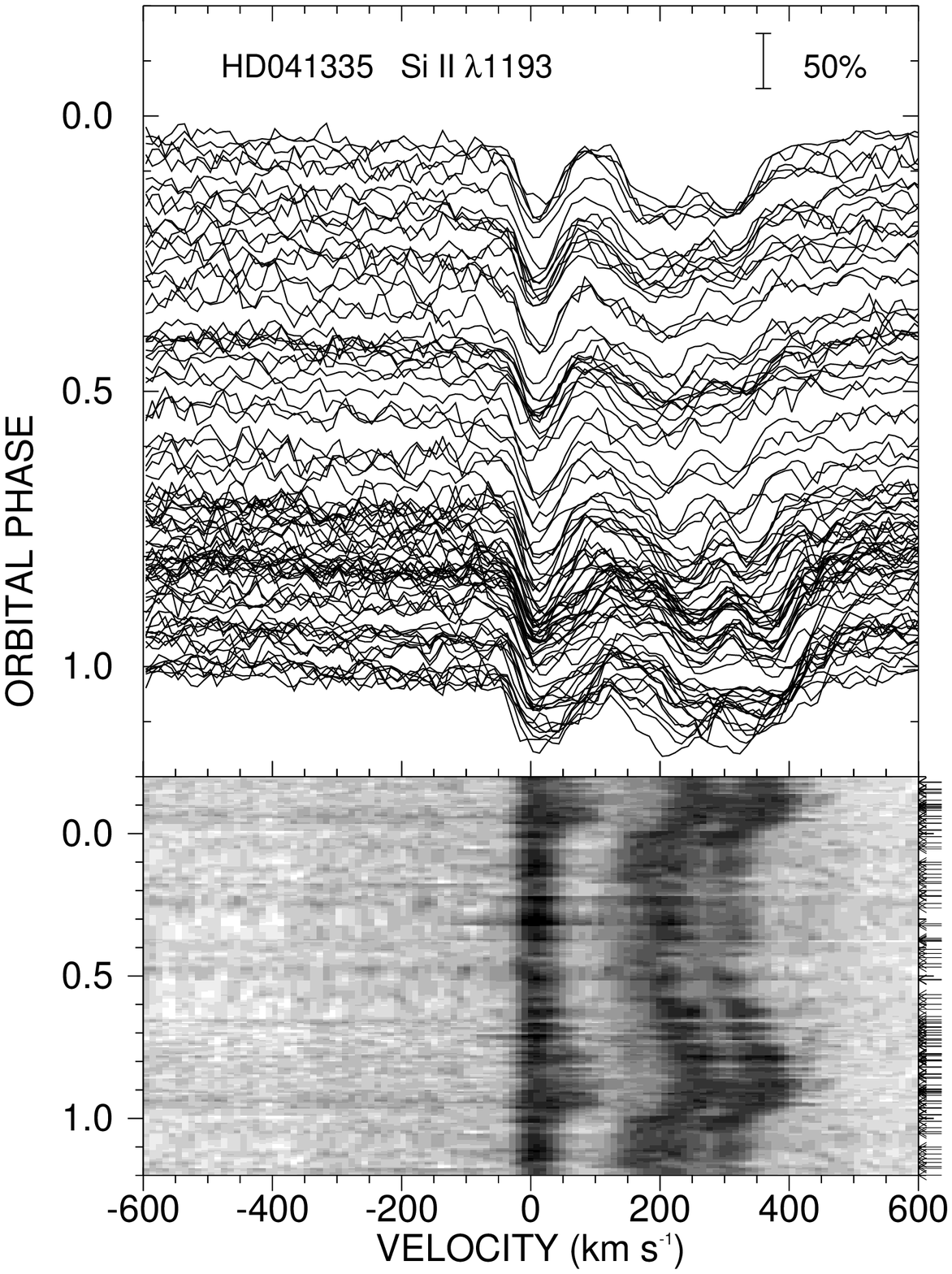}
\figsetgrpnote{The top panel shows the individual spectra of
\ion{Si}{2} $\lambda 1193.290$ plotted as a function of radial 
velocity and orbital phase (such that the continuum is aligned 
with the phase of observation).  The spectral flux depth relative
to the continuum is indicated by the scale bar in upper right. 
The lower panel shows the same interpolated in orbital phase 
and portrayed as a gray scale image (black corresponding to deepest
absorption and white to strongest continuum flux).  The actual
phases of observation are indicated by arrows on the right hand side. 
The lines of \ion{S}{3} $\lambda\lambda 1194.061, 1194.457$ appear offset 
to higher velocity.}
\figsetgrpend

\figsetgrpstart
\figsetgrpnum{3.2}
\figsetgrptitle{r2}
\figsetplot{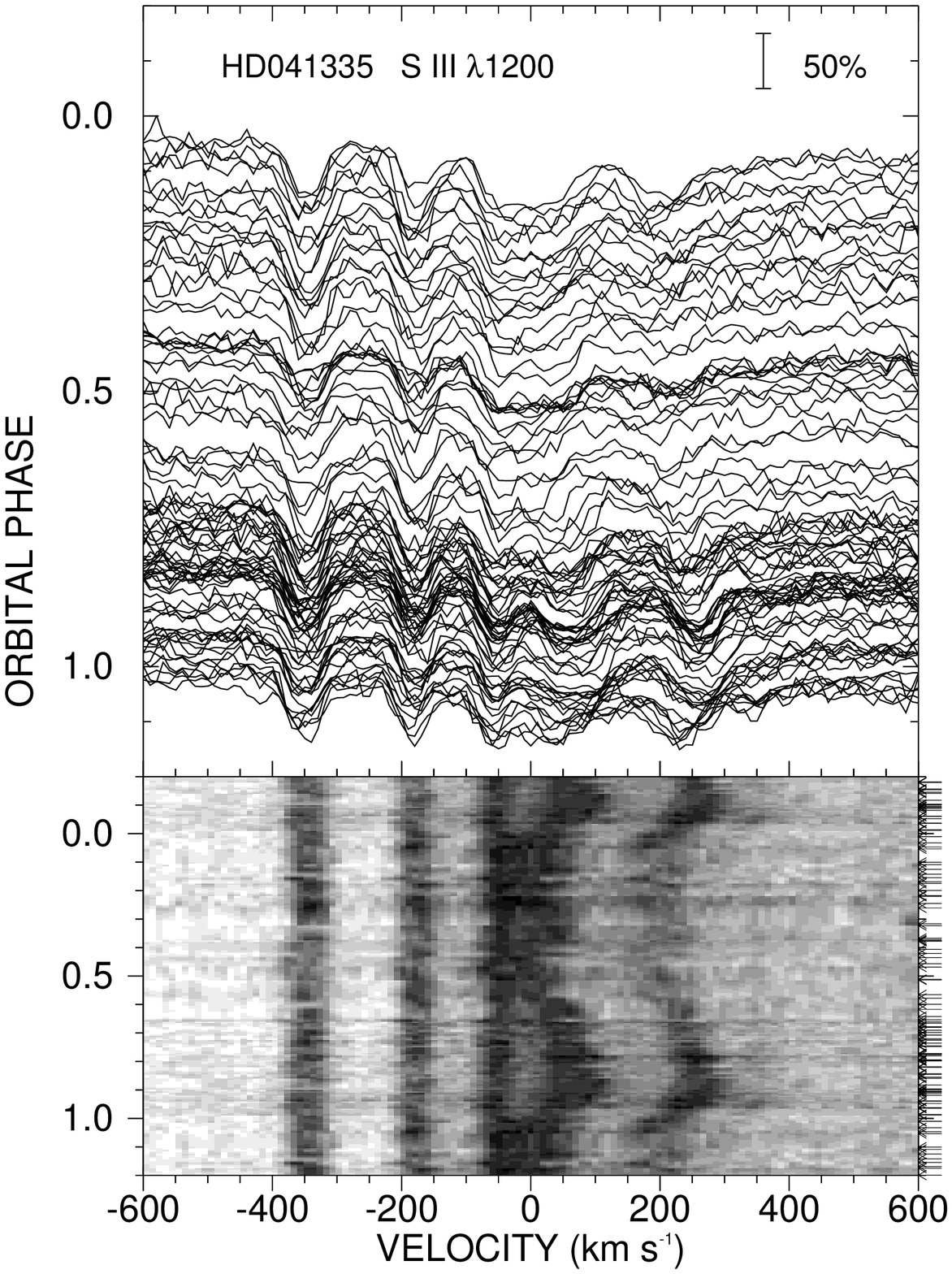}
\figsetgrpnote{The top panel shows the individual spectra of
\ion{S}{3} $\lambda 1200.970$ plotted as a function of radial 
velocity and orbital phase (such that the continuum is aligned 
with the phase of observation).  The spectral flux depth relative
to the continuum is indicated by the scale bar in upper right. 
The lower panel shows the same interpolated in orbital phase 
and portrayed as a gray scale image (black corresponding to deepest
absorption and white to strongest continuum flux).  The actual
phases of observation are indicated by arrows on the right hand side. 
The stationary lines of \ion{N}{1} $\lambda\lambda 1199.5,1200.2,1200.7$
appear offset to lower velocity, while \ion{S}{3} $\lambda 1201.730$
appears at higher velocity.}
\figsetgrpend

\figsetgrpstart
\figsetgrpnum{3.3}
\figsetgrptitle{r3}
\figsetplot{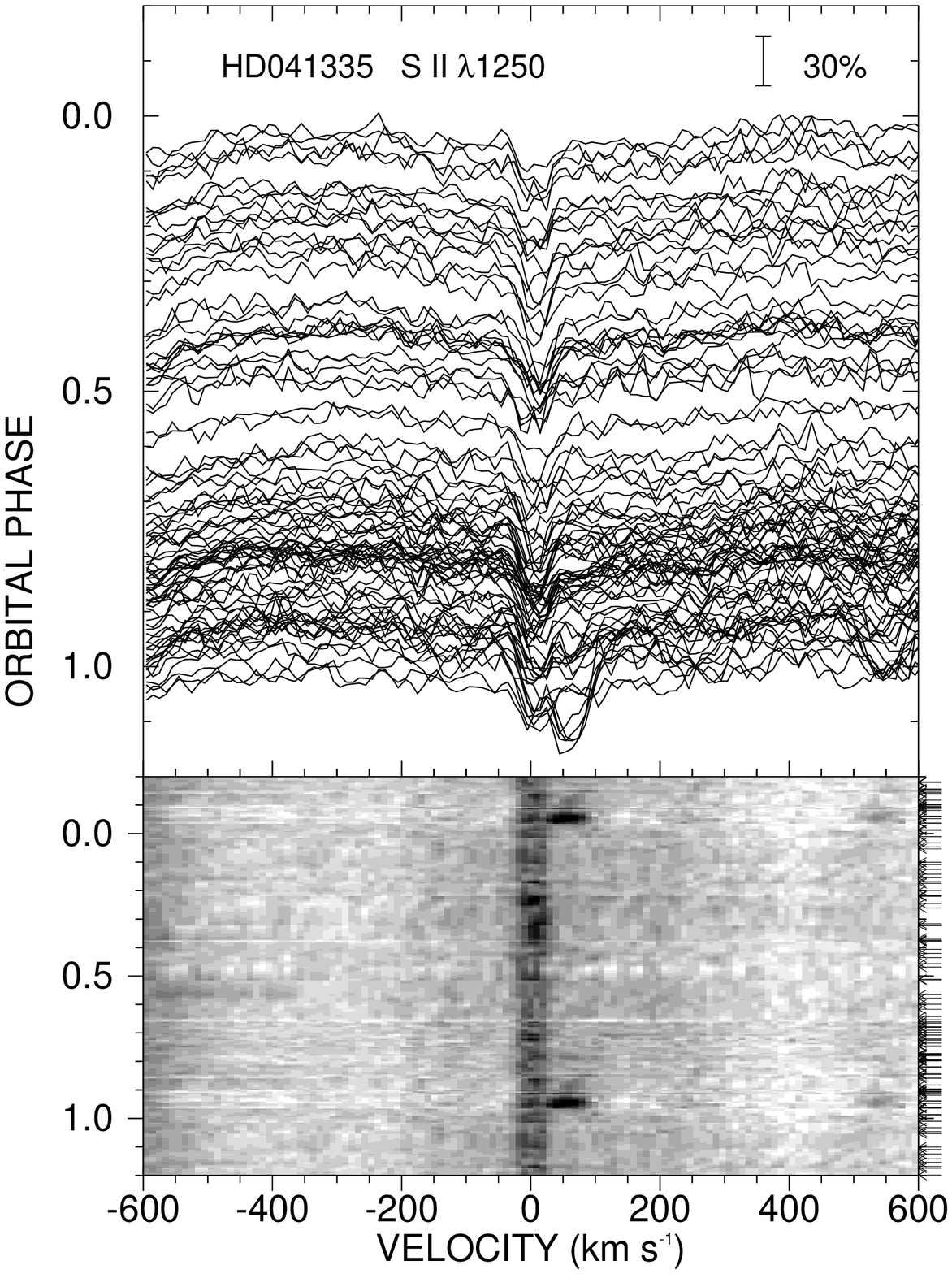}
\figsetgrpnote{The top panel shows the individual spectra of
\ion{S}{2} $\lambda 1250.583$ plotted as a function of radial 
velocity and orbital phase (such that the continuum is aligned 
with the phase of observation).  The spectral flux depth relative
to the continuum is indicated by the scale bar in upper right. 
The lower panel shows the same interpolated in orbital phase 
and portrayed as a gray scale image (black corresponding to deepest
absorption and white to strongest continuum flux).  The actual
phases of observation are indicated by arrows on the right hand side. 
The shell line is visible only close to orbital phase $\phi=0.95$.}
\figsetgrpend

\figsetgrpstart
\figsetgrpnum{3.4}
\figsetgrptitle{r4}
\figsetplot{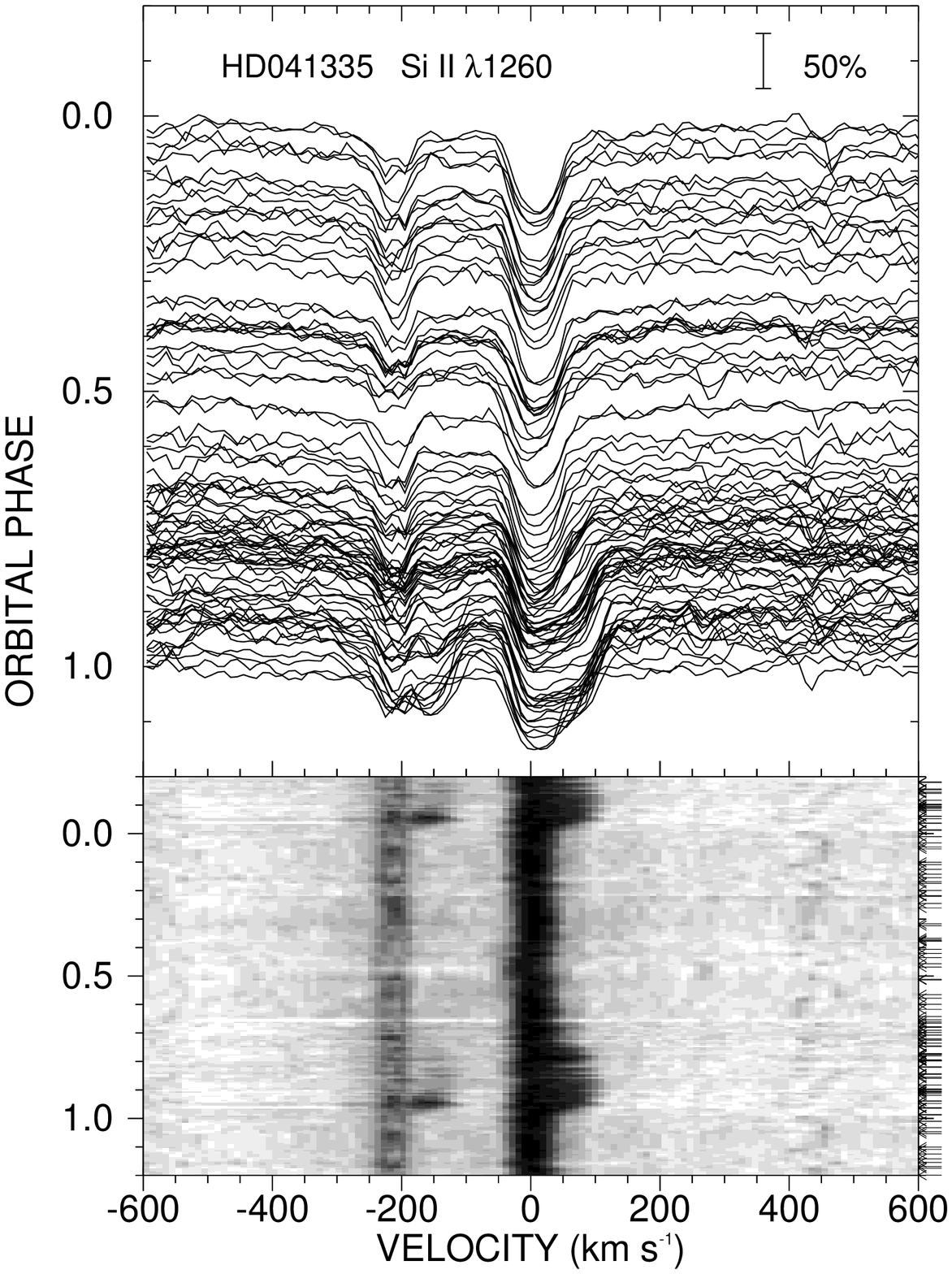}
\figsetgrpnote{The top panel shows the individual spectra of
\ion{Si}{2} $\lambda 1260.4223$ plotted as a function of radial 
velocity and orbital phase (such that the continuum is aligned 
with the phase of observation).  The spectral flux depth relative
to the continuum is indicated by the scale bar in upper right. 
The lower panel shows the same interpolated in orbital phase 
and portrayed as a gray scale image (black corresponding to deepest
absorption and white to strongest continuum flux).  The actual
phases of observation are indicated by arrows on the right hand side. 
The shell component of \ion{S}{2} $\lambda 1259.518$ (offset to 
lower velocity) is visible only close to orbital phase $\phi=0.95$.}
\figsetgrpend

\figsetgrpstart
\figsetgrpnum{3.5}
\figsetgrptitle{r5}
\figsetplot{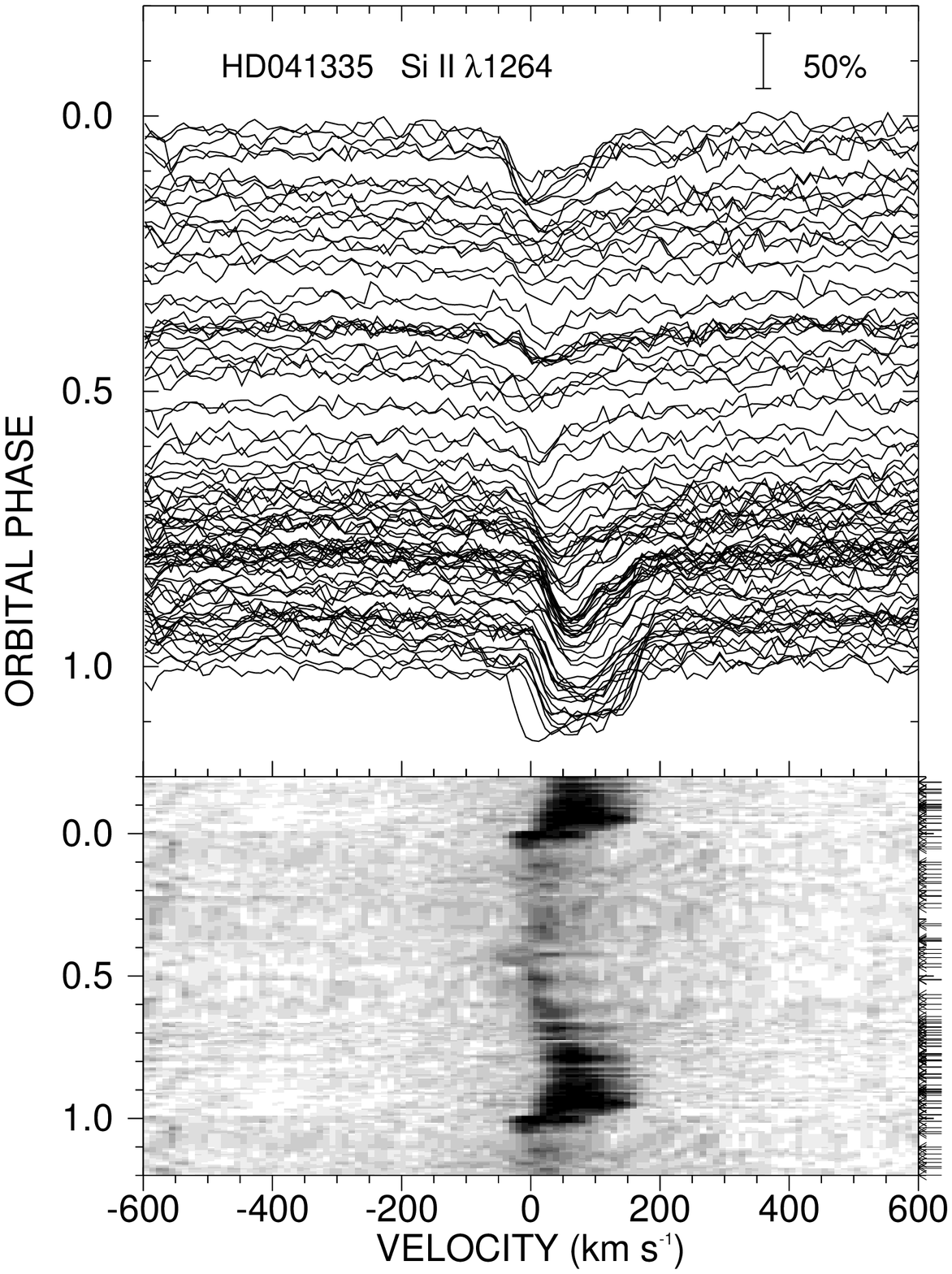}
\figsetgrpnote{The top panel shows the individual spectra of
\ion{Si}{2} $\lambda 1264.73$ plotted as a function of radial 
velocity and orbital phase (such that the continuum is aligned 
with the phase of observation).  The spectral flux depth relative
to the continuum is indicated by the scale bar in upper right. 
The lower panel shows the same interpolated in orbital phase 
and portrayed as a gray scale image (black corresponding to deepest
absorption and white to strongest continuum flux).  The actual
phases of observation are indicated by arrows on the right hand side.}
\figsetgrpend

\figsetgrpstart
\figsetgrpnum{3.6}
\figsetgrptitle{r6}
\figsetplot{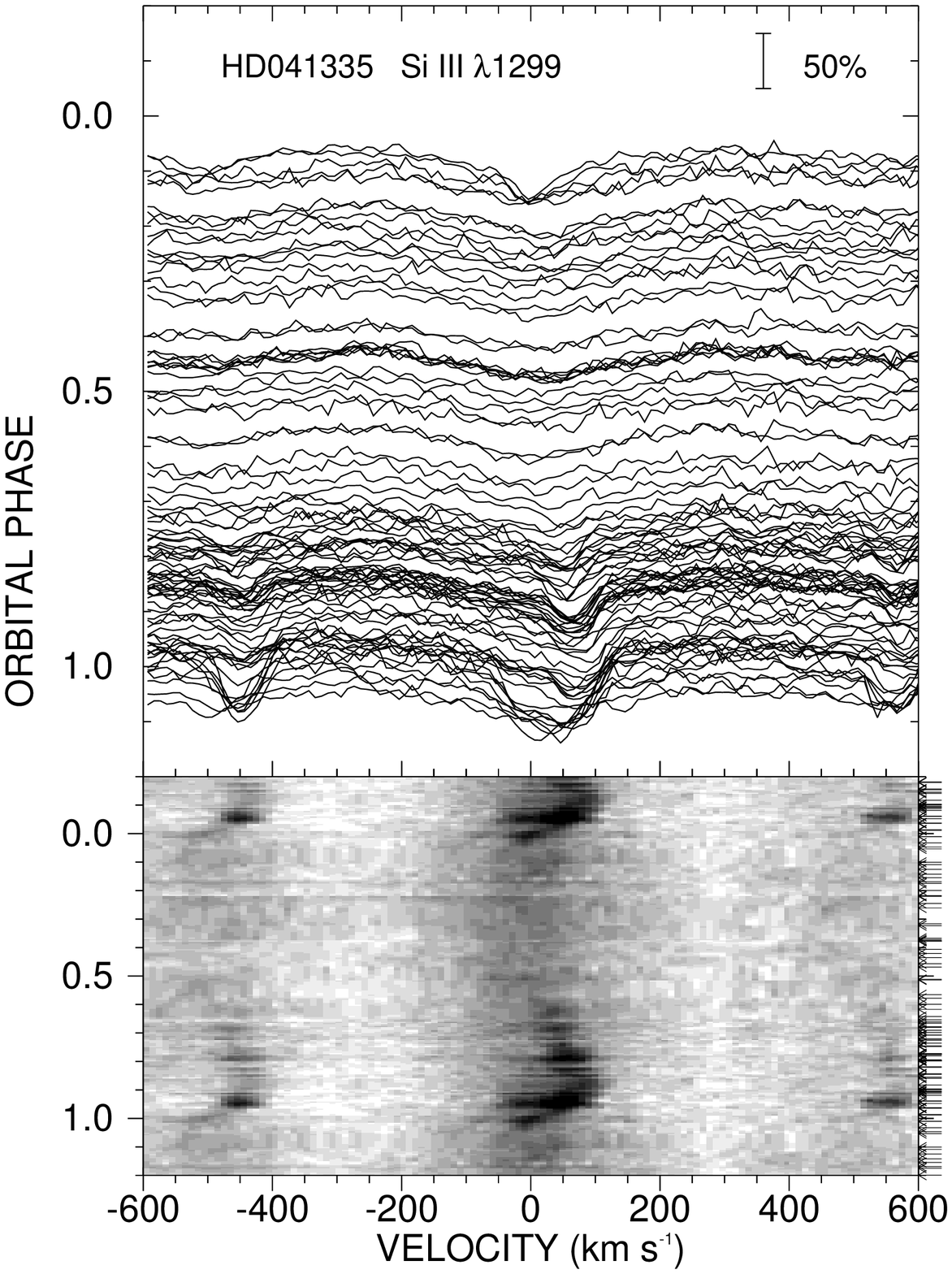}
\figsetgrpnote{The top panel shows the individual spectra of
\ion{Si}{3} $\lambda 1298.96$ plotted as a function of radial 
velocity and orbital phase (such that the continuum is aligned 
with the phase of observation).  The spectral flux depth relative
to the continuum is indicated by the scale bar in upper right. 
The lower panel shows the same interpolated in orbital phase 
and portrayed as a gray scale image (black corresponding to deepest
absorption and white to strongest continuum flux).  The actual
phases of observation are indicated by arrows on the right hand side. 
The lines of \ion{Si}{3} $\lambda\lambda 1296.73,1298.89,1301.15$ also 
appear as sudden shell features at orbital phase $\phi=0.95$.}
\figsetgrpend

\figsetgrpstart
\figsetgrpnum{3.7}
\figsetgrptitle{r7}
\figsetplot{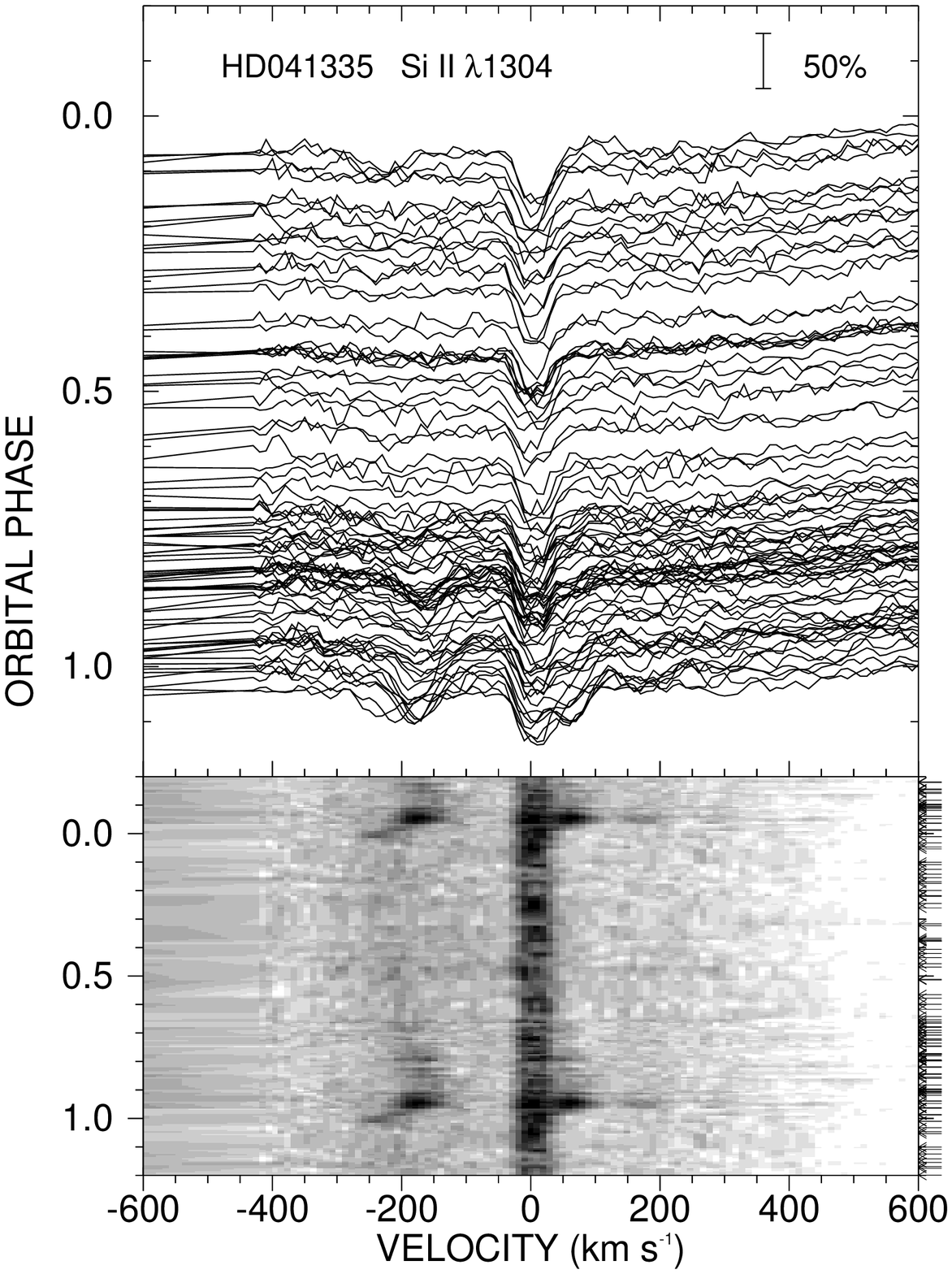}
\figsetgrpnote{The top panel shows the individual spectra of
\ion{Si}{2} $\lambda 1304.3711$ plotted as a function of radial 
velocity and orbital phase (such that the continuum is aligned 
with the phase of observation).  The spectral flux depth relative
to the continuum is indicated by the scale bar in upper right. 
The lower panel shows the same interpolated in orbital phase 
and portrayed as a gray scale image (black corresponding to deepest
absorption and white to strongest continuum flux).  The actual
phases of observation are indicated by arrows on the right hand side. 
The shell component appears suddenly at orbital phase $\phi=0.95$.
The flat region at left represents a linear interpolation across
the interstellar line of \ion{O}{1} $\lambda 1302$.}
\figsetgrpend

\figsetgrpstart
\figsetgrpnum{3.8}
\figsetgrptitle{r8}
\figsetplot{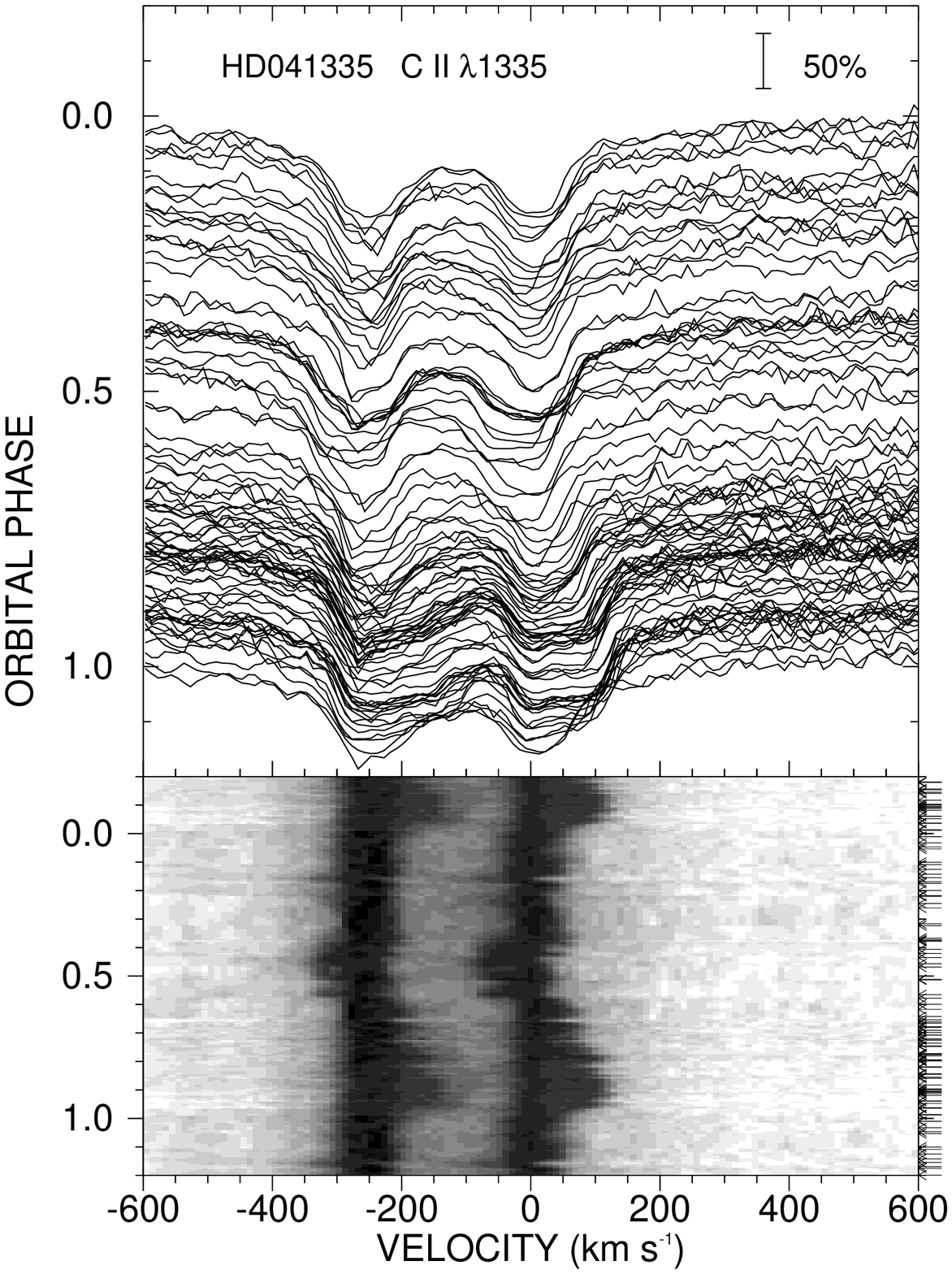}
\figsetgrpnote{The top panel shows the individual spectra of
\ion{C}{2} $\lambda 1335.708$ plotted as a function of radial 
velocity and orbital phase (such that the continuum is aligned 
with the phase of observation).  The spectral flux depth relative
to the continuum is indicated by the scale bar in upper right. 
The lower panel shows the same interpolated in orbital phase 
and portrayed as a gray scale image (black corresponding to deepest
absorption and white to strongest continuum flux).  The actual
phases of observation are indicated by arrows on the right hand side. 
The similar line of \ion{C}{2} $\lambda 1334.532$ appears offset 
to lower velocity.}
\figsetgrpend

\figsetgrpstart
\figsetgrpnum{3.9}
\figsetgrptitle{r9}
\figsetplot{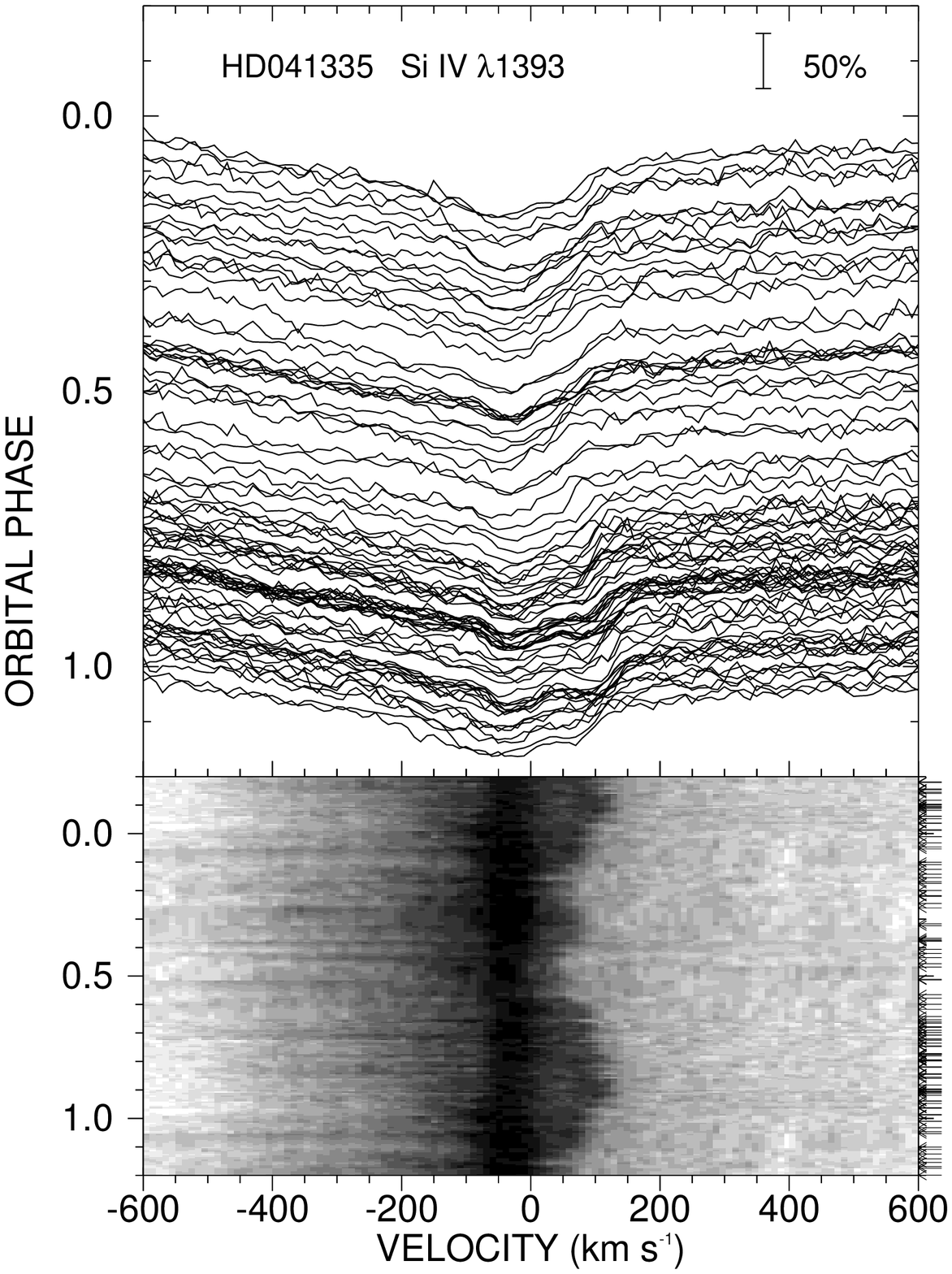}
\figsetgrpnote{The top panel shows the individual spectra of
\ion{Si}{4} $\lambda 1393.755$ plotted as a function of radial 
velocity and orbital phase (such that the continuum is aligned 
with the phase of observation).  The spectral flux depth relative
to the continuum is indicated by the scale bar in upper right. 
The lower panel shows the same interpolated in orbital phase 
and portrayed as a gray scale image (black corresponding to deepest
absorption and white to strongest continuum flux).  The actual
phases of observation are indicated by arrows on the right hand side. 
The feature is formed mainly in the wind of the Be star, but shell
variations are present on the high velocity side of the core.}
\figsetgrpend

\figsetgrpstart
\figsetgrpnum{3.10}
\figsetgrptitle{r10}
\figsetplot{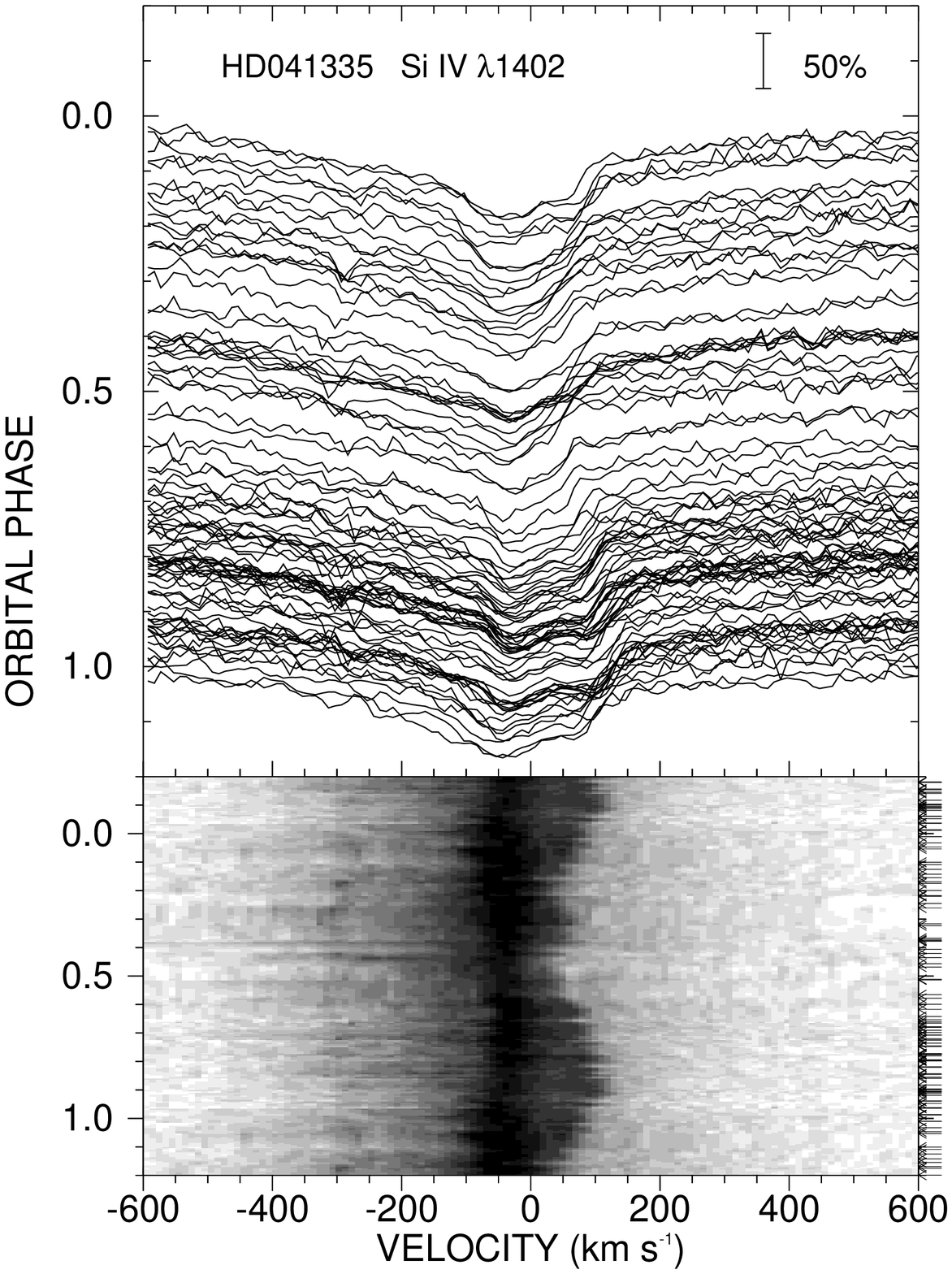}
\figsetgrpnote{The top panel shows the individual spectra of
\ion{Si}{4} $\lambda 1402.770$ plotted as a function of radial 
velocity and orbital phase (such that the continuum is aligned 
with the phase of observation).  The spectral flux depth relative
to the continuum is indicated by the scale bar in upper right. 
The lower panel shows the same interpolated in orbital phase 
and portrayed as a gray scale image (black corresponding to deepest
absorption and white to strongest continuum flux).  The actual
phases of observation are indicated by arrows on the right hand side. 
The feature is formed mainly in the wind of the Be star, but shell
variations are present on the high velocity side of the core.}
\figsetgrpend

\figsetgrpstart
\figsetgrpnum{3.11}
\figsetgrptitle{r11}
\figsetplot{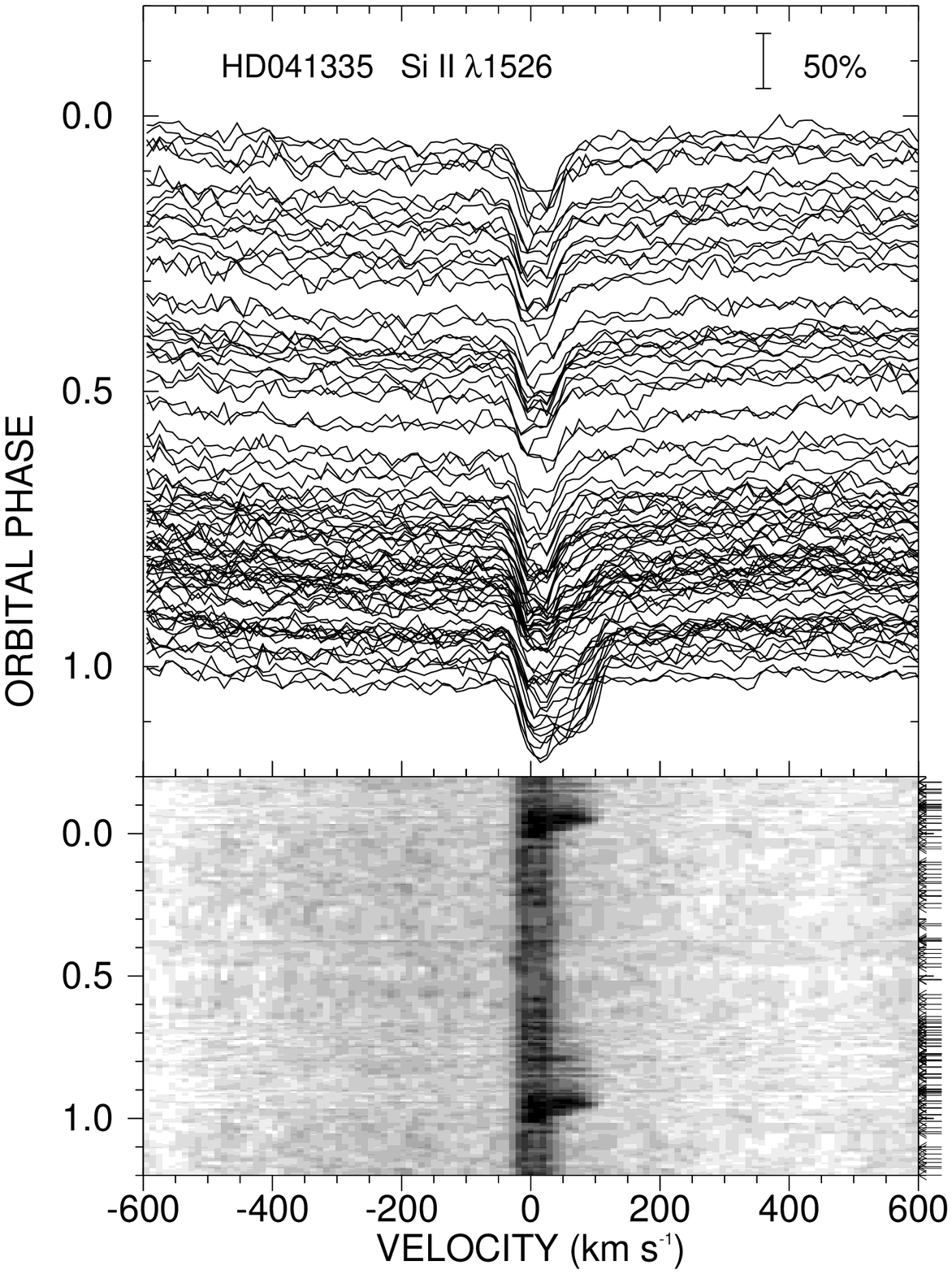}
\figsetgrpnote{The top panel shows the individual spectra of
\ion{Si}{2} $\lambda 1526.7071$ plotted as a function of radial 
velocity and orbital phase (such that the continuum is aligned 
with the phase of observation).  The spectral flux depth relative
to the continuum is indicated by the scale bar in upper right. 
The lower panel shows the same interpolated in orbital phase 
and portrayed as a gray scale image (black corresponding to deepest
absorption and white to strongest continuum flux).  The actual
phases of observation are indicated by arrows on the right hand side. 
The shell component appears suddenly at orbital phase $\phi=0.95$.}
\figsetgrpend

\figsetgrpstart
\figsetgrpnum{3.12}
\figsetgrptitle{r12}
\figsetplot{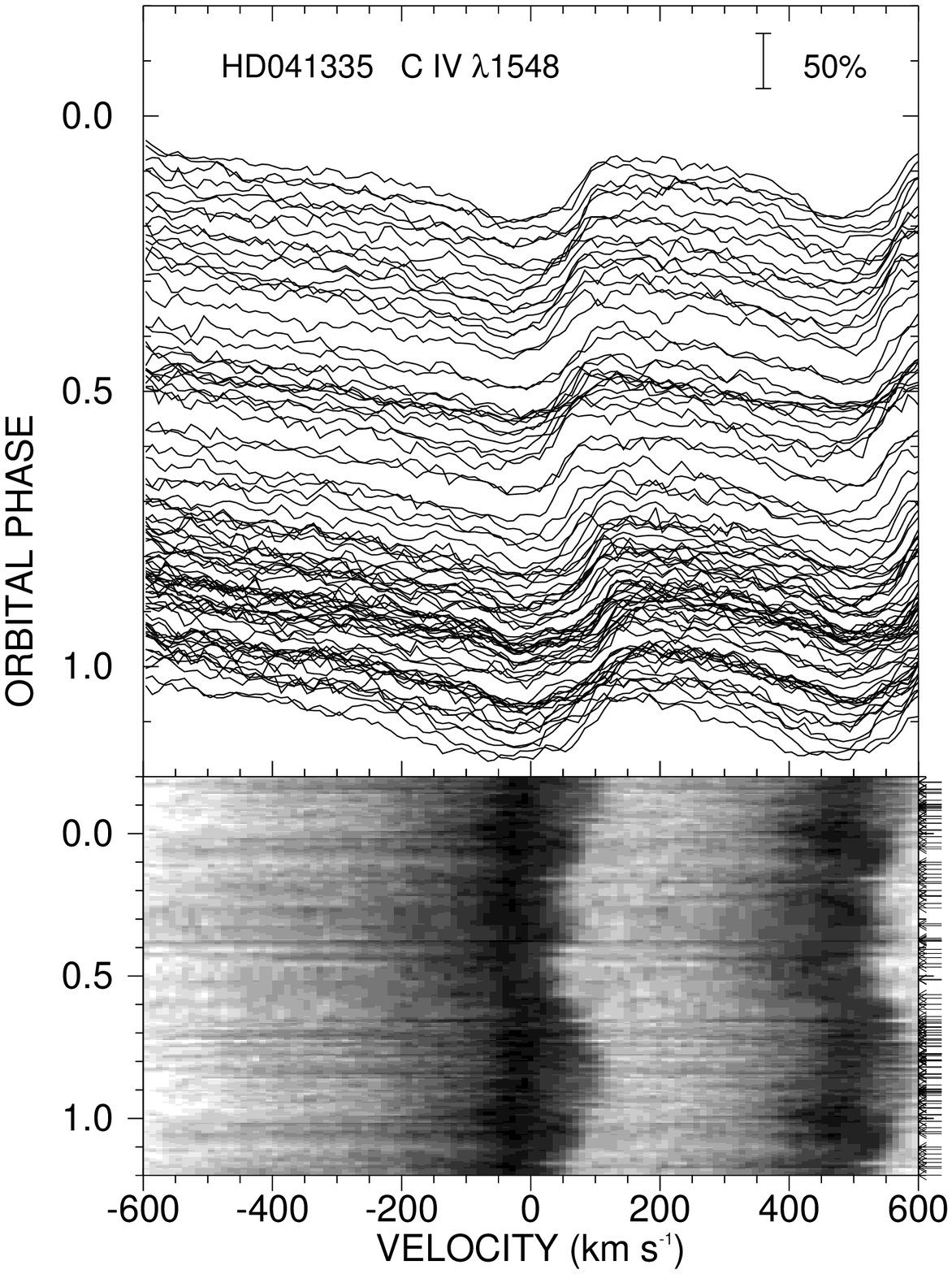}
\figsetgrpnote{The top panel shows the individual spectra of
\ion{C}{4} $\lambda 1548.202$ plotted as a function of radial 
velocity and orbital phase (such that the continuum is aligned 
with the phase of observation).  The spectral flux depth relative
to the continuum is indicated by the scale bar in upper right. 
The lower panel shows the same interpolated in orbital phase 
and portrayed as a gray scale image (black corresponding to deepest
absorption and white to strongest continuum flux).  The actual
phases of observation are indicated by arrows on the right hand side. 
The similar line of \ion{C}{4} $\lambda 1550.774$ appears offset 
to higher velocity.  Both features are formed mainly in the wind 
of the Be star, but shell variations are present on the high 
velocity side of the core.}
\figsetgrpend

\figsetgrpstart
\figsetgrpnum{3.13}
\figsetgrptitle{r13}
\figsetplot{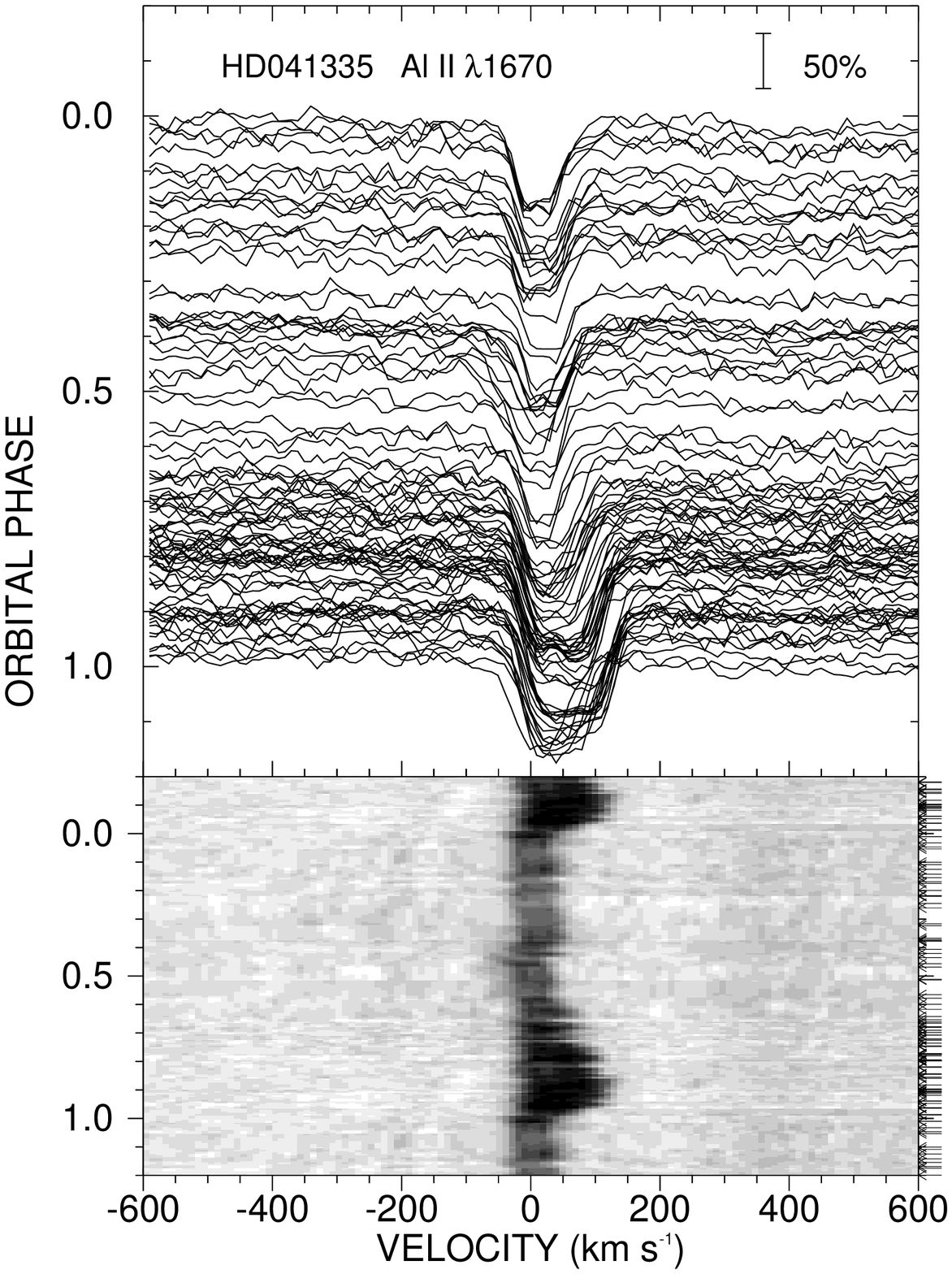}
\figsetgrpnote{The top panel shows the individual spectra of
\ion{Al}{2} $\lambda 1670.7867$ plotted as a function of radial 
velocity and orbital phase (such that the continuum is aligned 
with the phase of observation).  The spectral flux depth relative
to the continuum is indicated by the scale bar in upper right. 
The lower panel shows the same interpolated in orbital phase 
and portrayed as a gray scale image (black corresponding to deepest
absorption and white to strongest continuum flux).  The actual
phases of observation are indicated by arrows on the right hand side.}
\figsetgrpend

\figsetgrpstart
\figsetgrpnum{3.14}
\figsetgrptitle{r14}
\figsetplot{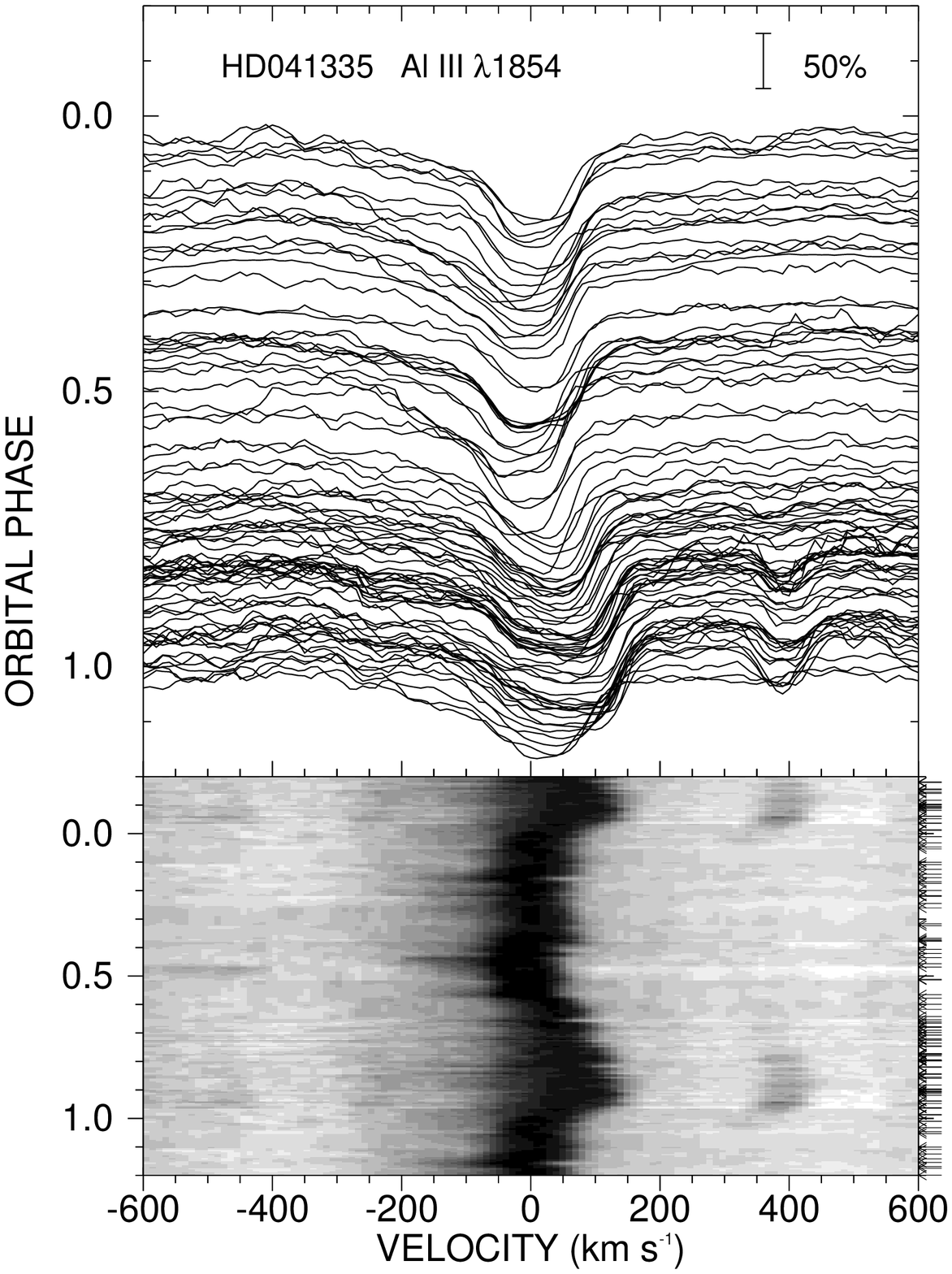}
\figsetgrpnote{The top panel shows the individual spectra of
\ion{Al}{3} $\lambda 1854.7164$ plotted as a function of radial 
velocity and orbital phase (such that the continuum is aligned 
with the phase of observation).  The spectral flux depth relative
to the continuum is indicated by the scale bar in upper right. 
The lower panel shows the same interpolated in orbital phase 
and portrayed as a gray scale image (black corresponding to deepest
absorption and white to strongest continuum flux).  The actual
phases of observation are indicated by arrows on the right hand side.}
\figsetgrpend

\figsetgrpstart
\figsetgrpnum{3.15}
\figsetgrptitle{r15}
\figsetplot{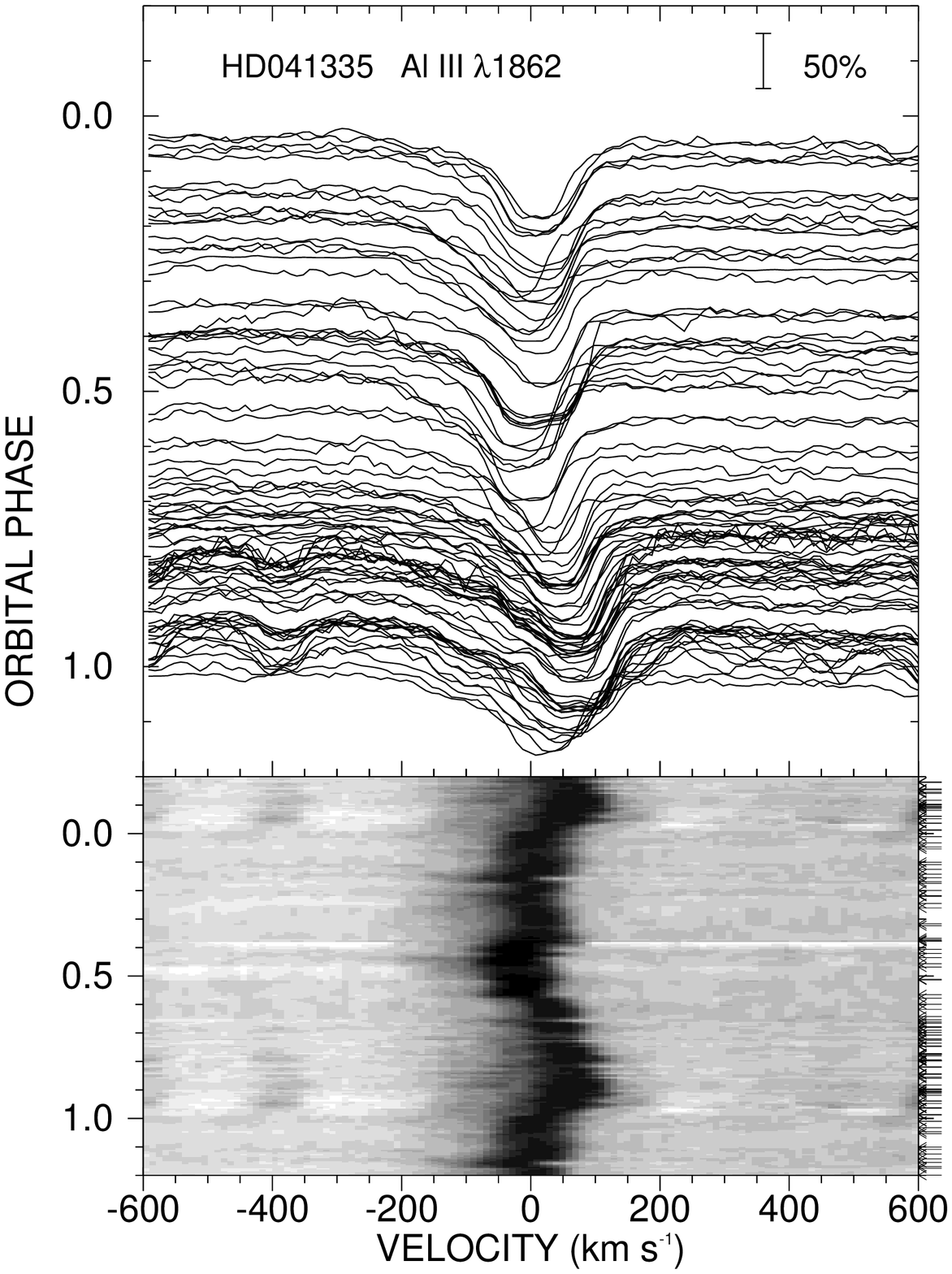}
\figsetgrpnote{The top panel shows the individual spectra of
\ion{Al}{3} $\lambda 1862.7895$ plotted as a function of radial 
velocity and orbital phase (such that the continuum is aligned 
with the phase of observation).  The spectral flux depth relative
to the continuum is indicated by the scale bar in upper right. 
The lower panel shows the same interpolated in orbital phase 
and portrayed as a gray scale image (black corresponding to deepest
absorption and white to strongest continuum flux).  The actual
phases of observation are indicated by arrows on the right hand side.}
\figsetgrpend

\figsetgrpstart
\figsetgrpnum{3.16}
\figsetgrptitle{r16}
\figsetplot{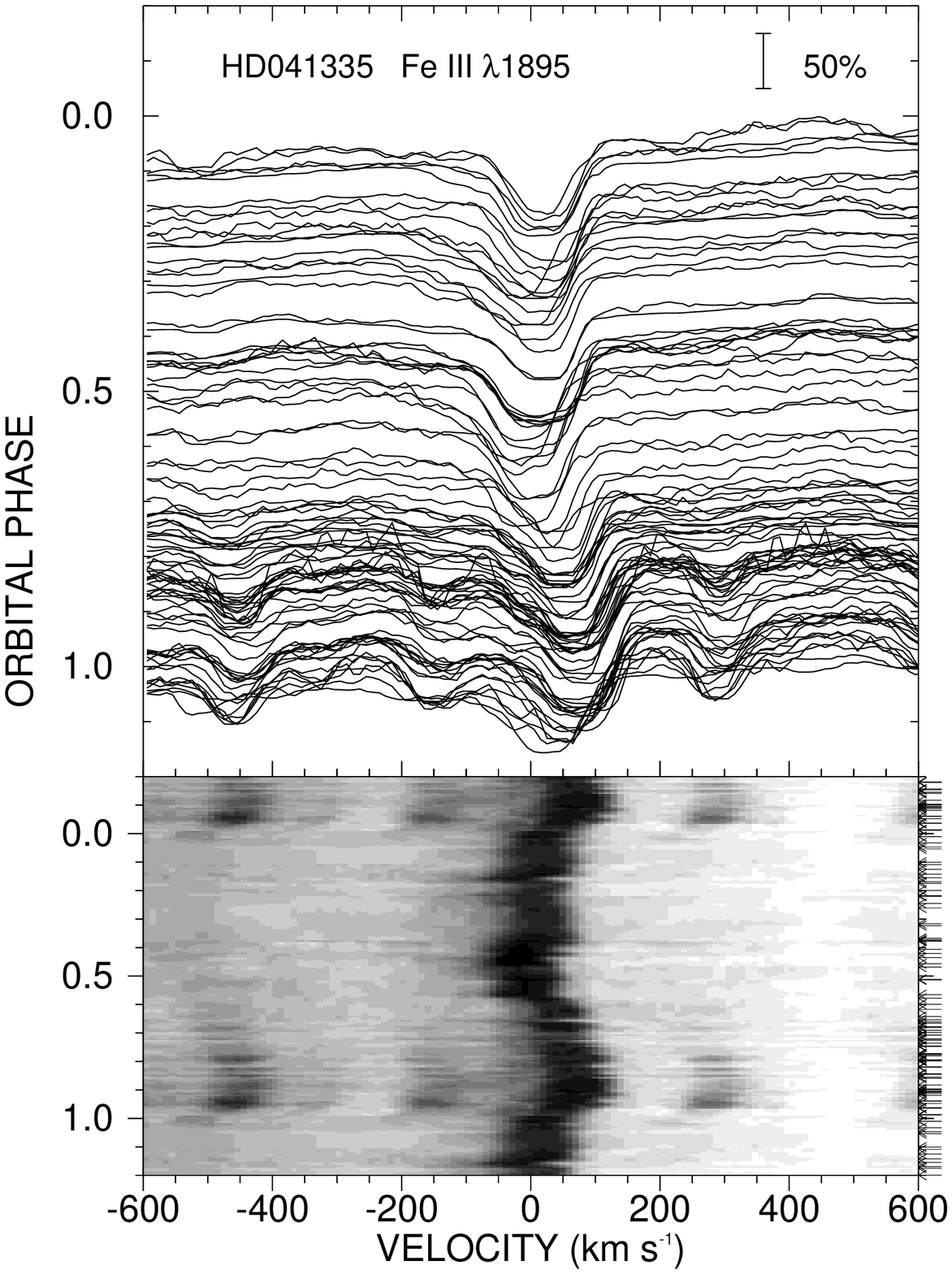}
\figsetgrpnote{The top panel shows the individual spectra of
\ion{Fe}{3} $\lambda 1895.46$ plotted as a function of radial 
velocity and orbital phase (such that the continuum is aligned 
with the phase of observation).  The spectral flux depth relative
to the continuum is indicated by the scale bar in upper right. 
The lower panel shows the same interpolated in orbital phase 
and portrayed as a gray scale image (black corresponding to deepest
absorption and white to strongest continuum flux).  The actual
phases of observation are indicated by arrows on the right hand side.
The lines of \ion{Fe}{3} $\lambda\lambda 1895.46, 1896.814$ also 
appear in the figure.}
\figsetgrpend

\figsetgrpstart
\figsetgrpnum{3.17}
\figsetgrptitle{r17}
\figsetplot{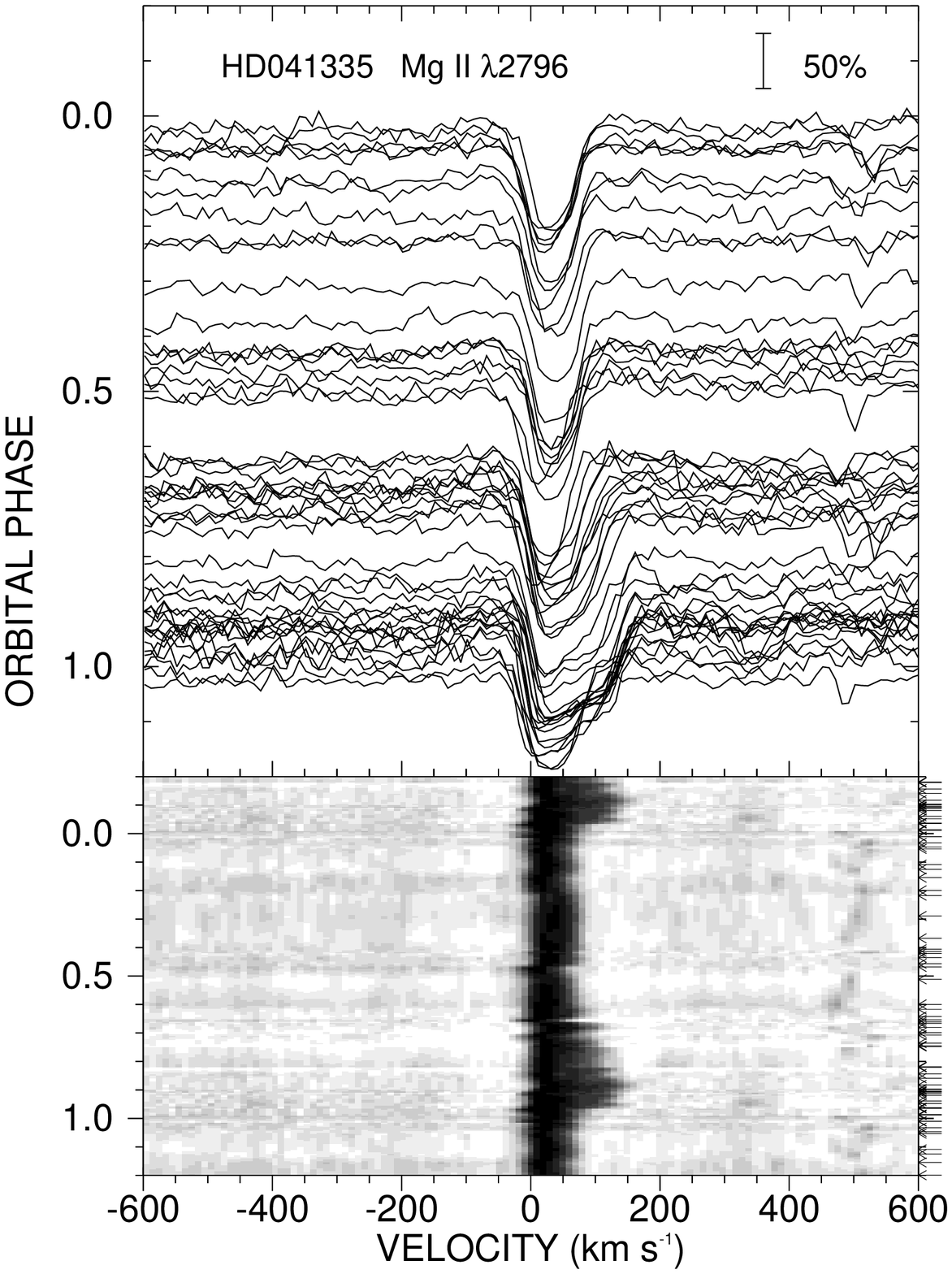}
\figsetgrpnote{The top panel shows the individual spectra of
\ion{Mg}{2} $\lambda 2796.352$ plotted as a function of radial 
velocity and orbital phase (such that the continuum is aligned 
with the phase of observation).  The spectral flux depth relative
to the continuum is indicated by the scale bar in upper right. 
The lower panel shows the same interpolated in orbital phase 
and portrayed as a gray scale image (black corresponding to deepest
absorption and white to strongest continuum flux).  The actual
phases of observation are indicated by arrows on the right hand side.}
\figsetgrpend

\figsetgrpstart
\figsetgrpnum{3.18}
\figsetgrptitle{r18}
\figsetplot{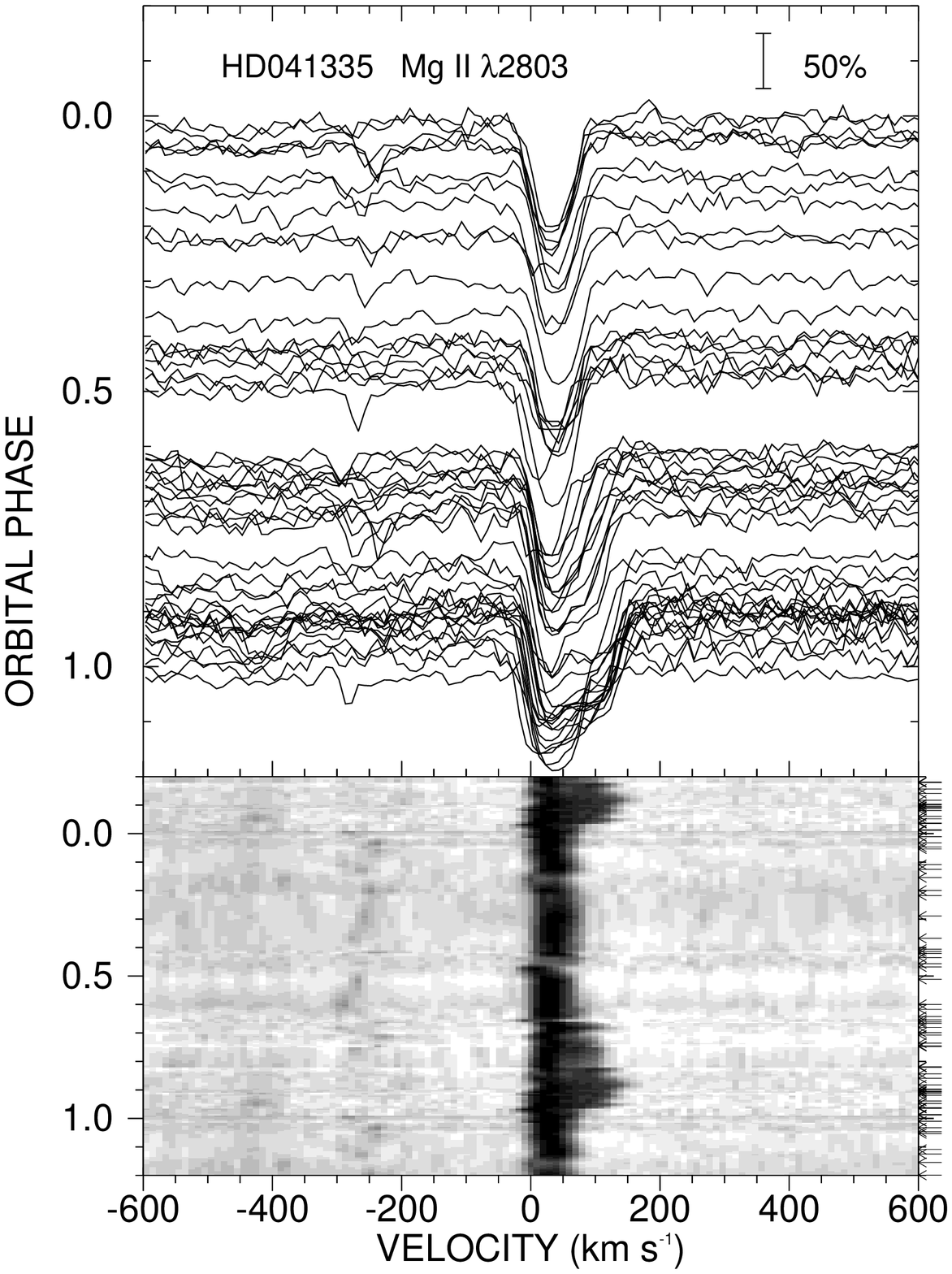}
\figsetgrpnote{The top panel shows the individual spectra of
\ion{Mg}{2} $\lambda 2803.531$ plotted as a function of radial 
velocity and orbital phase (such that the continuum is aligned 
with the phase of observation).  The spectral flux depth relative
to the continuum is indicated by the scale bar in upper right. 
The lower panel shows the same interpolated in orbital phase 
and portrayed as a gray scale image (black corresponding to deepest
absorption and white to strongest continuum flux).  The actual
phases of observation are indicated by arrows on the right hand side.}
\figsetgrpend

\figsetend

\clearpage

\setcounter{figure}{2}
\renewcommand{\thefigure}{\arabic{figure}.1}
\begin{figure}
\begin{center}
{\includegraphics[angle=0,height=16cm]{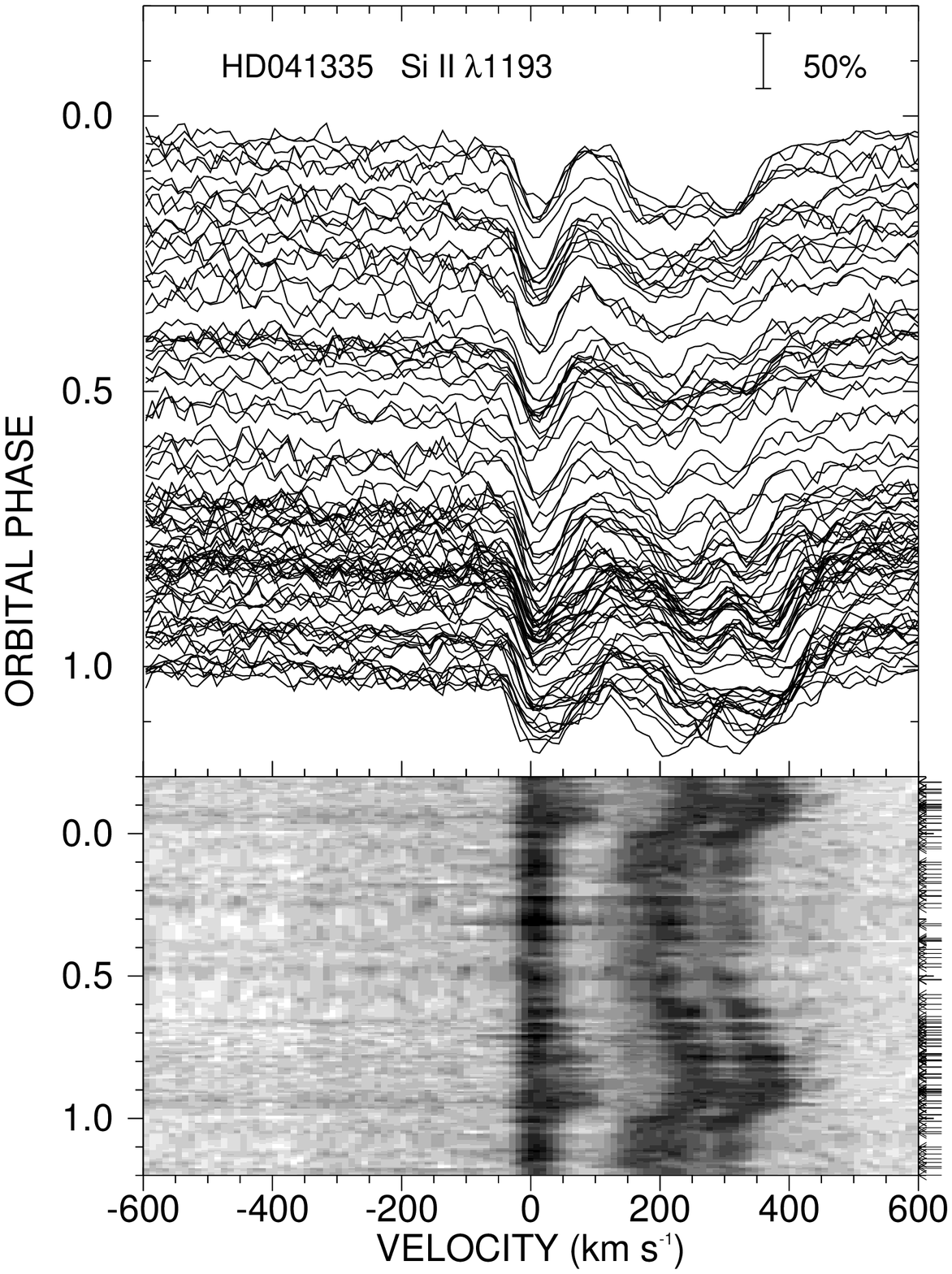}}
\end{center}
\caption{The top panel shows the individual spectra of
\ion{Si}{2} $\lambda 1193.290$ plotted as a function of radial 
velocity and orbital phase (such that the continuum is aligned 
with the phase of observation).  The spectral flux depth relative
to the continuum is indicated by the scale bar in upper right. 
The lower panel shows the same interpolated in orbital phase 
and portrayed as a gray scale image (black corresponding to deepest
absorption and white to strongest continuum flux).  The actual
phases of observation are indicated by arrows on the right hand side. 
The lines of \ion{S}{3} $\lambda\lambda 1194.061, 1194.457$ appear offset 
to higher velocity.}
\end{figure}

\clearpage

\setcounter{figure}{2}
\renewcommand{\thefigure}{\arabic{figure}.2}
\begin{figure}
\begin{center}
{\includegraphics[angle=0,height=16cm]{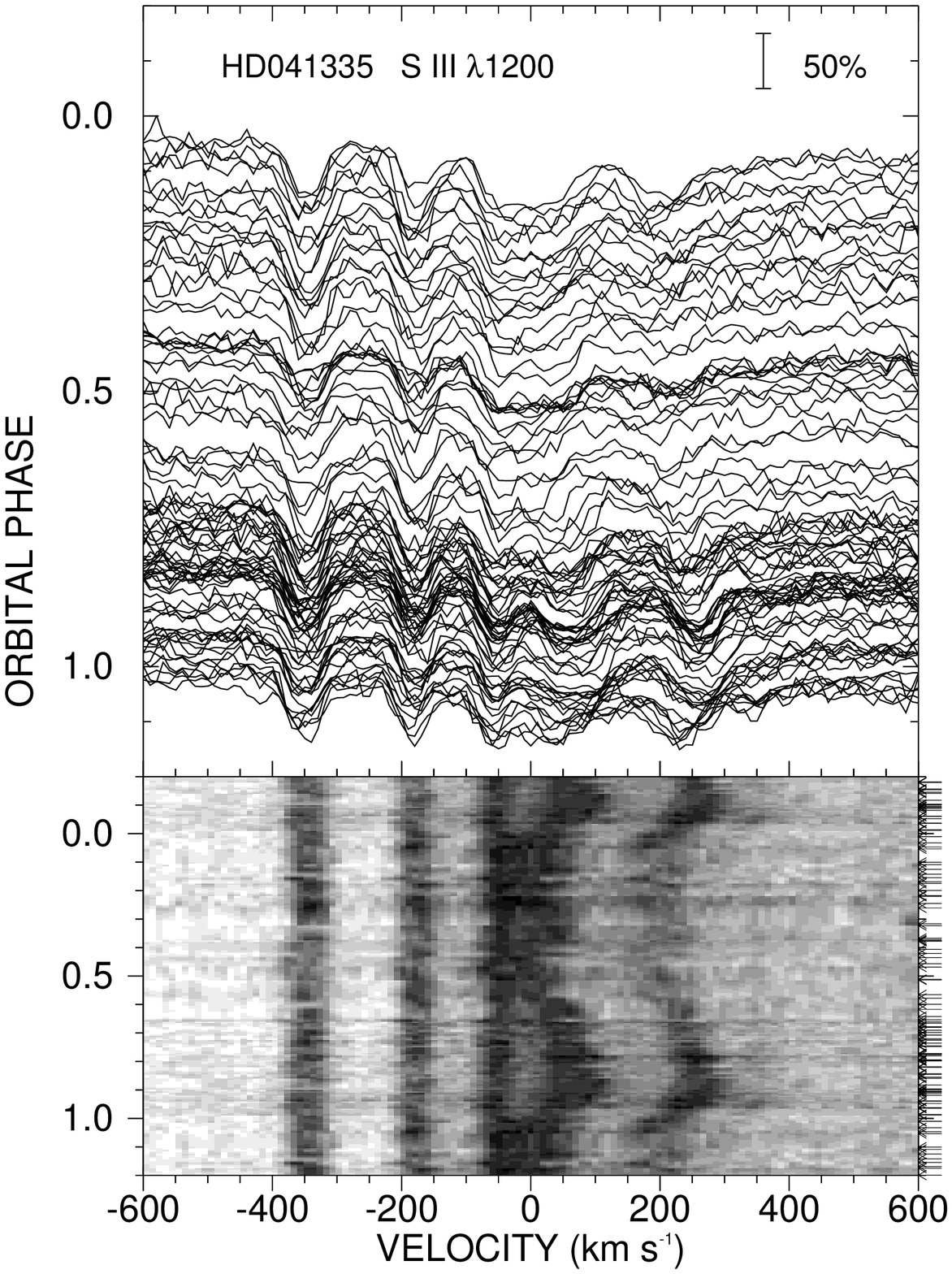}}
\end{center}
\caption{The top panel shows the individual spectra of
\ion{S}{3} $\lambda 1200.970$ plotted as a function of radial
velocity and orbital phase (such that the continuum is aligned
with the phase of observation).  The spectral flux depth relative
to the continuum is indicated by the scale bar in upper right.
The lower panel shows the same interpolated in orbital phase
and portrayed as a gray scale image (black corresponding to deepest
absorption and white to strongest continuum flux).  The actual
phases of observation are indicated by arrows on the right hand side.
The stationary lines of \ion{N}{1} $\lambda\lambda 1199.5,1200.2,1200.7$
appear offset to lower velocity, while \ion{S}{3} $\lambda 1201.730$
appears at higher velocity.}
\end{figure}

\clearpage

\setcounter{figure}{2}
\renewcommand{\thefigure}{\arabic{figure}.3}
\begin{figure}
\begin{center}
{\includegraphics[angle=0,height=16cm]{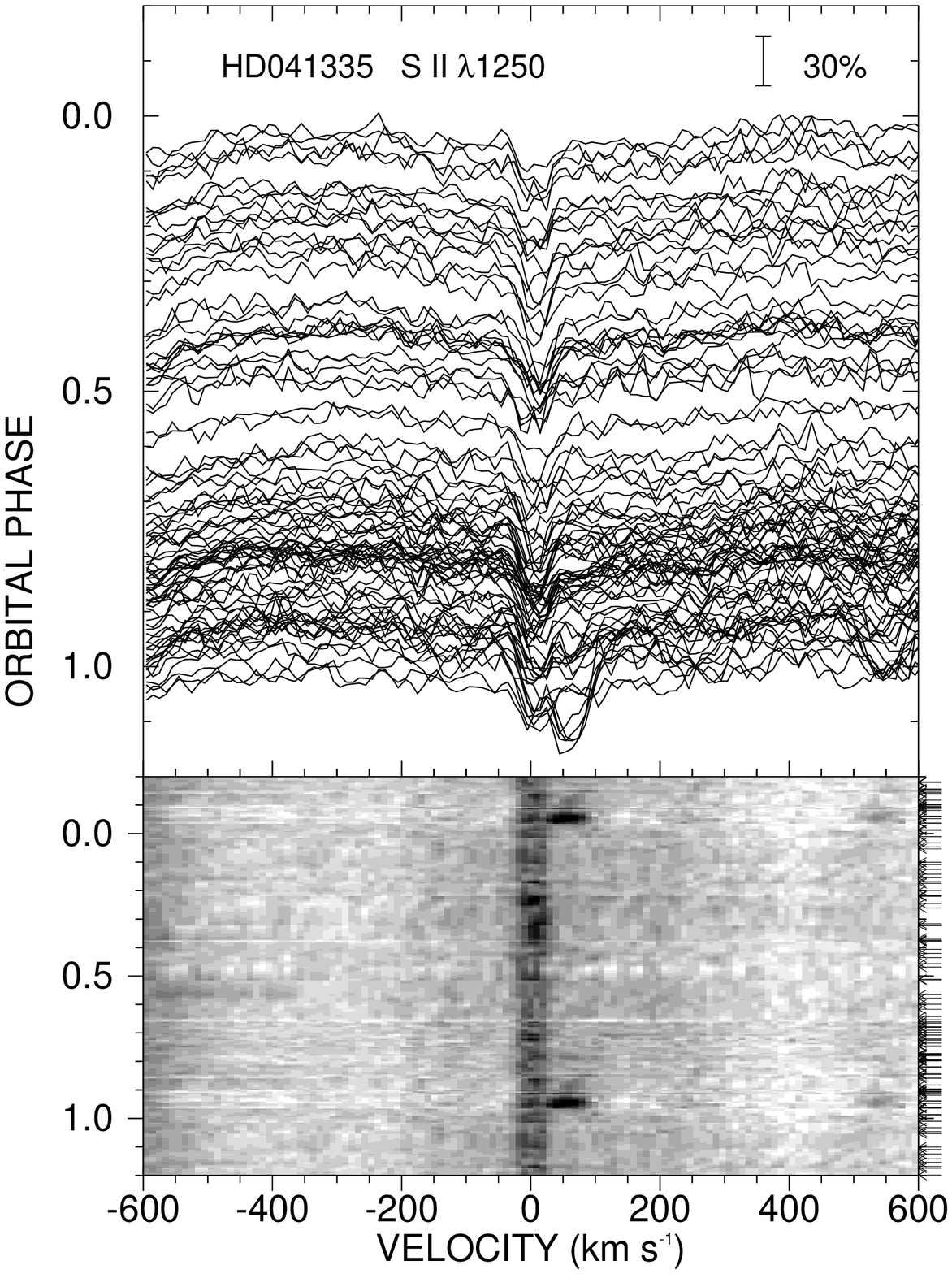}}
\end{center}
\caption{The top panel shows the individual spectra of
\ion{S}{2} $\lambda 1250.583$ plotted as a function of radial
velocity and orbital phase (such that the continuum is aligned
with the phase of observation).  The spectral flux depth relative
to the continuum is indicated by the scale bar in upper right.
The lower panel shows the same interpolated in orbital phase
and portrayed as a gray scale image (black corresponding to deepest
absorption and white to strongest continuum flux).  The actual
phases of observation are indicated by arrows on the right hand side.
The shell line is visible only close to orbital phase $\phi=0.95$.}
\end{figure}

\clearpage

\setcounter{figure}{2}
\renewcommand{\thefigure}{\arabic{figure}.4}
\begin{figure}
\begin{center}
{\includegraphics[angle=0,height=16cm]{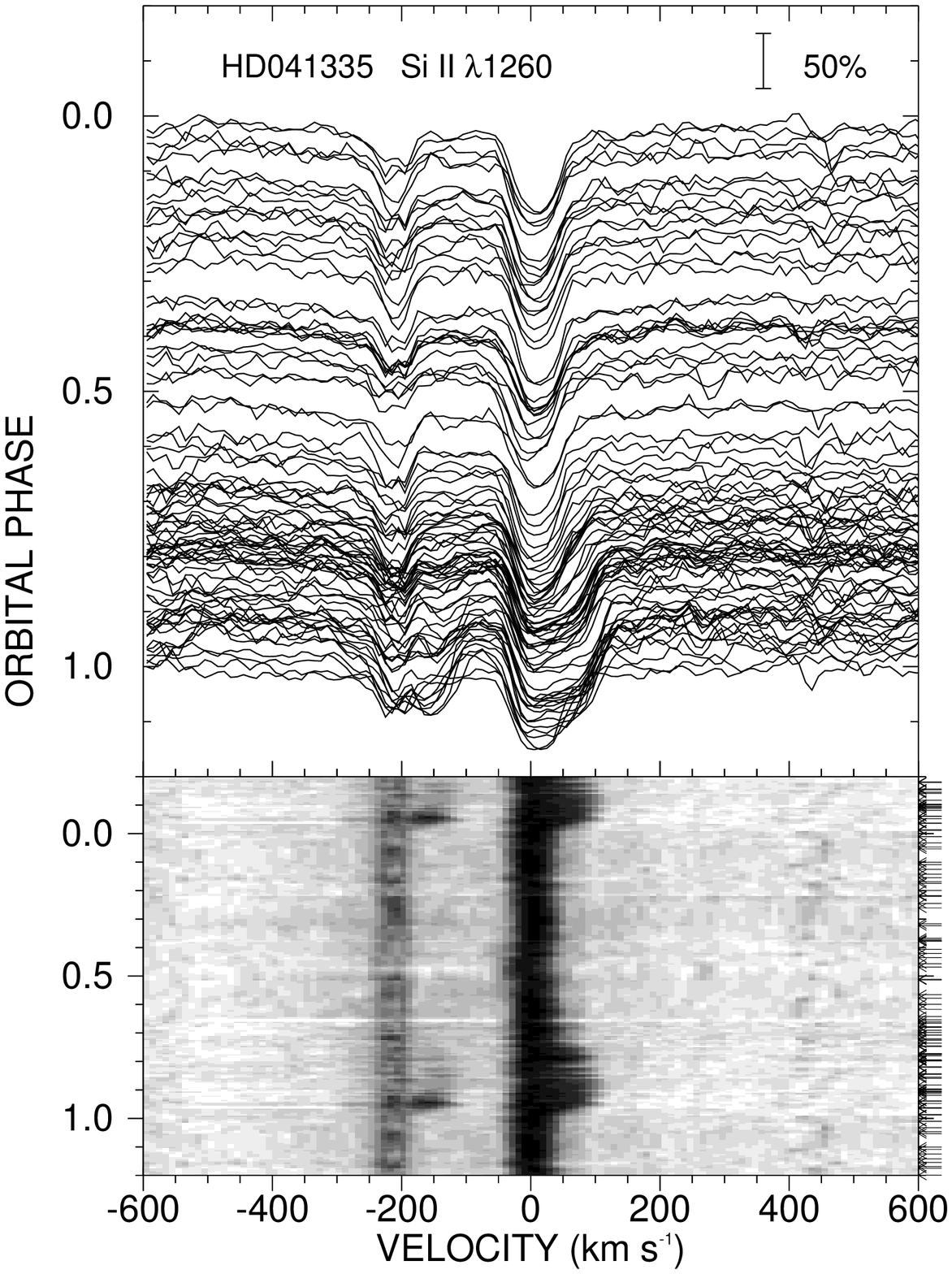}}
\end{center}
\caption{The top panel shows the individual spectra of
\ion{Si}{2} $\lambda 1260.4223$ plotted as a function of radial
velocity and orbital phase (such that the continuum is aligned
with the phase of observation).  The spectral flux depth relative
to the continuum is indicated by the scale bar in upper right.
The lower panel shows the same interpolated in orbital phase
and portrayed as a gray scale image (black corresponding to deepest
absorption and white to strongest continuum flux).  The actual
phases of observation are indicated by arrows on the right hand side.
The shell component of \ion{S}{2} $\lambda 1259.518$ (offset to
lower velocity) is visible only close to orbital phase $\phi=0.95$.}
\end{figure}

\clearpage

\setcounter{figure}{2}
\renewcommand{\thefigure}{\arabic{figure}.5}
\begin{figure}
\begin{center}
{\includegraphics[angle=0,height=16cm]{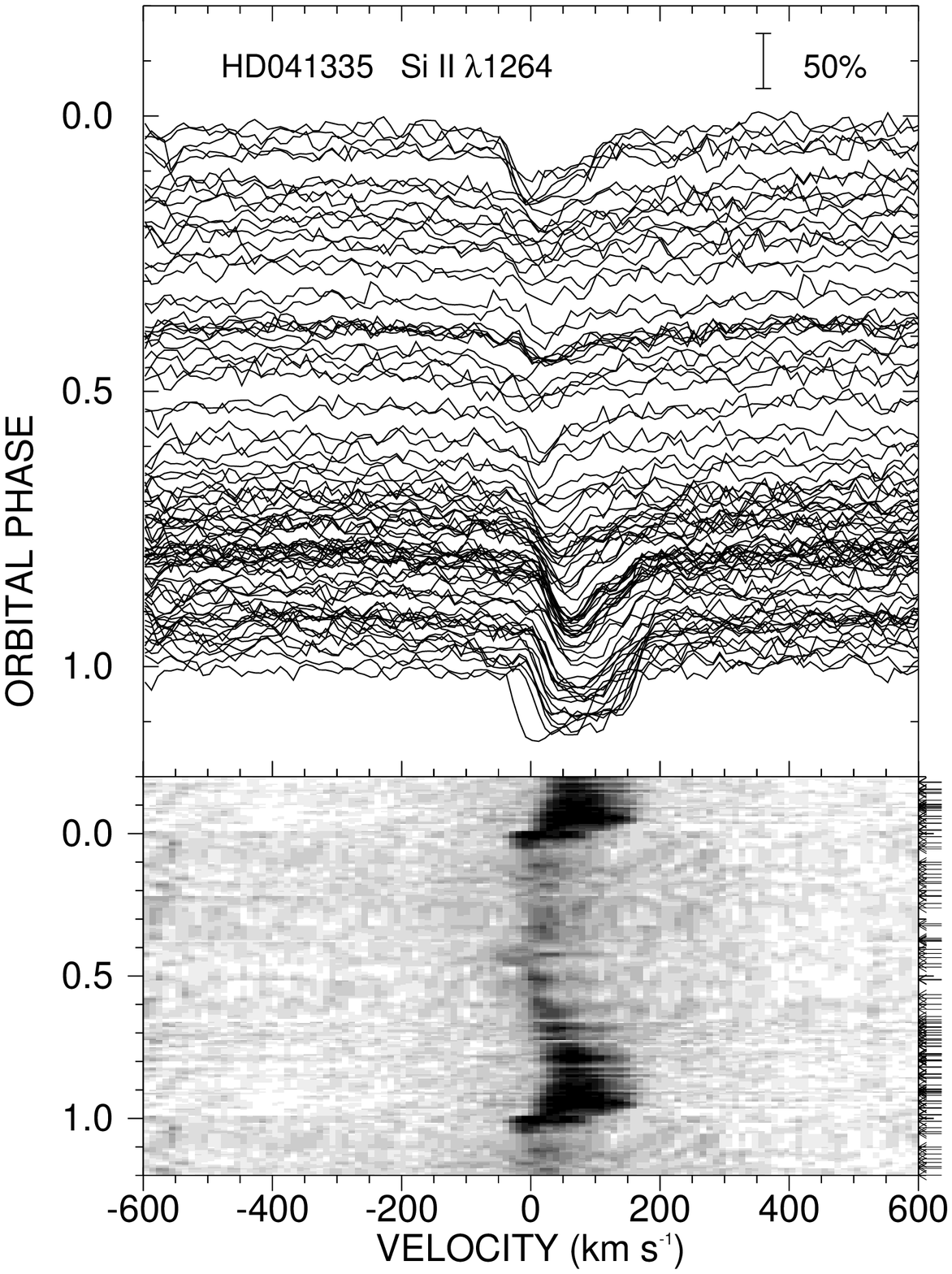}}
\end{center}
\caption{The top panel shows the individual spectra of
\ion{Si}{2} $\lambda 1264.73$ plotted as a function of radial
velocity and orbital phase (such that the continuum is aligned
with the phase of observation).  The spectral flux depth relative
to the continuum is indicated by the scale bar in upper right.
The lower panel shows the same interpolated in orbital phase
and portrayed as a gray scale image (black corresponding to deepest
absorption and white to strongest continuum flux).  The actual
phases of observation are indicated by arrows on the right hand side.}
\end{figure}

\clearpage

\setcounter{figure}{2}
\renewcommand{\thefigure}{\arabic{figure}.6}
\begin{figure}
\begin{center}
{\includegraphics[angle=0,height=16cm]{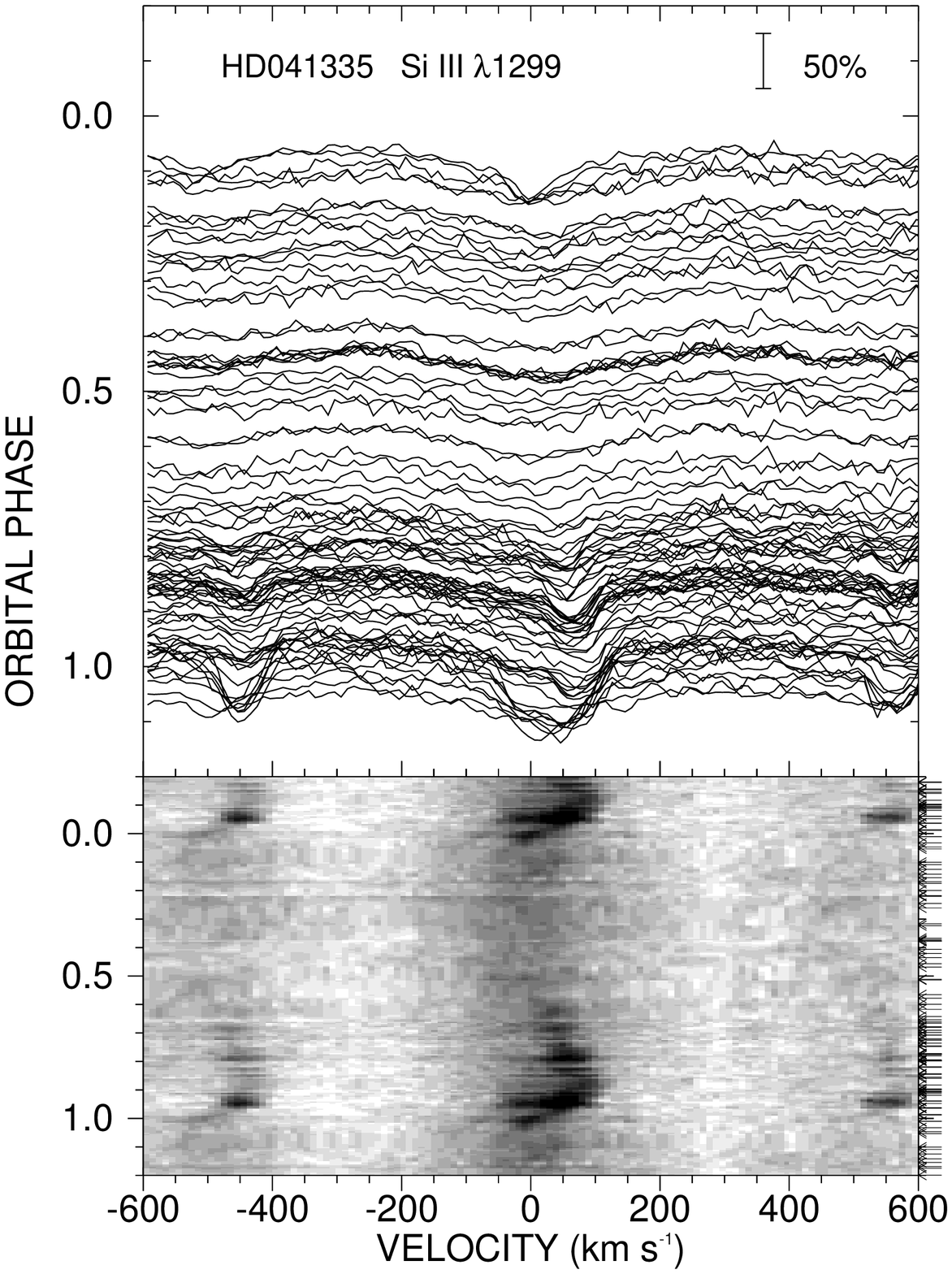}}
\end{center}
\caption{The top panel shows the individual spectra of
\ion{Si}{3} $\lambda 1298.96$ plotted as a function of radial
velocity and orbital phase (such that the continuum is aligned
with the phase of observation).  The spectral flux depth relative
to the continuum is indicated by the scale bar in upper right.
The lower panel shows the same interpolated in orbital phase
and portrayed as a gray scale image (black corresponding to deepest
absorption and white to strongest continuum flux).  The actual
phases of observation are indicated by arrows on the right hand side.
The lines of \ion{Si}{3} $\lambda\lambda 1296.73,1298.89,1301.15$ also
appear as sudden shell features at orbital phase $\phi=0.95$.}
\end{figure}

\clearpage

\setcounter{figure}{2}
\renewcommand{\thefigure}{\arabic{figure}.7}
\begin{figure}
\begin{center}
{\includegraphics[angle=0,height=16cm]{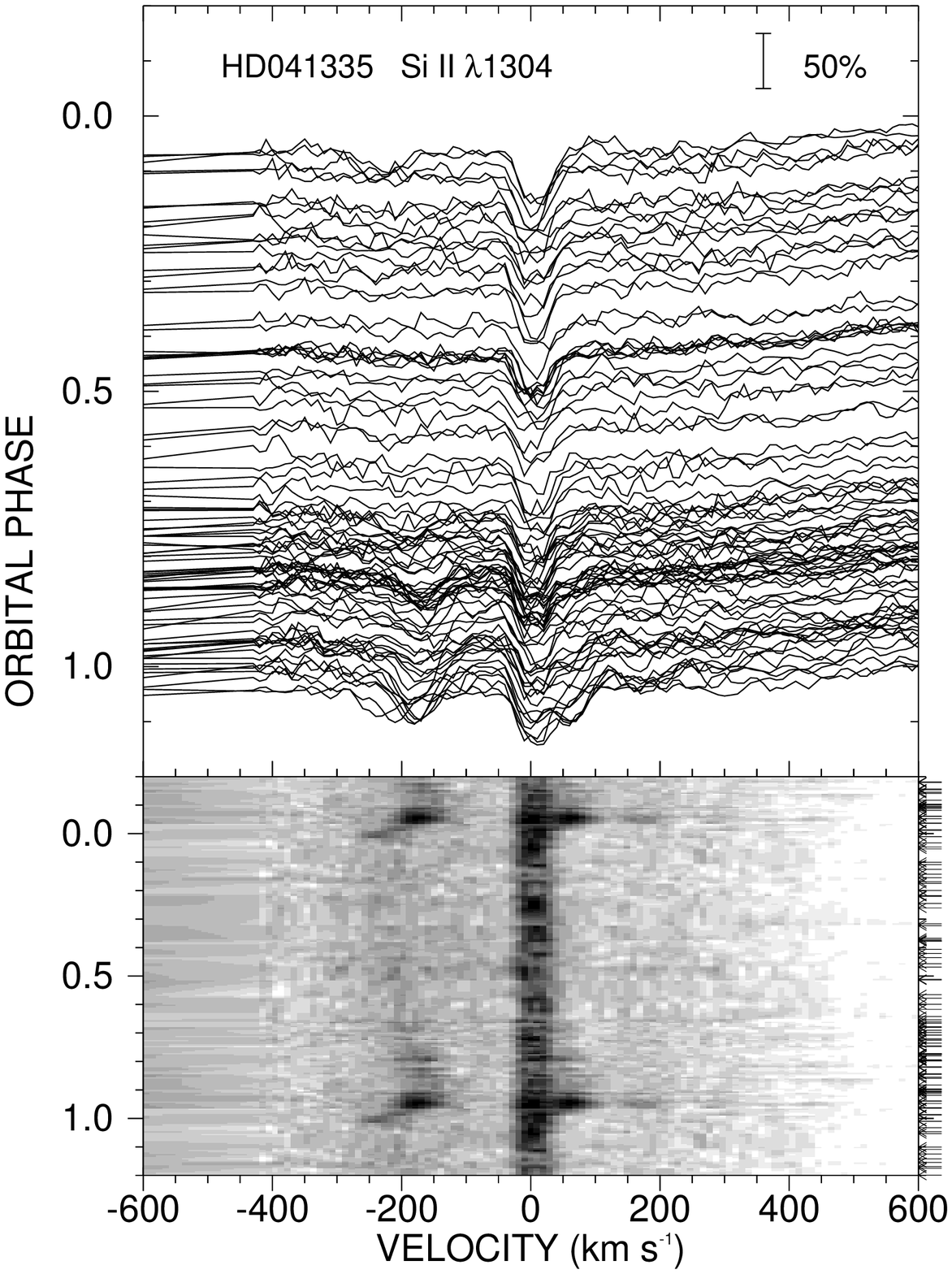}}
\end{center}
\caption{The top panel shows the individual spectra of
\ion{Si}{2} $\lambda 1304.3711$ plotted as a function of radial
velocity and orbital phase (such that the continuum is aligned
with the phase of observation).  The spectral flux depth relative
to the continuum is indicated by the scale bar in upper right.
The lower panel shows the same interpolated in orbital phase
and portrayed as a gray scale image (black corresponding to deepest
absorption and white to strongest continuum flux).  The actual
phases of observation are indicated by arrows on the right hand side.
The shell component appears suddenly at orbital phase $\phi=0.95$.
The flat region at left represents a linear interpolation across
the interstellar line of \ion{O}{1} $\lambda 1302$.}
\end{figure}

\clearpage

\setcounter{figure}{2}
\renewcommand{\thefigure}{\arabic{figure}.8}
\begin{figure}
\begin{center}
{\includegraphics[angle=0,height=16cm]{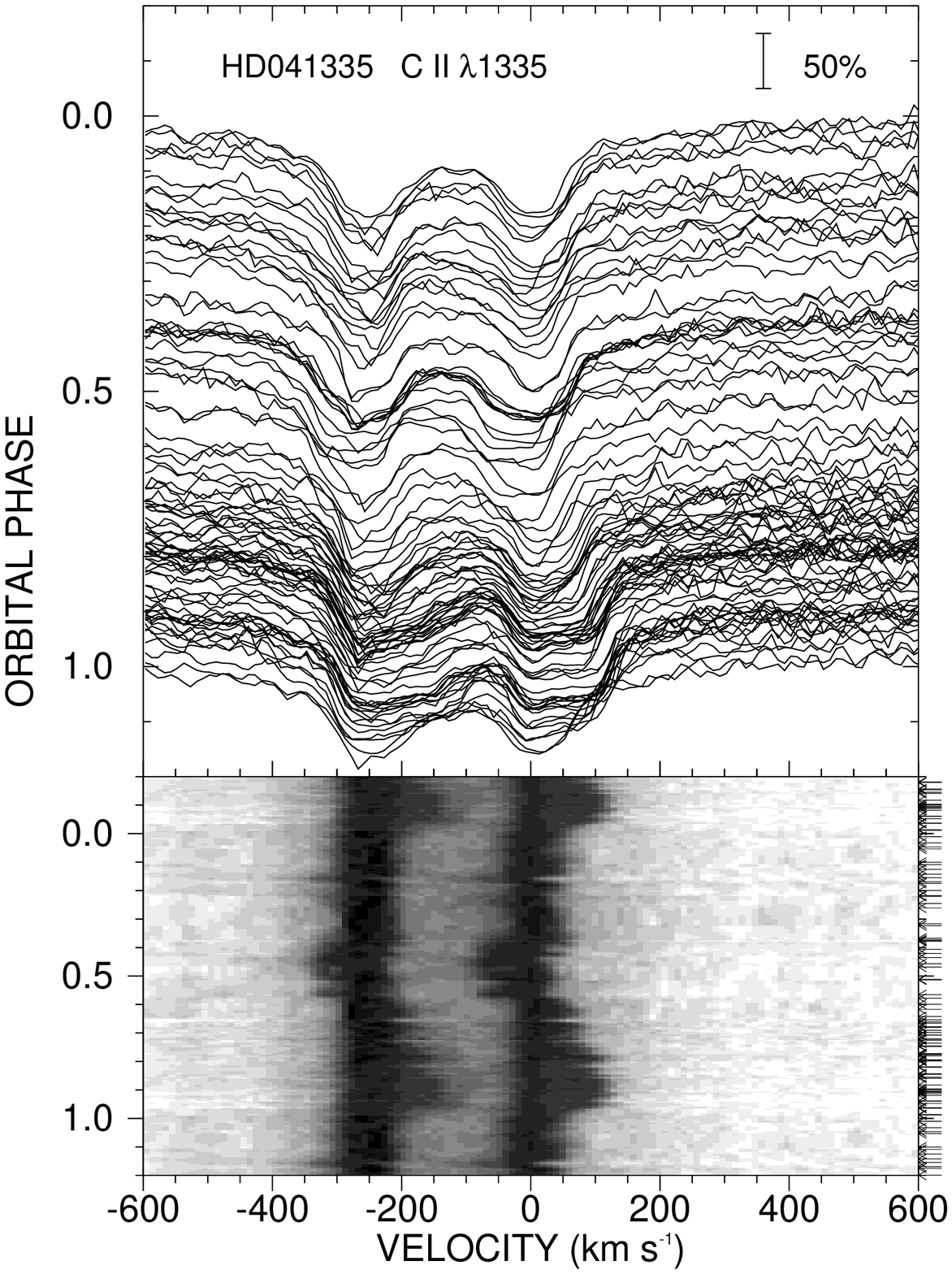}}
\end{center}
\caption{The top panel shows the individual spectra of
\ion{C}{2} $\lambda 1335.708$ plotted as a function of radial
velocity and orbital phase (such that the continuum is aligned
with the phase of observation).  The spectral flux depth relative
to the continuum is indicated by the scale bar in upper right.
The lower panel shows the same interpolated in orbital phase
and portrayed as a gray scale image (black corresponding to deepest
absorption and white to strongest continuum flux).  The actual
phases of observation are indicated by arrows on the right hand side.
The similar line of \ion{C}{2} $\lambda 1334.532$ appears offset
to lower velocity.}
\end{figure}

\clearpage

\setcounter{figure}{2}
\renewcommand{\thefigure}{\arabic{figure}.9}
\begin{figure}
\begin{center}
{\includegraphics[angle=0,height=16cm]{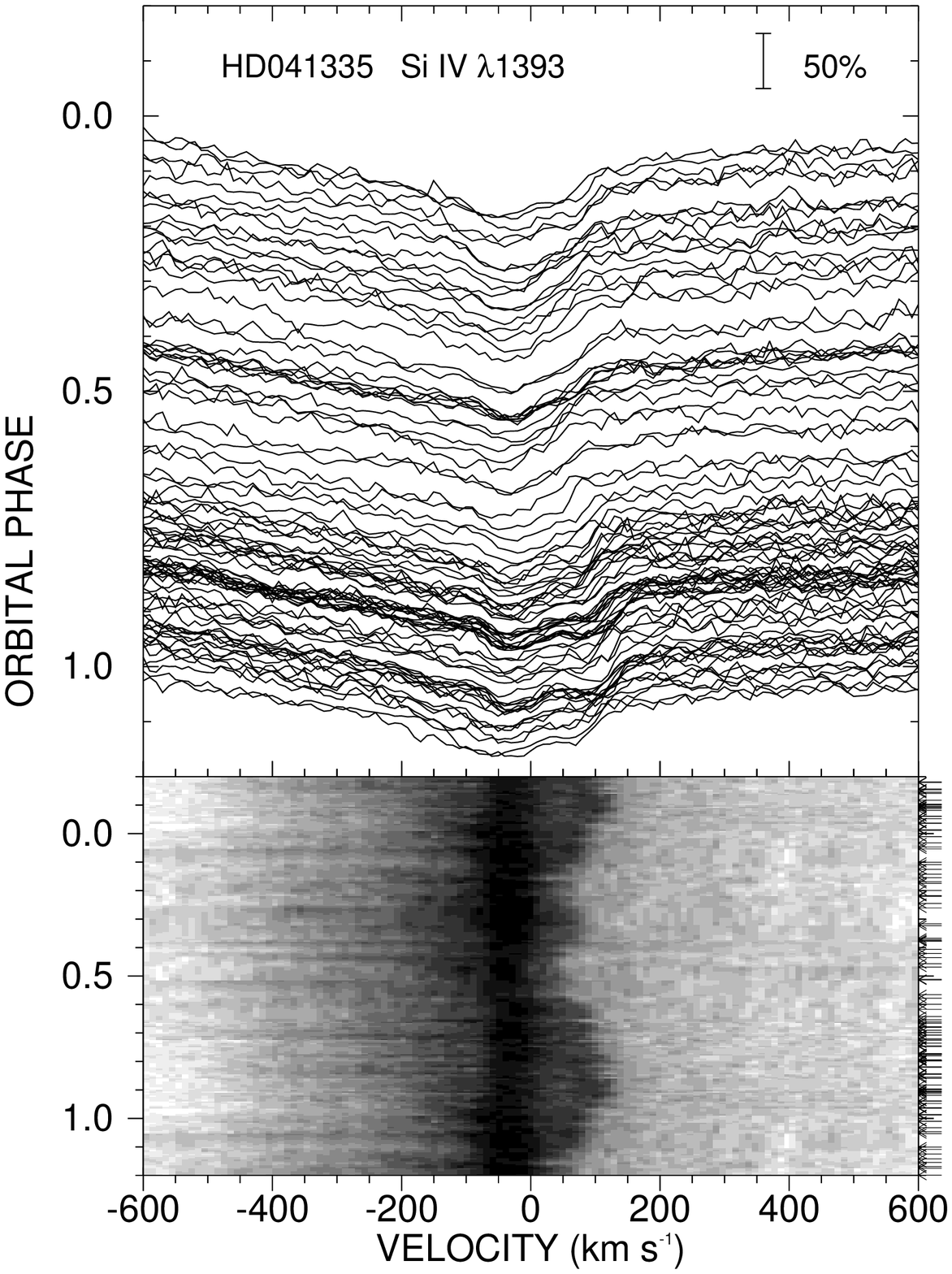}}
\end{center}
\caption{The top panel shows the individual spectra of
\ion{Si}{4} $\lambda 1393.755$ plotted as a function of radial
velocity and orbital phase (such that the continuum is aligned
with the phase of observation).  The spectral flux depth relative
to the continuum is indicated by the scale bar in upper right.
The lower panel shows the same interpolated in orbital phase
and portrayed as a gray scale image (black corresponding to deepest
absorption and white to strongest continuum flux).  The actual
phases of observation are indicated by arrows on the right hand side.
The feature is formed mainly in the wind of the Be star, but shell
variations are present on the high velocity side of the core.}
\end{figure}

\clearpage

\setcounter{figure}{2}
\renewcommand{\thefigure}{\arabic{figure}.10}
\begin{figure}
\begin{center}
{\includegraphics[angle=0,height=16cm]{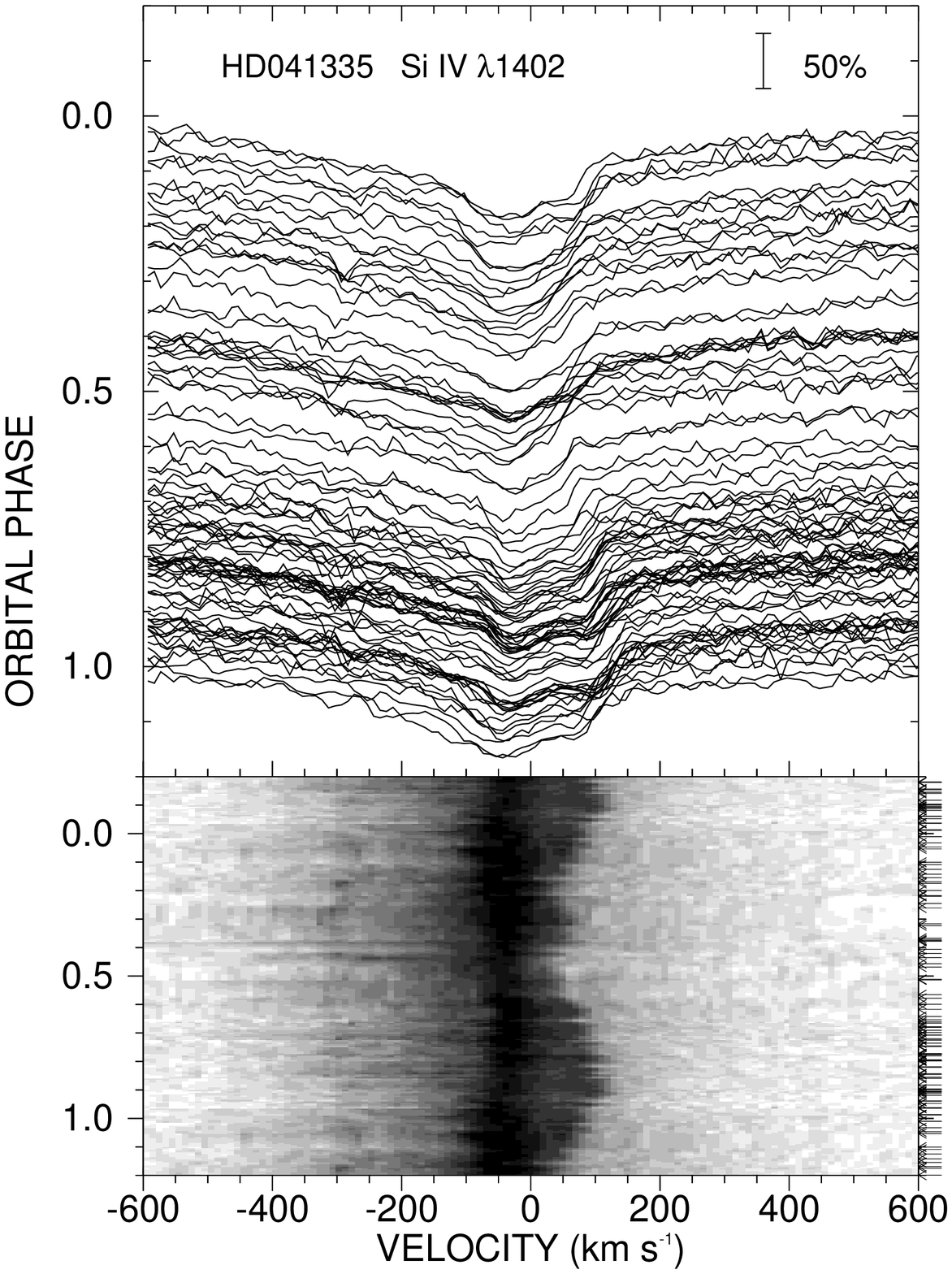}}
\end{center}
\caption{The top panel shows the individual spectra of
\ion{Si}{4} $\lambda 1402.770$ plotted as a function of radial
velocity and orbital phase (such that the continuum is aligned
with the phase of observation).  The spectral flux depth relative
to the continuum is indicated by the scale bar in upper right.
The lower panel shows the same interpolated in orbital phase
and portrayed as a gray scale image (black corresponding to deepest
absorption and white to strongest continuum flux).  The actual
phases of observation are indicated by arrows on the right hand side.
The feature is formed mainly in the wind of the Be star, but shell
variations are present on the high velocity side of the core.}
\end{figure}

\clearpage

\setcounter{figure}{2}
\renewcommand{\thefigure}{\arabic{figure}.11}
\begin{figure}
\begin{center}
{\includegraphics[angle=0,height=16cm]{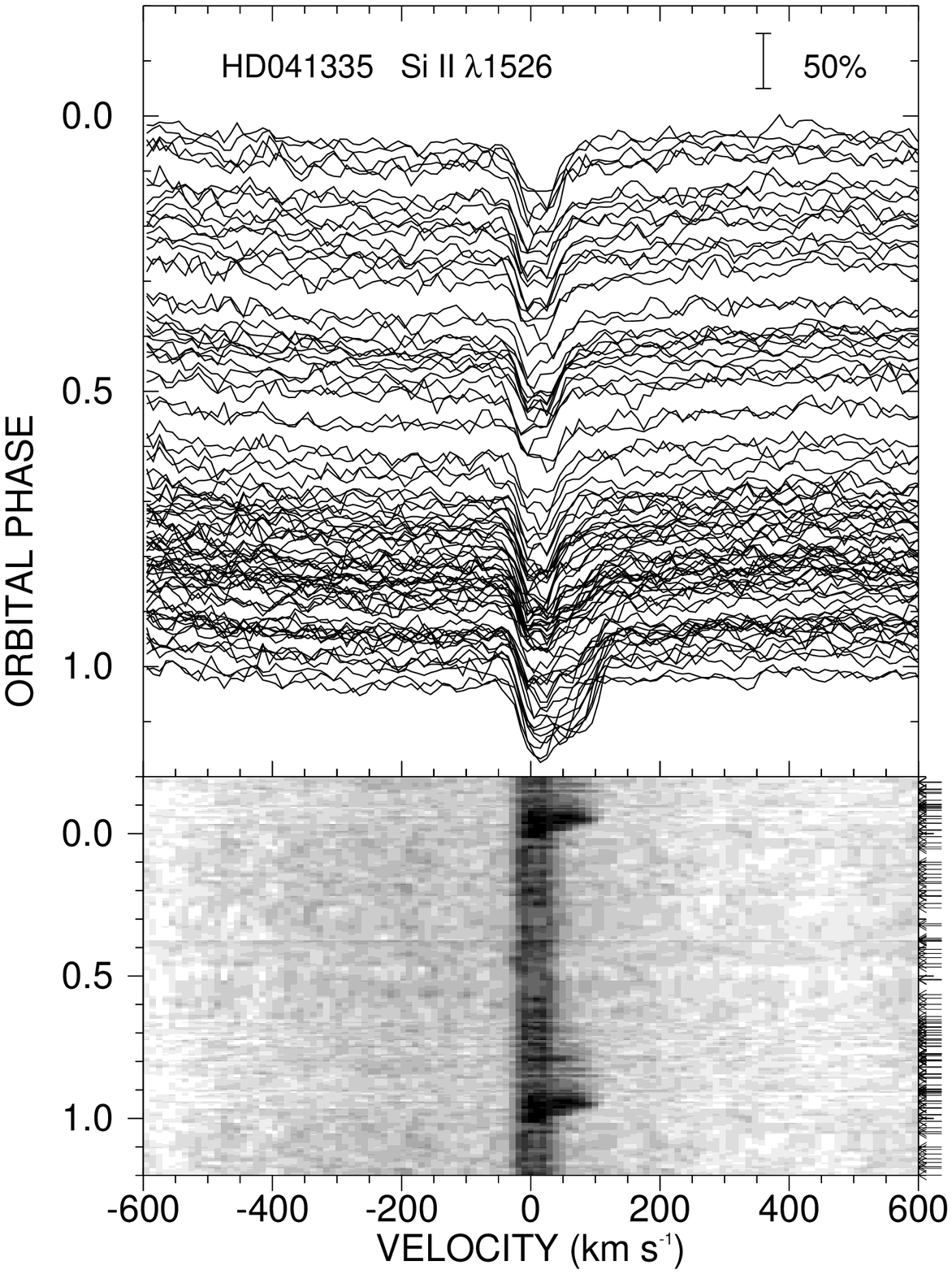}}
\end{center}
\caption{The top panel shows the individual spectra of
\ion{Si}{2} $\lambda 1526.7071$ plotted as a function of radial
velocity and orbital phase (such that the continuum is aligned
with the phase of observation).  The spectral flux depth relative
to the continuum is indicated by the scale bar in upper right.
The lower panel shows the same interpolated in orbital phase
and portrayed as a gray scale image (black corresponding to deepest
absorption and white to strongest continuum flux).  The actual
phases of observation are indicated by arrows on the right hand side.
The shell component appears suddenly at orbital phase $\phi=0.95$.}
\end{figure}

\clearpage

\setcounter{figure}{2}
\renewcommand{\thefigure}{\arabic{figure}.12}
\begin{figure}
\begin{center}
{\includegraphics[angle=0,height=16cm]{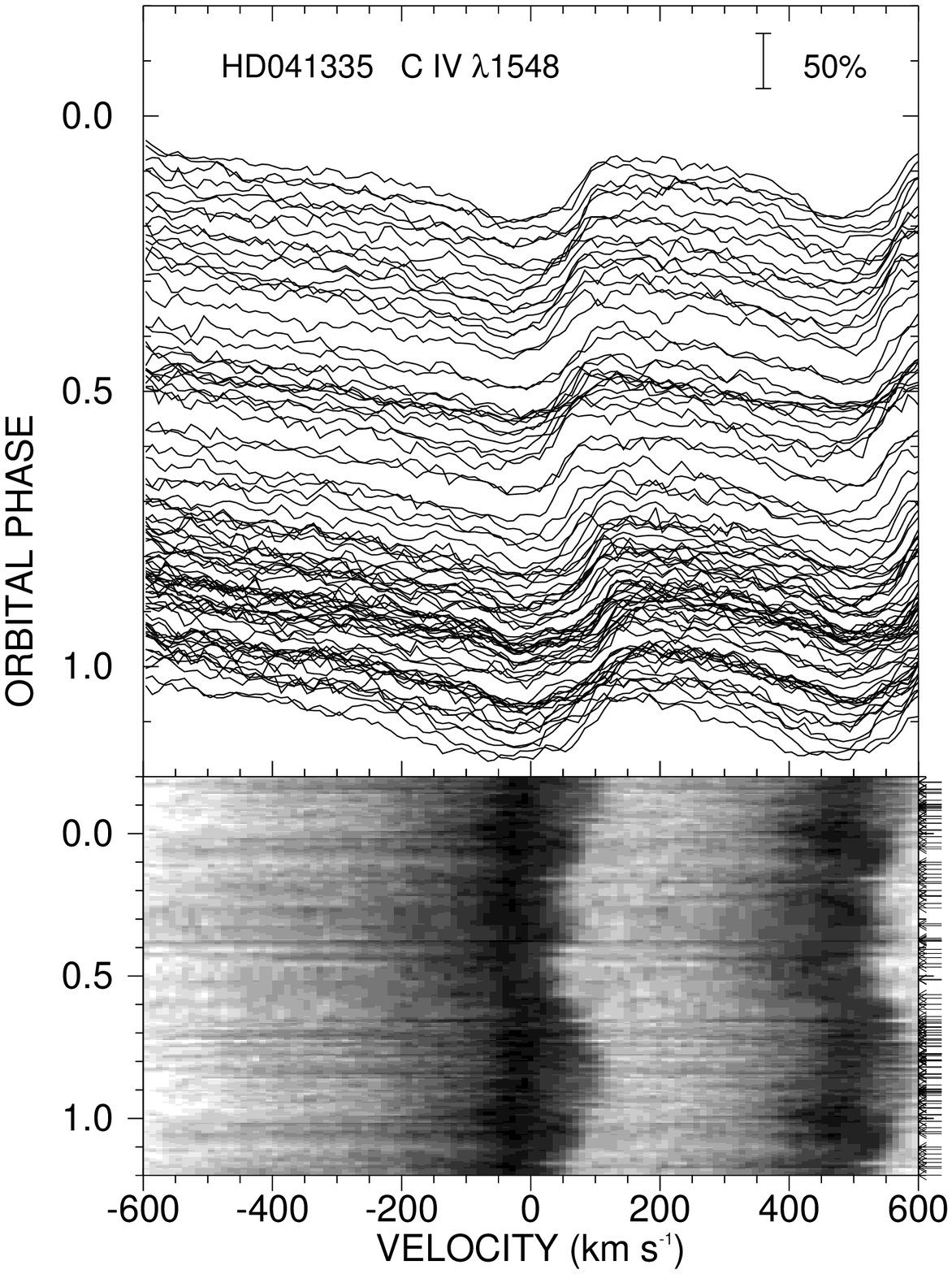}}
\end{center}
\caption{The top panel shows the individual spectra of
\ion{C}{4} $\lambda 1548.202$ plotted as a function of radial
velocity and orbital phase (such that the continuum is aligned
with the phase of observation).  The spectral flux depth relative
to the continuum is indicated by the scale bar in upper right.
The lower panel shows the same interpolated in orbital phase
and portrayed as a gray scale image (black corresponding to deepest
absorption and white to strongest continuum flux).  The actual
phases of observation are indicated by arrows on the right hand side.
The similar line of \ion{C}{4} $\lambda 1550.774$ appears offset
to higher velocity.  Both features are formed mainly in the wind
of the Be star, but shell variations are present on the high
velocity side of the core.}
\end{figure}

\clearpage

\setcounter{figure}{2}
\renewcommand{\thefigure}{\arabic{figure}.13}
\begin{figure}
\begin{center}
{\includegraphics[angle=0,height=16cm]{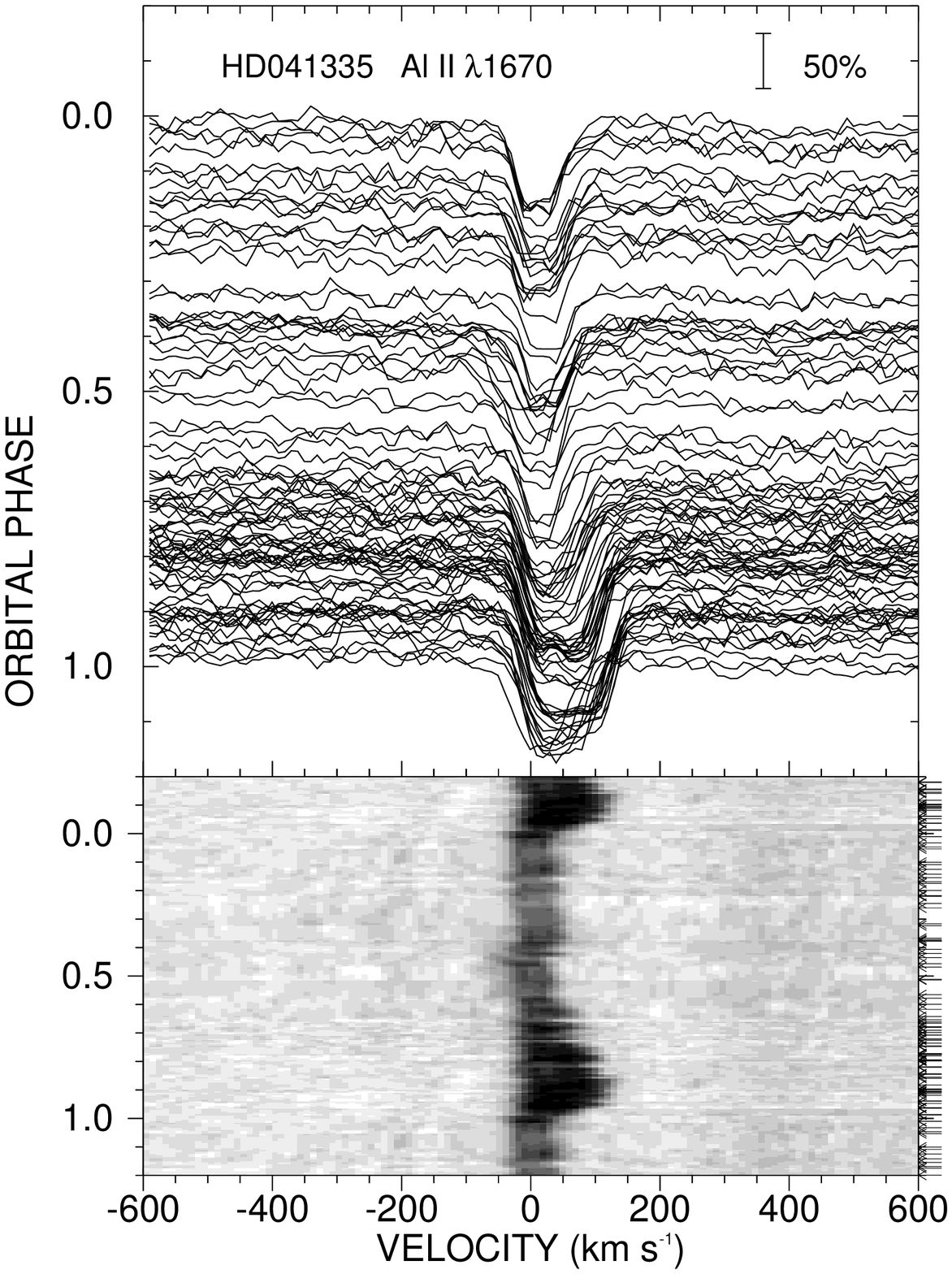}}
\end{center}
\caption{The top panel shows the individual spectra of
\ion{Al}{2} $\lambda 1670.7867$ plotted as a function of radial
velocity and orbital phase (such that the continuum is aligned
with the phase of observation).  The spectral flux depth relative
to the continuum is indicated by the scale bar in upper right.
The lower panel shows the same interpolated in orbital phase
and portrayed as a gray scale image (black corresponding to deepest
absorption and white to strongest continuum flux).  The actual
phases of observation are indicated by arrows on the right hand side.}
\end{figure}

\clearpage

\setcounter{figure}{2}
\renewcommand{\thefigure}{\arabic{figure}.14}
\begin{figure}
\begin{center}
{\includegraphics[angle=0,height=16cm]{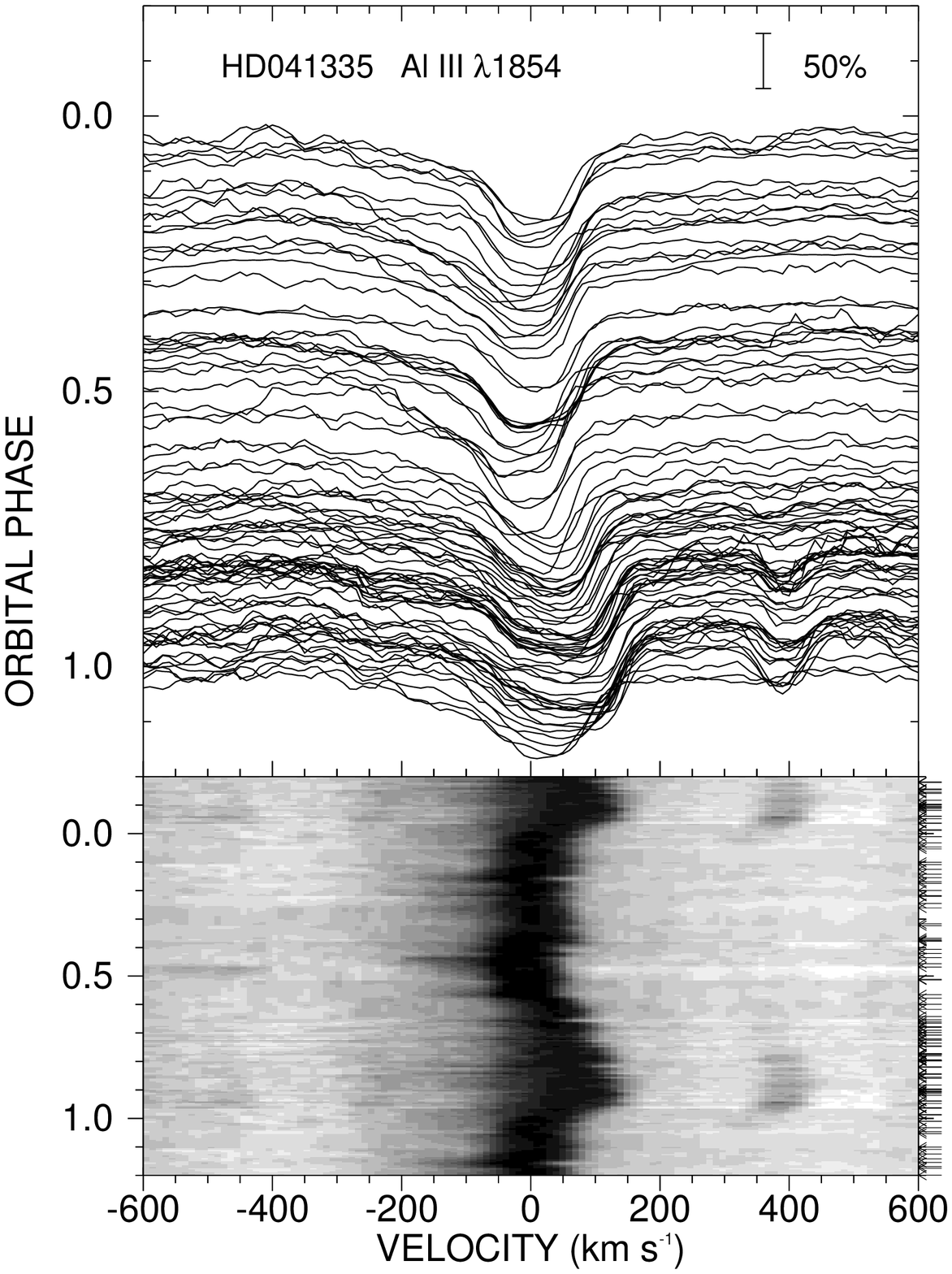}}
\end{center}
\caption{The top panel shows the individual spectra of
\ion{Al}{3} $\lambda 1854.7164$ plotted as a function of radial
velocity and orbital phase (such that the continuum is aligned
with the phase of observation).  The spectral flux depth relative
to the continuum is indicated by the scale bar in upper right.
The lower panel shows the same interpolated in orbital phase
and portrayed as a gray scale image (black corresponding to deepest
absorption and white to strongest continuum flux).  The actual
phases of observation are indicated by arrows on the right hand side.}
\end{figure}

\clearpage

\setcounter{figure}{2}
\renewcommand{\thefigure}{\arabic{figure}.15}
\begin{figure}
\begin{center}
{\includegraphics[angle=0,height=16cm]{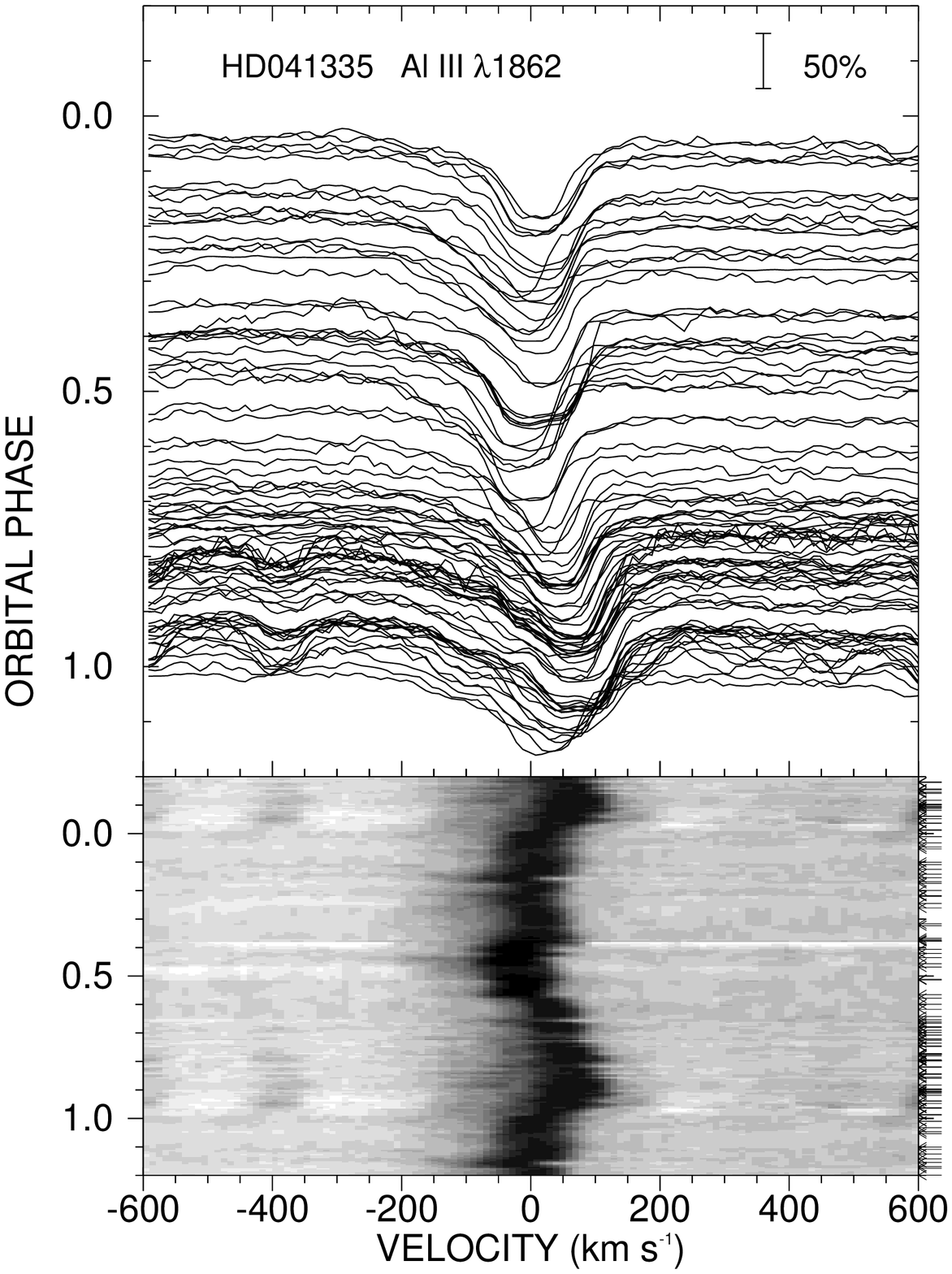}}
\end{center}
\caption{The top panel shows the individual spectra of
\ion{Al}{3} $\lambda 1862.7895$ plotted as a function of radial
velocity and orbital phase (such that the continuum is aligned
with the phase of observation).  The spectral flux depth relative
to the continuum is indicated by the scale bar in upper right.
The lower panel shows the same interpolated in orbital phase
and portrayed as a gray scale image (black corresponding to deepest
absorption and white to strongest continuum flux).  The actual
phases of observation are indicated by arrows on the right hand side.}
\end{figure}

\clearpage

\setcounter{figure}{2}
\renewcommand{\thefigure}{\arabic{figure}.16}
\begin{figure}
\begin{center}
{\includegraphics[angle=0,height=16cm]{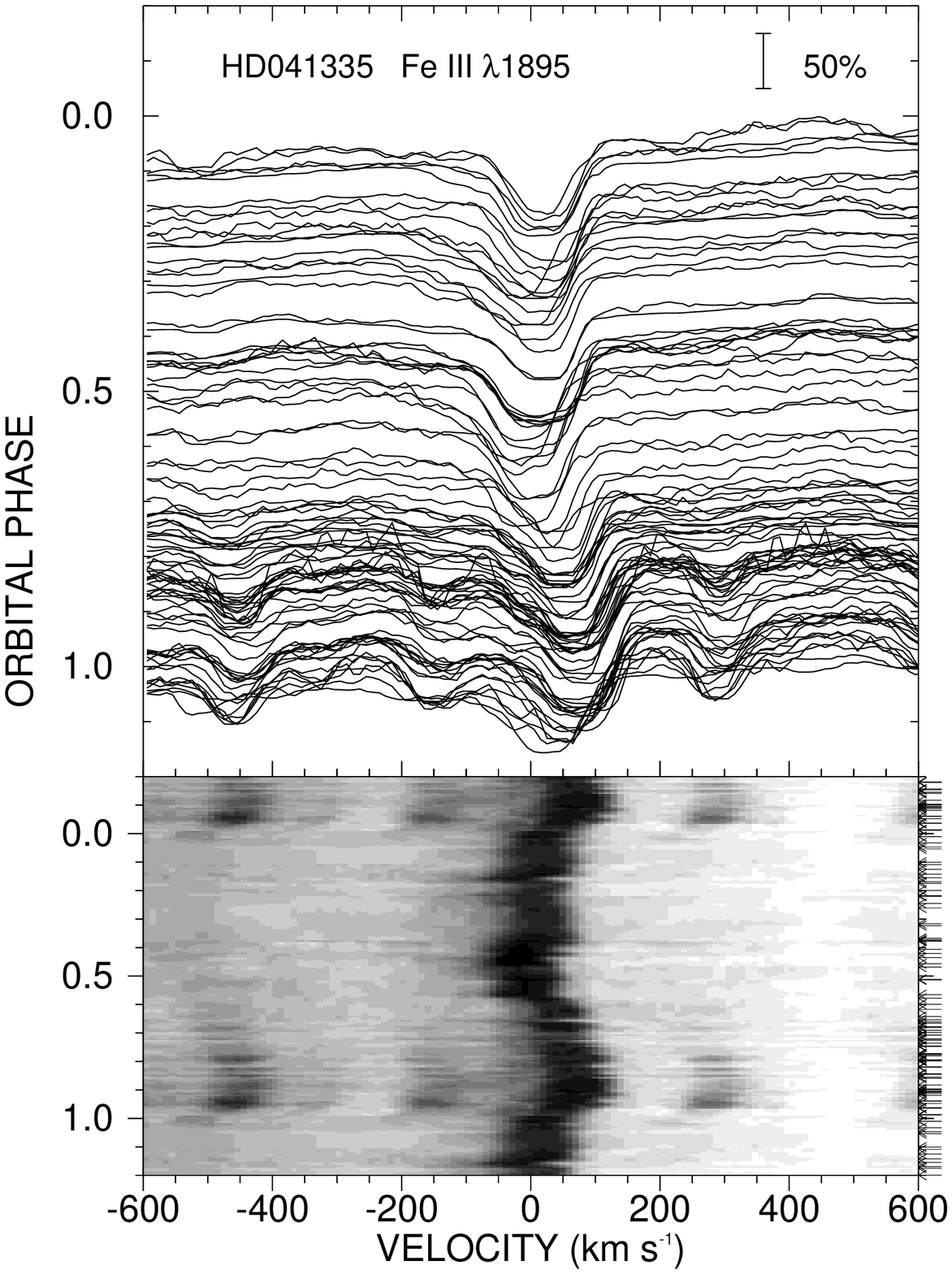}}
\end{center}
\caption{The top panel shows the individual spectra of
\ion{Fe}{3} $\lambda 1895.46$ plotted as a function of radial
velocity and orbital phase (such that the continuum is aligned
with the phase of observation).  The spectral flux depth relative
to the continuum is indicated by the scale bar in upper right.
The lower panel shows the same interpolated in orbital phase
and portrayed as a gray scale image (black corresponding to deepest
absorption and white to strongest continuum flux).  The actual
phases of observation are indicated by arrows on the right hand side.
The lines of \ion{Fe}{3} $\lambda\lambda 1895.46, 1896.814$ also
appear in the figure.}
\end{figure}

\clearpage

\setcounter{figure}{2}
\renewcommand{\thefigure}{\arabic{figure}.17}
\begin{figure}
\begin{center}
{\includegraphics[angle=0,height=16cm]{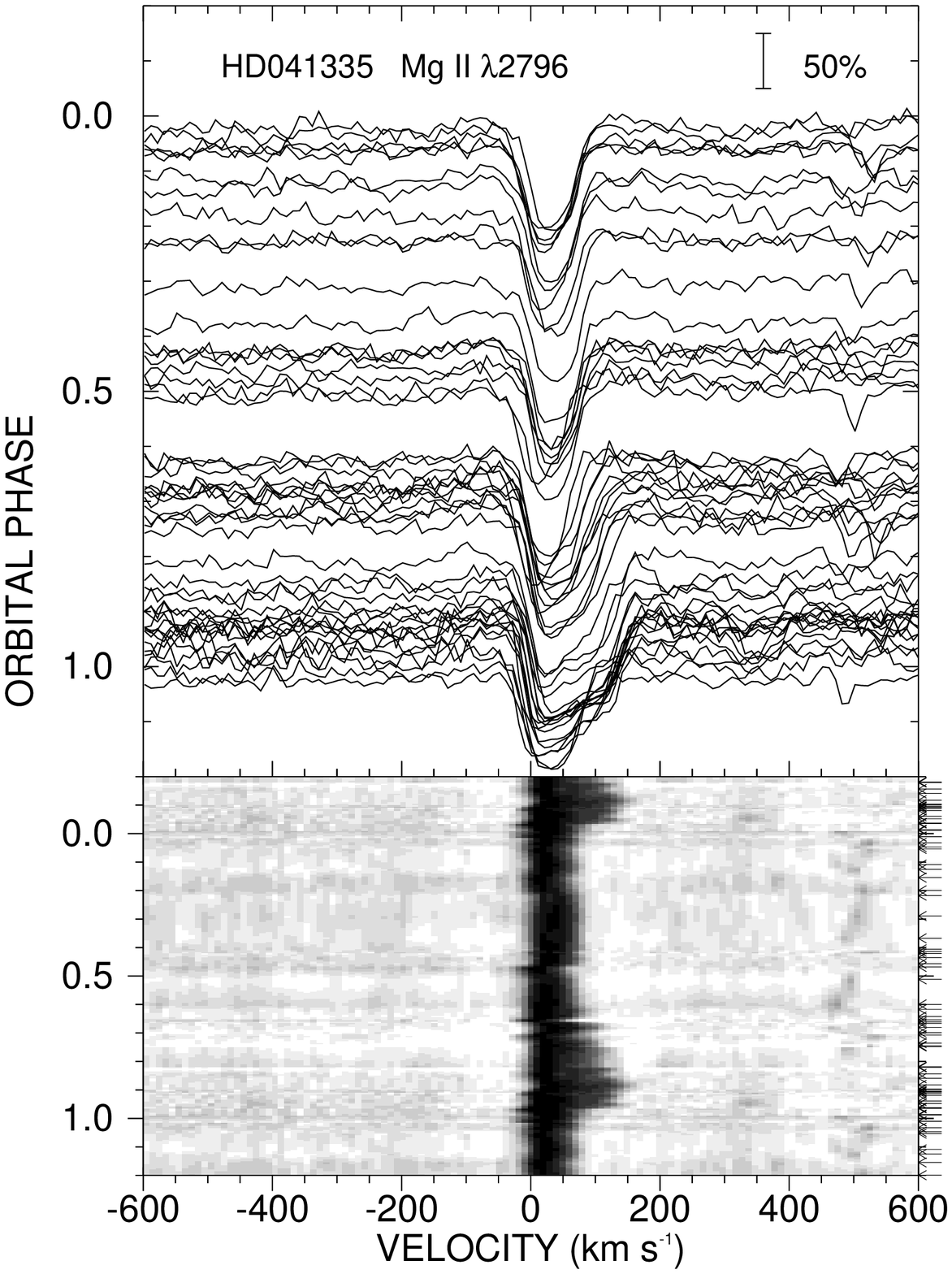}}
\end{center}
\caption{The top panel shows the individual spectra of
\ion{Mg}{2} $\lambda 2796.352$ plotted as a function of radial
velocity and orbital phase (such that the continuum is aligned
with the phase of observation).  The spectral flux depth relative
to the continuum is indicated by the scale bar in upper right.
The lower panel shows the same interpolated in orbital phase
and portrayed as a gray scale image (black corresponding to deepest
absorption and white to strongest continuum flux).  The actual
phases of observation are indicated by arrows on the right hand side.}
\end{figure}

\clearpage

\setcounter{figure}{2}
\renewcommand{\thefigure}{\arabic{figure}.18}
\begin{figure}
\begin{center}
{\includegraphics[angle=0,height=16cm]{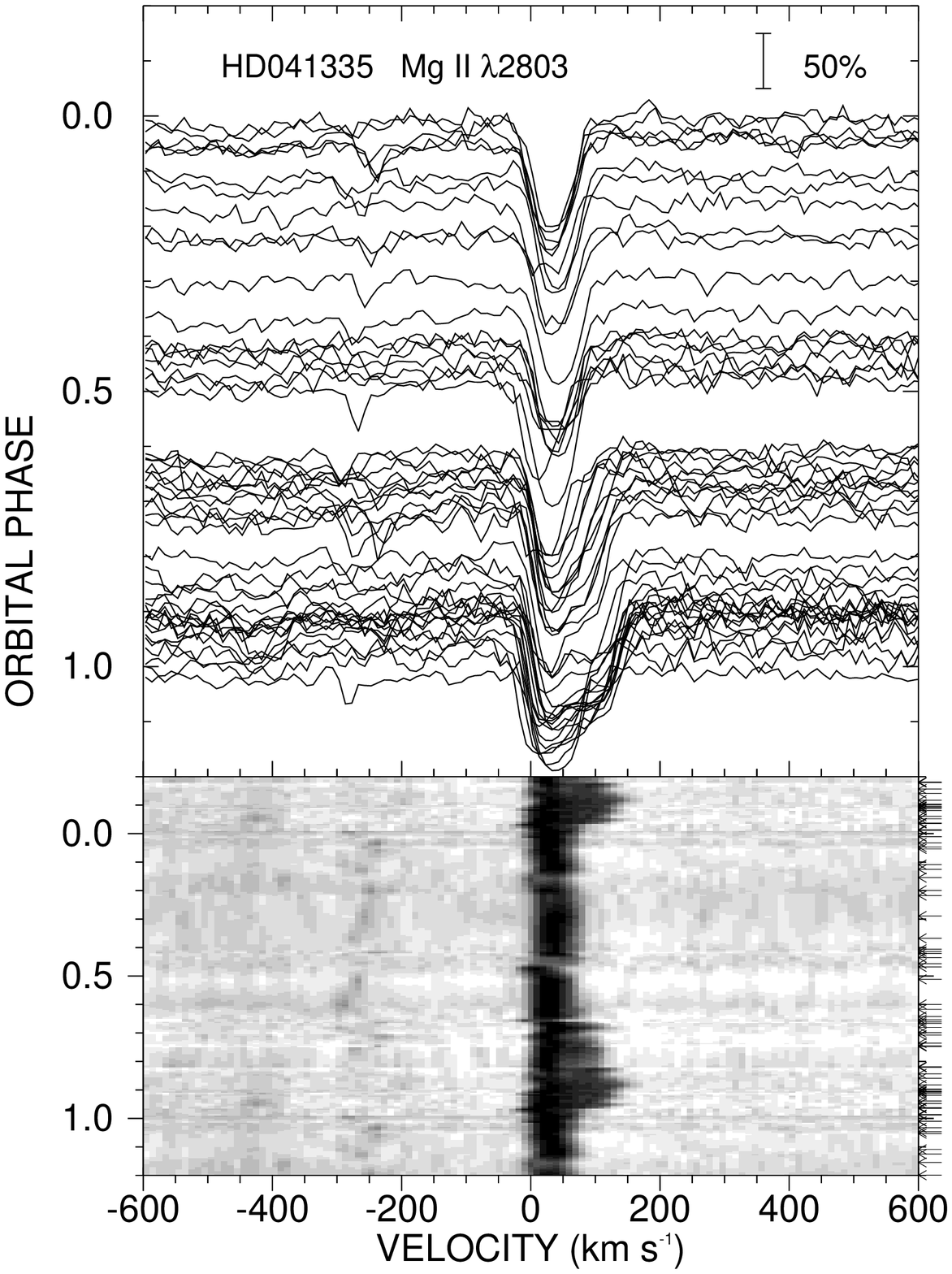}}
\end{center}
\caption{The top panel shows the individual spectra of
\ion{Mg}{2} $\lambda 2803.531$ plotted as a function of radial
velocity and orbital phase (such that the continuum is aligned
with the phase of observation).  The spectral flux depth relative
to the continuum is indicated by the scale bar in upper right.
The lower panel shows the same interpolated in orbital phase
and portrayed as a gray scale image (black corresponding to deepest
absorption and white to strongest continuum flux).  The actual
phases of observation are indicated by arrows on the right hand side.}
\end{figure}

\clearpage

We also observe several variations on the basic shell feature theme. 
Some of the weaker, lower opacity lines of low ionization species
only display a limited shell feature very close to the phase of maximum
absorption depth observed in \ion{Si}{2} $\lambda 1264$, $\phi=0.95$.
Examples include \ion{S}{2} $\lambda 1250$ (Fig.\ 3.3), 
\ion{Si}{2} $\lambda 1304$ (Fig.\ 3.7), and 
\ion{Si}{2} $\lambda 1526$ (Fig.\ 3.11).
The strong resonance lines formed in the stellar wind of the Be star are 
\ion{Si}{4} $\lambda 1393$ (Fig.\ 3.9), \ion{Si}{4} $\lambda 1402$ (Fig.\ 3.10), 
and \ion{C}{4} $\lambda\lambda 1548, 1550$ (Fig.\ 3.12).   
These lines have blue absorption wings that extend to $-600$ km~s$^{-1}$.
The cores of the wind lines also show evidence of the shell component 
(particularly on the less blended, positive velocity side of the line core)
that is is relatively strong over the phase range $\phi=0.6$ to 0.1.  
{\bf 
It is striking that the shell components appear over a wide range of atomic 
ionization states (for example, \ion{Si}{2}, \ion{Si}{3}, and \ion{Si}{4}).
This may be the result of the ionizing flux of the hot companion (Section 3)
and/or shock heating related to gas flows in the vicinity of the companion
(Section 5). 
}
Finally we observe another kind of blue-shifted, shell absorption feature around 
the opposite conjunction at $\phi=0.5$ that extends to $V_r = -80$ km~s$^{-1}$
in the \ion{C}{2} $\lambda\lambda 1334, 1335$ doublet (Fig.\ 3.8). 
This component was first noted by \citet{paterson1980} who 
suggested that it might originate in outflow through the 
external Lagrangian point in the direction away from the companion. 
This blue-shifted shell component is also observed in the lines 
\ion{Al}{3} $\lambda 1854$ (Fig.\ 3.14), \ion{Al}{3} $\lambda 1862$ (Fig.\ 3.15), 
and \ion{Fe}{3} $\lambda 1895$ (Fig.\ 3.16). 

% 3.1  f1193 Si II 1193.290 + S III 1194.061,1194.457
% 3.2  f1200 S III 1200.970, 1201.730 + ISM N I 1199.5,1200.2,1200.7
% 3.3  f1250 S II  1250.583
% 3.4  f1260 Si II 1260.4223 + S II 1259.518
% 3.5  f1264 Si II 1264.73 
% 3.6  f1299 Si III 1298.96, 1296.73,1298.89,1301.15
% 3.7  f1304 Si II 1304.3711  
% 3.8  f1335 C II  1335.708, 1334.532
% 3.9  f1393 Si IV 1393.755  wind
% 3.10 f1402 Si IV 1402.770  wind
% 3.11 f1526 Si II 1526.7071
% 3.12 f1548 C IV  1548.202,1550.774
% 3.13 f1670 Al II 1670.7867
% 3.14 f1854 Al III 1854.7164
% 3.15 f1862 Al III 1862.7895
% 3.16 f1895 Fe III 1895.46, 1895.46,1896.814
% 3.17 f2796 Mg II  2796.352
% 3.18 f2803 Mg II  2803.531

% AQ ionization energies
% C II   11.260 eV  reg  140  b0.5*
% C III  24.383     reg?      omit?
% C IV   47.888     reg  100
% Mg II   7.646     reg  140
% Al II   5.986     reg  130
% Al III 18.829     reg  140  b0.5:
% Si II   8.152 nar+reg  160  b0.0
% Si III 16.346     reg  120  b0.0
% Si IV  33.493   W+reg  120
% S II   10.360    *nar   90
% S III  23.338     reg  120  b0.0
% Fe III 16.188     reg  140  b0.5

\citet{peters1983} originally discussed the shell line appearance in the 
optical H Balmer lines, so for completeness, we show the orbital variations 
of the H$\alpha$ emission line in Figure~4.  This figure was constructed 
from the KPNO observations (Table~1) after deletion of several spectra 
where the emission was weaker than average.  The overall appearance 
is similar to that reported in the past \citep{hanuschik1996, pollmann2007}.
The variations are less pronounced in H$\alpha$ presumably because it forms 
over a large volume in the Be star disk, but the {\it primary shell phase} of increased, 
red-shifted absorption is clearly evident near $\phi=0.95$.  There are also hints of 
moving absorption sub-features that begin blue-shifted near $V_r=-100$ km~s$^{-1}$
at phases 0.0 and 0.5 that progress to line center half an orbit later. 
We discuss the possible origin of all these features in the next section.  

%\clearpage

% Figure 4

\setcounter{figure}{3}
\renewcommand{\thefigure}{\arabic{figure}}
\begin{figure}
\begin{center}
{\includegraphics[angle=0,height=16cm]{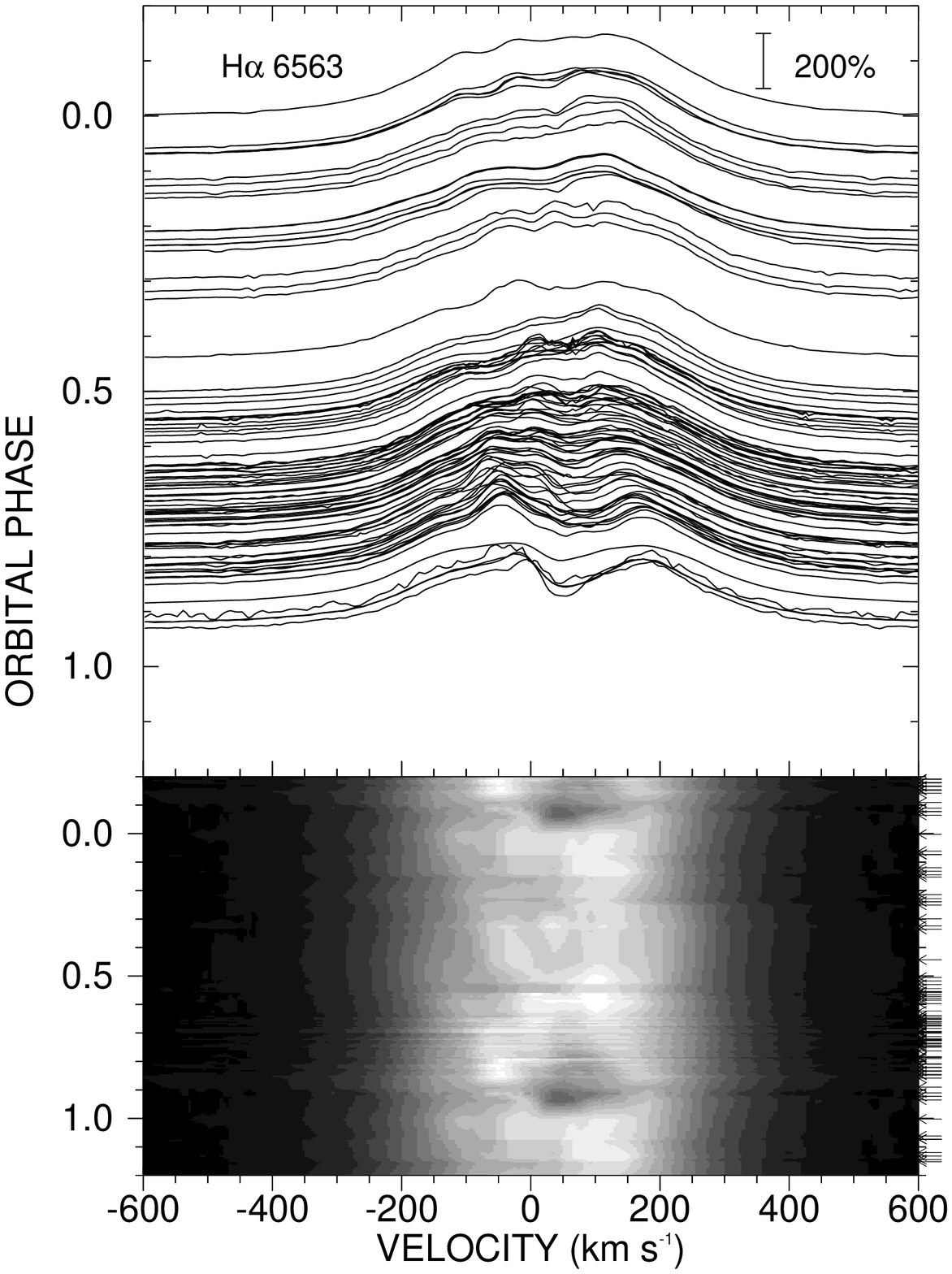}}
\end{center}
\caption{The top panel shows the individual spectra of
H$\alpha$ plotted as a function of radial
velocity and orbital phase (such that the continuum is aligned
with the phase of observation).  The spectral flux depth relative
to the continuum is indicated by the scale bar in upper right.
The lower panel shows the same interpolated in orbital phase
and portrayed as a gray scale image (black corresponding to lowest
absorption and white to strongest continuum flux).  The actual
phases of observation are indicated by arrows on the right hand side.
\label{fig4}}
\end{figure}

\clearpage

%%%%%%%%%%%%%%%%%%%%%%%%%%%%%%%%%%%%%%%%%%%%%%%%%%%%%%%%%%%%%%%%%%%%%%%%%%%%

\section{Circumbinary Disk Model}                                % Section 5

The fact that the orbital variations of the shell lines peak in strength near 
the conjunction phases indicates an origin related to the presence of 
the companion star.  \citet{peters1983} originally suggested that the 
companion was large, filling its Roche lobe, and that the primary shell
phase absorption occurred in a gas stream from the companion to the Be star. 
However, our detection of the FUV spectrum of the companion (Section 3)
shows that the companion is much smaller than its Roche lobe. 
Thus, here we present a different model that is based upon the idea 
that the companion creates a radial gap in the disk of the Be star and 
that the shell absorptions\footnote{In this discussion we focus on the 
phase-dependent variations in the shell lines.  A corresponding 
broad underlying absorption is also observed in some lines which 
may be formed through another mechanism.} 
are formed through the opacity of 
gas streams that are crossing the gap.  We refer to this scenario
as a circumbinary disk model in which the net outward motion of the 
disk gas extends beyond the dimensions of the binary. 

The elements of the circumbinary disk model are sketched in 
a polar view diagram of the system in Figure~5.  
We set the dimensions of the binary using the projected semimajor axis 
$a_1 \sin i$ from Table~3, the mass ratio $q = 0.07 \pm 0.02$
from Section~3, and an orbital inclination of $i=85^\circ$
(high enough to permit structures close to the disk to be projected 
against the disk of the Be star from our line of sight).  This results 
in a semimajor axis of $a = 0.82$~AU ($176 R_\odot$).  
Figure~5 shows the Be star at the origin (outlined for a radius of 
$5 R_\odot$) with the tiny companion at position $(x,y)=(+a,0)$.
Small arrows from each component indicate their relative orbital motion, 
although this is so small for the Be primary that only the arrowhead shows.  
The small plus sign on the axis joining the components marks the center of 
mass position. Tick marks appear on the periphery at intervals of $a/2$, 
and orientations of the view from Earth are marked by the orbital phases 
near the edges. The dashed line indicates the Roche lobes, 
and both stars are well within their respective Roche lobe boundaries. 

%\clearpage

% Figure 5

\begin{figure} 
\begin{center} 
{\includegraphics[angle=90,height=12cm]{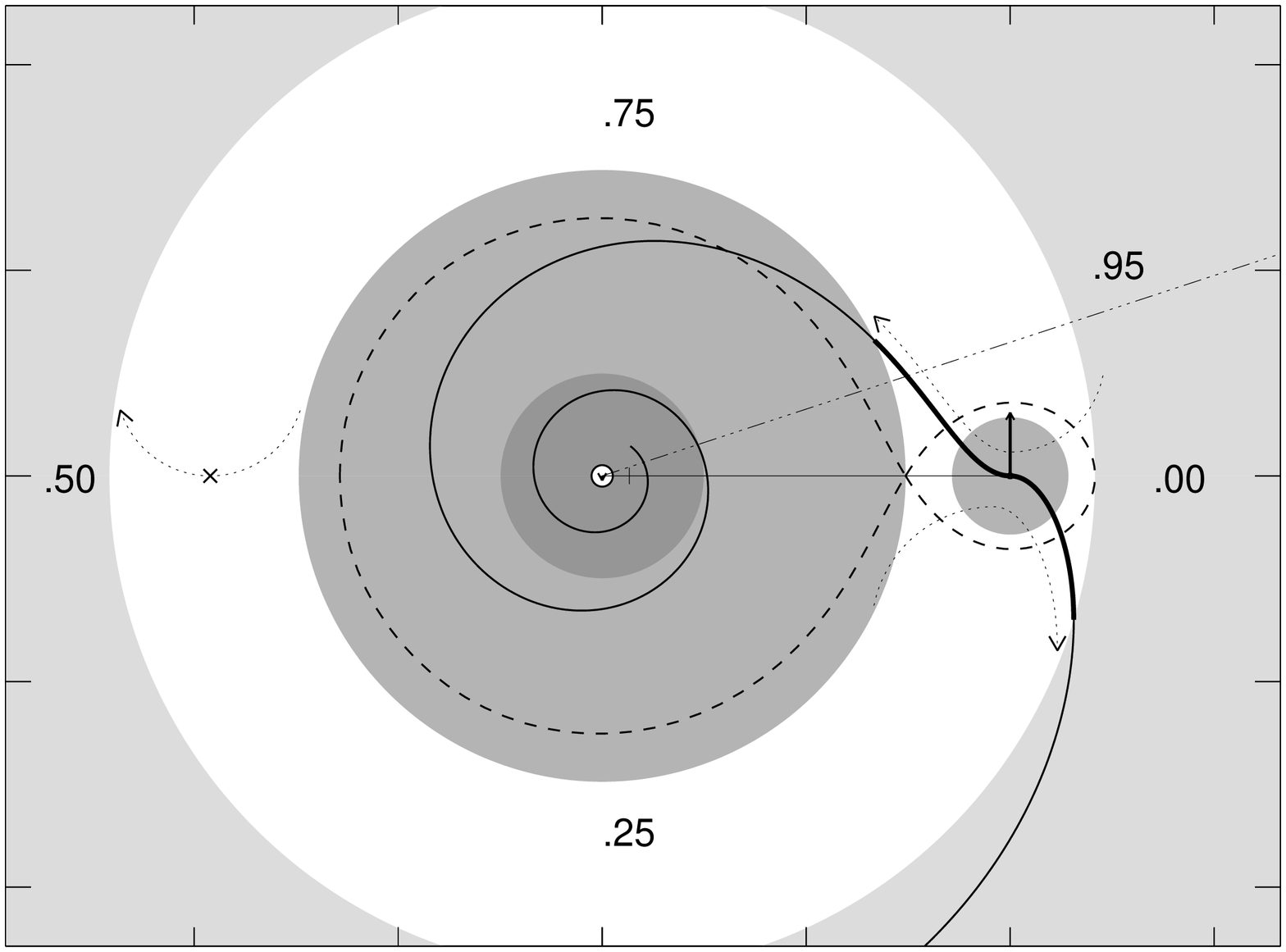}} 
\end{center} 
\caption{A sketch of the main components of HR~2142 as viewed from 
above the orbital plane. Tick marks on the axes occur at spacings of 
half the semimajor axis. The white circle immediately left of the 
center of mass (plus sign) represents the Be star, and the surrounding 
dark and medium gray zones represent the dense and extended regions of
its circumstellar disk.  The medium gray zone to the right represents
the accretion disk of the hot companion, while the outer light gray 
region represents the circumbinary disk.  The dashed line shows the 
Roche lobe boundaries, and we assume that the companion forms a gap 
in the disk that is equal in size to the long axis of the companion's
Roche lobe.  The spiral lines show the tidal shock location from 
the analytical expression from \citet{ogilvie2002}.  Dotted lines 
with arrows indicate suggestive flowlines across the gap in the vicinity
of the companion (right) and L3 (left, shown by a cross sign).  
Orbital phases for the direction of our line of sight are indicated 
around the periphery, and the dashed-dotted line shows the sight line 
to the Be star at $\phi=0.95$ when the shell lines are strongest. 
\label{fig5}} 
\end{figure}

The shaded regions indicate the Be star disk.  The darkest
gray portion close to the Be star illustrates the $9.5 R_\star$ 
boundary that we derive from the mean H$\alpha$ equivalent width,
$W_\lambda = -31$~\AA , and the method of \citet{grundstrom2006}. 
This radius represents the disk half-width at half-maximum of the intensity as
seen in the plane of the sky.  The disk extends beyond this 
radius with lower density and lower surface brightness (shown as
the medium gray shaded region).  

We next assume that the companion creates a gap in the disk in the same 
way as do planets in protoplanetary disks \citep{bate2003}. 
We adopt the simplest approximation that the size of the gap 
is equal to the companion's Roche lobe dimensions along the 
axis joining the stars.  The white region in Figure~5 illustrates the gap.  
Beyond the gap, we imagine that the disk gas accumulates and continues again 
with lower density and an outwardly directed flow (shown as the light gray 
region at the periphery of the diagram). 

Low mass companions orbiting within disks create tidal shocks in 
the disk gas through their gravitational perturbation of the 
Keplerian motion \citep{ogilvie2002,bate2003}.  We calculated 
the position of a one-armed spiral wake using the two-dimensional
approximation given by \citet{ogilvie2002} (see their equations 
13 and 24 for the outer and inner portions, respectively). 
These curves depend only on a parameter $\epsilon$ that is 
equal to the ratio of the gas sound speed to the circular 
Keplerian velocity of the companion around the primary. 
We assumed a sound speed of 14 km~s$^{-1}$ based upon an 
adopted disk gas temperature of 15~kK ($\approx 0.6 T_{\rm eff}$) 
and a mean molecular weight of $\mu = 0.66$.  Then for a 
companion velocity of $V({\rm relative}) \approx 100$ km~s$^{-1}$, 
we find $\epsilon=0.12$ (close to the $\epsilon=0.1$ case 
shown in Fig.\ 5 of \citealt{ogilvie2002}).  The tidal wakes are 
shown as solid lines that cross through the position of
the companion star.  We extended the tidal wake curve well into
the interior of the Be star disk, but we suspect that the 
wake density enhancement drops closer to the Be star 
because of gas turbulence in the disk.

The thickest portion of the wake marks those sections 
where we expect the density enhancements are the greatest 
and, hence, where the gas is located that may cause the shell
line episodes.   The dotted lines above and below the 
companion show the flow lines of gas crossing the gap due
to the gravitational pull of the companion.  These are 
parts of so-called ``horseshoe orbits'' that are well 
known in the restricted three-body problem \citep{dermott1981}.    
For example, gas at the outer edge of the gap ahead of the companion
will be pulled inwards by the companion's gravity.  Then 
by conservation of angular momentum, the azimuthal speed 
around the Be star increases, so the gas moves ahead of 
the companion and forms the wake at the inner edge of the 
gap.  In a similar way, gas at the inner edge of the gap 
that is approaching the companion (below it in the diagram) 
will be pulled outwards toward the companion.  Here  
conservation of angular momentum causes this gas to slow 
in azimuthal speed, and it falls behind the companion to 
strike the outer gap edge at the outer wake.   
Small arrows attached to each dotted line indicate the 
relative sense of motion in the frame of the binary. 

Thus, we suspect that the main red-shifted shell feature
that we see around phase $\phi = 0.95$ is due to our view 
of the Be star as seen through 
the inner tidal wake where gas is moving inward towards the 
Be star (and away from us).  This is probably the strongest
and longest duration shell phase because the gas is striking
more dense gas in the inner region and over a larger wake 
angle relative to the Be star.  Then the second blue-shifted
shell phase occurs for a short time after phase 0.0 as 
we view the Be star flux through the outward moving gas 
in the outer wake.  The portions of the wake 
that are most important for the shell line formation are 
indicated with extra line thickness. 

We suggest that the {\it primary shell phase} absorptions occur as 
we view the Be star through inward crossing gas that creates shocks 
and high density regions in the vicinity of the tidal wake
spiral arm.  Beginning shortly after orbital phase $\phi=0.5$ 
we see the tidal wake gas projected against the Be star with 
a modest velocity towards the Be star (and away from us), and
as the companion moves into the foreground, we observe progressively
larger inflow velocities that peak near $\phi=0.95$ (shown by 
the dashed-dotted line of sight in Fig.~5).  This line of sight corresponds 
approximately to the location of the intersection of the tidal wake 
with the outer boundary of the inner disk.  Three-dimensional 
simulations of gas motions in the vicinity of the tidal wake 
and companion by \citet{bate2003} indicate that the inner shock 
region may attain a significant distance above the orbital plane
that amounts to several disk scale heights.  Consequently, 
we expect that shock structures in the disk could be seen projected against 
the Be star even if the orbital inclination is somewhat less than $90^\circ$.
 
The models by \citet{bate2003} show that some gap crossing gas will
flow into an accretion disk around the companion, and we have 
indicated a possible accretion disk around the companion as the 
medium shaded gray circle.  Such an accretion disk may partially 
account for the apparent obscuration of the flux of the hot companion. 
Inward crossing gas ahead of the companion
might encounter the outer regions of the accretion disk and attain
a velocity comparable to the local Keplerian motion around the companion. 
For example, at a distance of $0.08 a$ ahead of the companion, the 
accretion disk orbital velocity would be $\approx 100$ km~s$^{-1}$
directed towards the Be star, which is comparable to the observed 
maximum red-shift velocity of the shell lines. 

We suspect that the inner disk gas crosses the gap to the outer circumbinary disk 
along the axis joining the stars where the gravitational potential is most favorable. 
The {\it secondary shell phase} that occurs immediately after orbital phase
$\phi = 0.0$ probably results as we view the Be star through outwardly 
moving (blue-shifted) dense gas at the intersection of the tidal wake and 
the inner boundary of the outer disk.  Similarly, the short episode of 
blue-shifted absorption observed near phase $\phi = 0.5$ may result from 
gas that crosses the gap near the L3 Lagrangian point and forms a 
density enhancement where it strikes the outer disk.   We speculate that
the weak, blue-shifted and migrating sub-features in H$\alpha$ that begin at each 
conjunction originate in tidal wakes in the outer circumbinary disk.  

{\bf
Our schematic model in Figure~5 must be only an approximation of the 
actual disk structure, and fortunately there are a number of numerical 
simulations by \citet{okazaki2002} and \citet{panoglou2016} that use three-dimensional,
smoothed particle hydrodynamical codes to model the disks of Be stars 
in binary systems.  These show that the presence of the companion 
creates a disk truncation radius that is somewhat smaller than the 
Roche radius of the Be star $R_L$, i.e., between $0.7 R_L$ for truncation at 
the 4:1 period resonance ratio \citep{okazaki2002} and $0.8 R_L$ for truncation 
at the 3:1 period resonance ratio \citep{panoglou2016} depending on the disk viscosity.  
Consequently, the inner gap boundary plotted in Figure~5 may be somewhat large. 
The inner tidal wake shown in Figure~5 is also found in the circular orbit
simulations of \citet{panoglou2016} (described there as a spiral arm), and the 
geometry of wake in the simulations (for a ratio of sound speed to companion 
orbital velocity of $\epsilon = 0.08$) agrees well with the analytical 
formula (compare Fig.~4 of \citealt{panoglou2016} with Fig.~5 of \citealt{ogilvie2002}).
However, the numerical models do not show evidence of an outer disk, 
i.e., instead of a gap there is simply one truncation of the Be star disk. 
This is partially due to the nature of the simulations that are focused on 
the inner structure of the disk and not on the gas flows in the vicinity 
of the secondary (where the timescales are too short for efficient 
numerical resolution).  Nevertheless, we anticipate that future simulations
will document the outward leakage of disk gas, which carries away the angular 
momentum shed by the Be star, in such a way that gas is preferentially lost along 
the axis joining the stars into a circumbinary disk (for example, as is 
predicted in Roche lobe overflow systems; \citealt{nazarenko2006}). 
}

%%%%%%%%%%%%%%%%%%%%%%%%%%%%%%%%%%%%%%%%%%%%%%%%%%%%%%%%%%%%%%%%%%%%%%%%%%%%

\section{Conclusions}                                            % Section 6

Our investigation of the UV and optical spectrum of HR~2142 has led to a 
new understanding of the nature of this binary system and its circumbinary disk. 
We used the revised spectroscopic orbit of the Be star to establish the 
orbital geometry and to predict the radial velocity of the faint companion 
at the times of observation.  These predicted Doppler shifts were applied 
to cross-correlation functions of the far-UV spectra of HR~2142 with a
model spectrum for a hot companion, and the shift-and-added CCFs reveal 
a faint peak caused by the spectral features of the companion.  
The companion is a hot and small subdwarf, the stripped down remains of 
the originally more massive star in this system that transferred both
mass and angular momentum to the Be star.  Our current results 
lead to mass estimates of $9 M_\odot$ and $0.7 M_\odot$ for the 
Be star and hot companion, respectively.  Future observations of the
orbital motion of the distant tertiary companion of HR~2142 
\citep{esa1997,horch2002,oudmaijer2010} will help establish the total system mass.

HR~2142 is now the fourth Be+sdO system where UV spectroscopy has led
to the detection of the spectral contribution of the hot companion  
\citep{gies1998,peters2008,peters2013}.  The hot subdwarf is small, 
$R_2/R_\odot > 0.13$, and it has a relatively low luminosity, 
$\log L/L_\odot > 1.7$.  In fact, the companion may have a mass and 
radius typical of a white dwarf \citep{provencal1998}, although it is
much hotter than most.  Binary evolutionary models may help determine 
if the hot companion of HR~2142 is still in a core He burning stage 
or some more advanced stage (A.\ Schootemeijer et al., in preparation). 

We have documented the orbital variations of the shell line absorptions in 
the UV spectrum and H$\alpha$, and we suggest that these are formed mainly 
in dense gas regions where a tidal wake caused by the companion intersects
with boundaries of a gap in the disk of the Be star.  Shocks are formed 
in these regions as gas crossing the gap strikes the dense gas of the disk, 
and the resulting shock heating helps to explain the presence of the 
shell components in high ionization species like \ion{Si}{4} and \ion{C}{4}.
It may be possible in the future to detect the gap in the disk and the 
tidal wake regions through high angular resolution observations with 
optical long baseline interferometry. 
{\bf Furthermore, the heated tidal 
wake regions may be responsible for the emission line formation in the vicinity
of the companion that is observed in some Be+sdO binaries (in addition  
to direct heating by the flux of the companion; \citealt{hummel2001}). 
The \ion{He}{1} $\lambda 6678$ emission line shows evidence of localized
heating in the cases of $\phi$ Per \citep{stefl2000}, FY~CMa \citep{peters2008},
and 59~Cyg \citep{peters2013}, so we might expect similar emission variations
in the spectrum of HR~2142.  There is some evidence of a variable emission 
component that fills in the wide absorption profile of \ion{He}{1} $\lambda 6678$
in our spectra, but it is much weaker than observed in the other Be+sdO 
systems, perhaps due to the lower luminosity of the subdwarf. 
}

{\bf
It was the remarkable orbital variations in the shell line spectrum that first 
drew our attention to HR~2142 \citep{peters1971,peters1972}.  If the circumbinary
disk model of gap crossing gas is applicable to other Be+sdO binaries, then 
we might naively expect that they should also display such shell line 
variations in their spectra.  However, to our knowledge the appearance 
of shell lines near Be star superior conjunction has only been reported 
for the sudden appearance of \ion{Fe}{4} lines in the FUV spectrum of 
$\phi$~Per \citep{gies1998}.  We speculate that tidal wakes may occur in 
other Be binaries, but that they only present observable shell lines if 
(1) the Be star's disk is particularly massive and has substantial density 
out to the vicinity of the companion, (2) the orbital inclination is close 
to $i=90^\circ$ so that the disk perturbations caused by the tidal wakes reach
to a vertical displacement that occults our line of sight to the Be star, and 
(3) the companion is not so massive that it creates a gap that is too wide 
to facilitate gas migration across the gap.  HR~2142 may represent a
fortuitous conjunction in which all three conditions appear to be met. 
} 
The example of HR~2142 illustrates the importance of binary Be stars as 
laboratories for investigations of disk dynamical processes.

%%%%%%%%%%%%%%%%%%%%%%%%%%%%%%%%%%%%%%%%%%%%%%%%%%%%%%%%%%%%%%%%%%%%%%%%%%%%

\acknowledgments

Some of the data presented in this paper were obtained from the 
Mikulski Archive for Space Telescopes (MAST). STScI is operated by the 
Association of Universities for Research in Astronomy, Inc., under 
NASA contract NAS5-26555. Support for MAST for non-HST data is provided 
by the NASA Office of Space Science via grant NNX09AF08G and by other 
grants and contracts.
This work has made use of the BeSS database, operated at LESIA, 
Observatoire de Meudon, France: http://basebe.obspm.fr.
Our work was supported in part by NASA grant NNX10AD60G (GJP) and 
by the National Science Foundation under grant AST-1411654 (DRG). 
Institutional support has been provided from the GSU College
of Arts and Sciences, the Research Program Enhancement
fund of the Board of Regents of the University System of Georgia
(administered through the GSU Office of the Vice President
for Research and Economic Development), and by the USC Women in
Science and Engineering (WiSE) program (GJP).  

%%%%%%%%%%%%%%%%%%%%%%%%%%%%%%%%%%%%%%%%%%%%%%%%%%%%%%%%%%%%%%%

Facilities: \facility{IUE, KPNO:CFT}

%%%%%%%%%%%%%%%%%%%%%%%%%%%%%%%%%%%%%%%%%%%%%%%%%%%%%%%%%%%%%%%

% Bibliography

\bibliographystyle{apj}
\bibliography{apj-jour,ms}

%%%%%%%%%%%%%%%%%%%%%%%%%%%%%%%%%%%%%%%%%%%%%%%%%%%%%%%%%%%%%%%

\end{document}